\documentclass[12pt]{article}
\usepackage{jheppub}
\pdfoutput=1
\usepackage{amsmath,bbm,array,amsfonts,graphicx,wrapfig,lscape,float,mathtools,multirow,longtable,amsthm}
\usepackage[dvipsnames]{xcolor}
\usepackage{stackrel}
\usepackage[all]{xy}
\usepackage{graphicx}
\usepackage{caption}
\usepackage{subcaption}
\usepackage[bottom]{footmisc}
\usepackage{multirow}
\usepackage{mathrsfs}
\usepackage{tikz}
\usepackage{bm}
\usepackage{color}
\usepackage{enumerate}
\usetikzlibrary{decorations.pathreplacing}
\usepackage{diagbox}
\numberwithin{equation}{section}
\numberwithin{figure}{section}
\numberwithin{table}{section}
\usepackage{float}
\usepackage{pgfplots}
\pgfplotsset{compat=1.14}
\usepackage{longtable}
\usepackage{makecell}
\usepackage{pdflscape}
\usepackage[utf8]{inputenc}

\usepackage[autostyle=true]{csquotes}
\newtheorem{definition}{Definition}[section]
\newtheorem{theorem}{Theorem}[section]

\newtheorem{lemma}[theorem]{Lemma}
\newtheorem{proposition}[theorem]{Proposition}
\newtheorem{conjecture}[theorem]{Conjecture}

\usepackage[toc,page]{appendix}

\def\i{\iota}

\newenvironment{dedication}
{
	\cleardoublepage
	\thispagestyle{empty}
	\vspace*{\stretch{1}}
	\hfill\begin{minipage}[t]{0.66\textwidth}
		\raggedright
	}%
	{
	\end{minipage}
	\vspace*{\stretch{3}}
	\clearpage
}

	\title{Dessins d'Enfants, Seiberg-Witten Curves and Conformal Blocks}
	
	\author[a]{Jiakang Bao,}
	\author[b]{Omar Foda,}
	\author[a,c,d,e]{Yang-Hui He,}
	\author[a]{Edward Hirst,}
	\author[f]{James Read,}
	\author[g]{Yan Xiao,}
	\author[h]{Futoshi Yagi}
	
	\affiliation[a]{
		Department of Mathematics, City, University of London, EC1V 0HB, UK}
	\affiliation[b]{
		School of Mathematics and Statistics, University of Melbourne, Royal Parade, Parkville, VIC 3010, Australia}
	\affiliation[c]{
		Merton College, University of Oxford, OX1 4JD, UK}
	\affiliation[d]{
		London Institute of Mathematical Sciences, 35a South St, Mayfair, London, W1K 2XF, UK}
	\affiliation[e]{
		School of Physics, NanKai University, Tianjin, 300071, P.R. China}
	\affiliation[f]{
	Pembroke College, University of Oxford, OX1 1DW, UK}
    \affiliation[g]{
	Department of Physics, Tsinghua University, Beijing, 100084, China}
    \affiliation[h]{
	School of Mathematics, Southwest Jiaotong University, West Zone, High-Tech District, Chengdu, Sichuan, 611756, China}
	
	\emailAdd{jiakang.bao@city.ac.uk}
	\emailAdd{omar.foda@unimelb.edu.au}
	\emailAdd{hey@maths.ox.ac.uk}
	\emailAdd{edward.hirst@city.ac.uk}
	\emailAdd{james.read@pmb.ox.ac.uk}
	\emailAdd{steven1025xiao@gmail.com}
	\emailAdd{futoshi\_yagi@swjtu.edu.cn}
	
	\preprint{
		\begin{flushright}
			
		\end{flushright}
	}

	\abstract{We show how to map Grothendieck's dessins d'enfants to algebraic curves as Seiberg-Witten curves, then use the mirror map and the AGT map to obtain the corresponding 4d $\mathcal{N}=2$ supersymmetric instanton partition functions and 2d Virasoro conformal blocks. We explicitly demonstrate the 6 trivalent dessins with 4 punctures on the sphere. We find that the parametrizations obtained from a dessin should be related by certain duality for gauge theories. Then we will discuss that some dessins could correspond to conformal blocks satisfying certain rules in different minimal models.
	}

\begin{document}
	\maketitle
	
	\begin{dedication}
		\begin{flushright}
		\LARGE{Dedicated to the memory of\\
			our dear friend,\\
			Professor Omar Foda,\\
			a gentleman and a scholar $\dots$}
		\end{flushright}
	\end{dedication}

\section{Introduction and Summary}\label{intro}
Consider a 4-point conformal block (CB) in a 2d conformal field theory (CFT) based on $\mathcal{W}_2\times\mathcal{H}$, where $\mathcal{W}_2$ is the Virasoro algebra and $\mathcal{H}$ is the Heisenberg algebra. Using the Alday-Gaiotto-Tachikawa (AGT) correspondence \cite{Alday:2009aq}, this is identified with an instanton partition function in an $\mathcal{N}=2$ supersymmetric Yang-Mills (SYM) theory, with an SU(2) gauge group and four fundamental hypers.

The low energy physics of this gauge theory is described in terms of a Seiberg-Witten (SW) curve and the SW differential on it \cite{Seiberg:1994rs,Seiberg:1994aj}. Then in \cite{Nekrasov:2002qd}, a method of instanton counting was introduced to find these low energy solutions to SW theories. Later, the S-duality for $\mathcal{N}=2$ supersymmetric systems was studied in \cite{Gaiotto:2009we}. In recent works \cite{He:2012kw,He:2012jn,He:2013lha,He:2015vua,He:2015yoa,Ashok:2006br,He:2020eva}, connections between Grothendieck's dessins d'enfants on the one hand and 4d $\mathcal{N}=2$ SYM on the other were studied. In this note, we further explore these connections and extend them to 2d conformal field theory. We focus on six specific trivalent dessins with 4 punctures on the sphere, which, as we will see, are related to a simple and important class of 4d $\mathcal{N}=2$ SYM theories and conformal blocks in 2d conformal field theory.

From these dessins, we obtain algebraic curves that we interpret as SW curves of 4d SU(2) $\mathcal{N}=2$ $N_f=4$, SYM theories. These curves are given in terms of six parameters, four mass parameters $(\mu_1,\mu_2,\mu_3,\mu_4)$, a parameter $\zeta$ and a modulus $U$. We write these curves in the form that appears in \cite{Eguchi:2009gf,Kozcaz:2010af}, and use their mirror map to translate the above parameters to those characterizing the 4d instanton partition function of a 4d $\mathcal{N}=2$ gauge theory. In particular, we map the modulus $U$ to the Coulomb parameter $a$. Following that, we use the AGT dictionary to interpret the result in 2d CFT terms.

Let us take a closer look at the six parameters for the SU(2) gauge theory. With $N_f=4$, the theory has an $\text{SO}(8)\supset\text{SU}(2)^4$ flavour symmetry. Then the mass parameters of the four hypers could be indentified as the charges of the primaries in Liouville theory under AGT correspondence \cite{Alday:2009aq}. Following \cite{Bershtein:2014qma,Alkalaev:2014sma}, the AGT map could also lead to diagonal minimal models by further restrictions on the partition pairs. As usual, we would arrange the poles of the SW curves at $z=0,1,\infty$ and $\zeta$. This $\zeta$ is nothing but the UV gauge coupling $\tau$ via $\zeta=\text{e}^{2\pi i\tau}$. For each dessin, we find that $\zeta$ could have several different values but these values enjoy certain triality.

Recall that the Coulomb parameter $a$ denotes the vev of the adjoint scalar $\phi$, or equivalently, $a$ could be obtained by integrating the SW differential along the so-called $A$-cycle on SW curve. Such supersymmetric vacua can be gauge invariantly parametrized by $u=\langle\text{tr}\phi^2\rangle/2=a^2$ up to quantum corrections. As we will discuss in \S\ref{topostring2SW}, the parameter $U$, which will appear in the parametrization of the curve, is linear in the Coulomb moduli $u$. In fact, as we will see, each dessin gives a family of solutions for the gauge theory parameters, and indeed, we would have the same corresponding dessin under the change $m_i\rightarrow km_i$, $a\rightarrow ka$, $U\rightarrow k^2U$ for $k^2\in\mathbb{R}$. This is consistent with their mass dimensions.

The above discussions can go the other way as well. Starting from the CBs in CFTs, we can write down the Nekrasov partition functions under the AGT dictionary. This 4d partition function can also be lifted to 5d, which leads to topological string partition functions and SW curves. As the SW differential and the Strebel differential from the dessin side are both quadratic, the gauge theories are naturally related to dessins.

Since the instanton partition functions with extra conditions on the Young tableaux pairs could be mapped to conformal blocks in minimal models \cite{Bershtein:2014qma,Alkalaev:2014sma}, we can then check whether (the parametrizations from) the dessins could correspond to such CBs in minimal models. As we will see in \S\ref{Gamma3mm}$\sim$\S\ref{general}, such map is not one-to-one. A dessin could correspond to one or more possible CBs in multiple minimal models. These CBs, albeit in different minimal models, would satisfy certain (fixed) rules for the dessin. There might also exist dessins that do not give rise to minimal models. As we are focusing on SU(2) gauge theory with 4 flavours, we will show that
\begin{proposition}
	There is a subset of trivalent dessins with four punctures on the sphere such that each dessin therein corresponds to the (external and internal) states of a family of 4-point conformal blocks for (A-series) minimal models.
\end{proposition}
As we will discuss in \S\ref{general}, we may also conjecture that these dessins would contain the full information for certain CBs. In principle, there are countably infinite such dessins although only a small part of them have been studied in details. As we will show, not all the dessins would give minimal models. For those in the above subset, we will also determine the families of CBs they correspond to. In particular, we will illustrate this proposition with six well-known dessins as examples. Notice that we can only say that this correspondence is not for all dessins. On the minimal model side, it is still not determined whether this subset of dessins could recover all the CBs or just part of them.

Now that the dessins are rigid, it would be interesting to understand whether the parametrizations from these dessins are special for gauge theories and (R)CFTs in future. It would also be natural to extend this for dessins with more faces and other gauge theories. So far, we only get the values for $\zeta$ from the dessins which are related by triality as briefly mentioned above, and the calculations should be non-perturbative. More details are worth studying for future works. It would be nice to even know more about the bulk of AdS through holographic duality. On the other hand, we would like to know whether the physical theories could in turn help us learn more about the dessins. For example, we know that not all dessins can give CBs in minimal models, but what kind of dessins have this correspondence is still unknown. Moreover, the Gorenthedieck-Teichmuller group, which is related to the Galois group $\Gamma(\bar{\mathbb{Q}}:\mathbb{Q})$ and dessins, may be related to the monodromy of CBs in RCFT \cite{Degiovanni:1994gr}. This might lead to deeper connections between the mathematics and the physics.

The paper is outlined as follows. In \S\ref{CB2dessin}, we start from the CFT side and review the AGT correspondence to get the corresponding partition functions. Then from A-model topological strings, we obtain the SW curve for SU(2) with 4 flavours and thence the dessins. In \S\ref{dessin2CB}, we reverse the discussion and contemplate six of the dessins which would yield specific parametrizations for SW curves. Then we will study if these parametrizations would give conformal blocks in minimal models following the AGT dictionary. In the appendices, we give some background on brane systems as well as elliptic curves and elliptic functions.

This work is dedicated to the memory of Professor Omar Foda, who was instrumental in initiating this project.

\section{From Conformal Blocks to Dessins d'Enfants}\label{CB2dessin}
Before we derive the results in 2d CFT from the 6 dessins with 4 punctures on the sphere, we give a brief review of different subjects including CBs, topological strings, SW curves and dessins, following a route map from CBs to dessins.

\subsection{From 2d Conformal Blocks to 4d Instanton Partition Functions}\label{CB2Zinst}
The connection between 2d Liouville CFTs and SU(2) supersymmetric gauge theories in 4d with $\mathcal{N}=2$ was first raised in \cite{Alday:2009aq}. The free parameters of the two areas are naturally mapped to each other under AGT correspondence. Later in \cite{Bershtein:2014qma,Alkalaev:2014sma}, the AGT correspondence was also extended to minimal models by further restrictions on the partitions/Young diagrams. We first start with the CFT side under the Coulomb-gas formalism.

\paragraph{Conformal Blocks} Conformal blocks form a basis of the vertex operator (VO) algebra, used when performing a particular operator product expansion (OPE) of a correlation function. They are a key ingredient in the conformal bootstrap approach to calculating these correlators in 2d CFTs. Global conformal Ward identities of the CFT allow 2-point functions to be completely determined, whilst 3-point functions to have fixed results up to their respective structure constants. Thus when calculating an $N\geq4$-point function the recursion of applying OPEs allows expression of the correlator in terms of these simpler 3-point function structure constants, and conformal blocks.

More specifically, an OPE amounts to summation over all representations of the vertex operator algebra. In the common case where this algebra factorises into two Virasoro algebras the sum includes all combinations of the left and right Virasoro algebras' representations for the CFT.
Each term in the sum has a product of two conformal blocks, one in each of the term's left and right representations respectively.

Conformal blocks in general are sums over the states in their representation, they're functions of the fields' positions \& conformal dimensions, the conformal dimension of the basis expansion, and the central charge of the algebra. Blocks over primary fields have simpler properties, whilst those including descendent fields can be determined with use of local Ward identities. If the correlator in question includes a degenerate field then BPZ equations need to be enforced, which can simplify conformal block computations \cite{Poland:2018epd}.

In the classic example of 4-point functions, the global Ward identities allow M\"obius transformation, mapping 3 of the 4 complex coordinates to $\{0,\zeta,1,\infty\}$, leaving a single `cross-ratio' coordinate $\zeta$ for the conformal blocks to be a function of. Explicitly
\begin{align}
\langle\prod_{i=1}^4 V_i \rangle\; = \sum_{R,\overline{R}} C_{12R}C_{34\overline{R}} B_R(\zeta)B_{\overline{R}}(\zeta)\label{fourpt}
\end{align}
for Virasoro operators $V_i$. The sum is over the fields' representations in the left and right Virasoro algebras, denoted $R, \overline{R}$; with the sum including structure constants, $C_{ijk}$, and conformal blocks, $B$. These highest-weight representations of the Virasoro algebra are described in terms of Verma modules. They are generated by primary states and are irreducible in the absence of degenerate fields.

Since fields in a correlator can be permuted without change to the result, this translates into allowing different OPEs of the same correlator as different conformal block bases are used for expansion. The equivalence of these OPEs leads to `crossing symmetry' and introduces additional consistency constraints which allow structure constants and block dimensions to be calculated.
This is the conformal bootstrap methodology, and leaves calculation of conformal blocks as the final ingredient for computing correlators \cite{BELAVIN1984333}.

Conformal blocks are traditionally computed via Zamolodchikov recursion methods, however in the cases of degenerate fields in the correlators the BPZ equations provide a shortcut to finding them. In special cases these blocks can be expressed simply - for example a 4-point function on the sphere with one degenerate field (with a 2nd order null vector) can be expressed in terms of hypergeometric functions.

The AGT correspondence then makes a connection between the conformal dimensions of the fields in the correlator, and the coordinate $\zeta$, with parameters arising in Nekrasov instanton partition functions, as subsequently described. With the help of this correspondence we then compute the conformal blocks associated to the 6 dessins considered in this study.

\paragraph{The Nekrasov Partition Function} For generic vev $a$, the general $\mathcal{N}=2$ low energy effective action reads
\begin{equation}
	S_\text{eff}=\int\text{d}^4x\text{d}^4\theta\mathcal{F}(\Psi)(+\text{c.c.}),
\end{equation}
where $\Psi$ is the $\mathcal{N}=2$ V-plet, and the holomorphic function $\mathcal{F}$ is known as the \emph{prepotential}. First conjectured in \cite{Nekrasov:2002qd} and then proven in \cite{Nekrasov:2003rj}, the prepotential can be solved by
\begin{equation}
	\mathcal{F}=\lim\limits_{\epsilon_{1,2}\rightarrow0}\epsilon_1\epsilon_2\log Z_\text{Nek},
\end{equation}
where $\epsilon_i$'s are known as the deformation parameters, and $Z_\text{Nek}$ is the \emph{Nekrasov partition function}, which reads $Z_\text{Nek}=Z_\text{tree}Z_\text{1-loop}Z_\text{inst}$, where $Z_\text{tree/1-loop}$ is the tree/1-loop level partition function and $Z_\text{inst}$ denotes the contribution from instantons.

We will now focus on the instanton partition function $Z_\text{inst}$. For SU(2) quiver theories, the Coulomb branches are parametrized by the Coulomb moduli $\vec{a}=(a_1,a_2)=(a,-a)$. Each Coulomb modulus is associated with a Young tableau $Y$, in which every box is labelled by a pair $s=(i,j)$ to denote its position. Hence, the instanton partition function depends on $\vec{Y}=(Y_1,Y_2)$, the vev $a$, and possibly the mass $m$ of matter in the theory. Let us define \cite{Kozcaz:2010af}
\begin{equation}
	E(a,Y_1,Y_2,s)\equiv a-\epsilon_1L_{Y_2}(s)+\epsilon_2(A_{Y_1}(s)+1)
\end{equation}
with
\begin{equation}
	L_{Y_2}(s)=k_i-j,~A_{Y_1}(s)=k_j'-i,
\end{equation}
where $k_i$ is the length of $i^\text{th}$ row of $Y_2$, and $k_j'$ is the height of $j^\text{th}$ column of $Y_2$. Let $I,J$ label the gauge nodes. Then
\begin{eqnarray}
	&&z_\text{bifund}(a^I,\vec{Y}^I;a^J,\vec{Y}^J;m)\nonumber\\
	&\equiv&\prod_{i,j=1}^2\left(\prod_{s\in Y_i^I}(E(a_i^I-a_j^J,Y_i^I,Y_j^J,s)-m)\prod_{s\in Y_j^J}(\epsilon-E(a_j^J-a_i^I,Y_j^J,Y_i^I,s)-m)\right),\nonumber\\
\end{eqnarray}
where $\epsilon=\epsilon_1+\epsilon_2$. For (anti-)fundamentals,
\begin{equation}
	z_\text{fund}(\vec{a},\vec{Y},m)\equiv\prod_{i=1}^2\prod_{s\in Y_i}(\phi(a_i,s)-m+\epsilon),~z_\text{antifund}(\vec{a},\vec{Y},m)\equiv z_\text{fund}(\vec{a},\vec{Y},\epsilon-m),
\end{equation}
where $\phi(a_i,s)=a_i+\epsilon_1(i-1)+\epsilon_2(j-1)$. For adjoint chiral and vector multiplets,
\begin{equation}
	z_\text{adj}(\vec{a},\vec{Y},m)\equiv z_\text{bifund}(\vec{a},\vec{Y};\vec{a},\vec{Y};m),~z_\text{vec}(\vec{a},\vec{Y})\equiv\frac{1}{z_\text{adj}(\vec{a},\vec{Y},0)}.
\end{equation}

\paragraph{The AGT correspondence} The (chiral) VO can be written as $V_\alpha=:\text{e}^{2\alpha\phi}:$ for some free scalar $\phi$. If we introduce some background charge $Q$, by considering the OPE between stress tensor and the VO, we get the conformal dimension of $V_\alpha$, which reads $\Delta_\alpha=\alpha(Q-\alpha)$. Likewise, the OPE between stress tensors yields the central charge $c=1+6Q^2$.

Now we are ready to bridge the CBs and instanton partition functions. Originally, this was done for Liouville theory in \cite{Alday:2009aq}. We can fix the scale by setting $b=\frac{\epsilon_1}{\sqrt{\epsilon_1\epsilon_2}}$, where $Q=b+\frac{1}{b}$ and $b$ is the parameter coming from the Liouville potential. Therefore, $Q=\frac{\epsilon_1+\epsilon_2}{\sqrt{\epsilon_1\epsilon_2}}\equiv\frac{\epsilon}{\sqrt{\epsilon_1\epsilon_2}}$. Consider a quiver consisting of an SU(2) gauge group with 2 SU(2) antifundamentals and 2 SU(2) fundamentals with mass parameters $\mu_{1,2}$ and $\mu_{3,4}$ respectively. Then the instanton partition function reads
\begin{equation}
	Z_\text{inst}=\sum_{Y_{1,2}}\frac{\text{e}^{2\pi\text{i}\tau(|Y_1|+|Y_2|)}}{\left(1-\text{e}^{2\pi\text{i}\tau}\right)^{\frac{1}{2}(\mu_1+\mu_2)(2\epsilon-(\mu_3+\mu_4))}}z_\text{vec}(\vec{a},\vec{Y})z_\text{matter},\label{Zinst4flav}
\end{equation}
where the denominator correpsonds to the decoupling of a U(1) factor, and
\begin{equation}
	z_\text{matter}=z_\text{antifund}(\vec{a},\vec{Y},\mu_1)z_\text{antifund}(\vec{a},\vec{Y},\mu_2)z_\text{fund}(\vec{a},\vec{Y},\mu_3)z_\text{fund}(\vec{a},\vec{Y},\mu_4).
\end{equation}
The instanton number $|Y_i|$ is the number of boxes in $Y_i$. Then under the following AGT dictionary,
\begin{eqnarray}
	&&\frac{\mu_1}{\sqrt{\epsilon_1\epsilon_2}}=\alpha_2+\alpha_1-\frac{Q}{2},~
	\frac{\mu_2}{\sqrt{\epsilon_1\epsilon_2}}=\alpha_2-\alpha_1+\frac{Q}{2},~
	\frac{\mu_3}{\sqrt{\epsilon_1\epsilon_2}}=\alpha_3+\alpha_4-\frac{Q}{2},\nonumber\\
	&&\frac{\mu_4}{\sqrt{\epsilon_1\epsilon_2}}=\alpha_3-\alpha_4+\frac{Q}{2},~
	\frac{a}{\sqrt{\epsilon_1\epsilon_2}}=\alpha_\text{int}-\frac{Q}{2},~
	\text{e}^{2\pi\text{i}\tau}=\zeta,\label{agt}
\end{eqnarray}
the instanton partition function is equal to $\mathcal{B}_{\alpha_\text{int}}(\alpha_i|\zeta)$,
where the conformal block from $\langle V_{\alpha_1}V_{\alpha_2}V_{\alpha_3}V_{\alpha_4}\rangle$ as in \eqref{fourpt} can be written as $B=\zeta^{\Delta_{\alpha_\text{int}}-\Delta_{\alpha_1}-\Delta_{\alpha_2}}\mathcal{B}_{\alpha_\text{int}}(\alpha_i|\zeta)$ and ``int'' stands for (the VO in) the intermediate channel. One may check this perturbatively, and at level $|Y|_\text{max}$, $\mathcal{B}_{\alpha_\text{int}}$ and $Z_\text{inst}$ should agree up to $\mathcal{O}\left(\zeta^{|Y|_\text{max}+1}\right)$ \cite{Alday:2009aq,rodger2013pedagogical}. Notice that when we have $c=1$ CBs, viz, $Q=0$, the AGT relation is simplified to
\begin{equation}
\frac{\mu_1}{\sqrt{\epsilon_1\epsilon_2}}=\alpha_1+\alpha_2,~
\frac{\mu_2}{\sqrt{\epsilon_1\epsilon_2}}=\alpha_1-\alpha_2,~
\frac{\mu_3}{\sqrt{\epsilon_1\epsilon_2}}=\alpha_3+\alpha_4,~
\frac{\mu_4}{\sqrt{\epsilon_1\epsilon_2}}=\alpha_3-\alpha_4,~
\frac{a}{\sqrt{\epsilon_1\epsilon_2}}=\alpha_\text{int}.
\end{equation}

It is also possible to build a similar correspondence between gauge theories and minimal models. In \cite{Bershtein:2014qma}, it was shown that $Z_\text{inst}$ should recover the CBs for minimal models if we put further restrictions to the Young tableaux pairs known as the \emph{Burge condition}. For a minimal model, we may write the central charge as
\begin{equation}
	c=1-6\frac{(p'-p)^2}{p'p},
\end{equation}
where $p'$, $p$ are coprime integers and $p'>p>1$. The spectrum is finite and all the VOs have conformal dimensions
\begin{equation}
	\Delta_{r,s}=\frac{(p'r-ps)^2-(p'-p)^2}{4p'p}
\end{equation}
for integers $r,s$ and $1\leq r<p$, $1\leq s<p'$. In other words, they should all live in the Kac table. Then the instanton partition function leads to well-defined A-series minimal models under \eqref{agt} if the partitions $Y_{1,2}$ are restricted to be \emph{Burge pairs}, that is, they satisfy \cite{Bershtein:2014qma}
\begin{equation}
	Y_{2,R}-Y_{1,R+s-1}\geq1-r,~Y_{1,R}-Y_{2,R+p-s-1}\geq1-p'+r,
\end{equation}
where $Y_{i,R}$ denotes (the number of boxes in) the $R^\text{th}$ row of $Y_i$. In particular, the deformation parameters can now be written in terms of the screening charges as
\begin{equation}
	\frac{\epsilon_1}{\sqrt{\epsilon_1\epsilon_2}}=-\text{i}\sqrt{\frac{p}{p'}},~\frac{\epsilon_2}{\sqrt{\epsilon_1\epsilon_2}}=\text{i}\sqrt{\frac{p'}{p}}.
\end{equation}

\subsection{From 4d to 5d Instanton Partition Functions and A-Model Topological String Partition Functions}\label{Zinst2topostring}

For type II string/M-theory (whose brane configurations are discussed in Appendix \ref{brane}) compactified on a Calabi-Yau 3-fold, the amplitudes at genus $g$ correspond to the A-model string amplitudes of the CY$_3$ which enumerates the holomorphic functions from genus $g$ Riemann surfaces to the CY$_3$ \cite{Antoniadis:1993ze,Bershadsky:1993cx}. The topological amplitudes for toric CY threefolds can be computed by \emph{topological vertices} introduced in \cite{Aganagic:2003db,Iqbal:2007ii,Awata:2005fa}. A topological vertex is a trivalent vertex as the (black) dual graph of the (grey) toric diagram:
\begin{equation}
	\tikzset{every picture/.style={line width=0.75pt}} 
	\begin{tikzpicture}[x=0.75pt,y=0.75pt,yscale=-1,xscale=1]
	\draw [color={rgb, 255:red, 155; green, 155; blue, 155 }  ,draw opacity=1 ]   (110.8,60.2) -- (170.8,120.2) ;
	\draw [color={rgb, 255:red, 155; green, 155; blue, 155 }  ,draw opacity=1 ]   (110.8,60.2) -- (110.5,120.17) ; 
	\draw [color={rgb, 255:red, 155; green, 155; blue, 155 }  ,draw opacity=1 ]   (110.5,120.17) -- (170.8,120.2) ;
	\draw [line width=1.5]    (130.17,100.5) -- (130.17,159.17) ;
	\draw [line width=1.5]    (130.17,100.5) -- (70.17,99.83) ;
	\draw [line width=1.5]    (170.17,59.83) -- (130.17,100.5) ;
	\draw (132.17,103.5) node [anchor=north west][inner sep=0.75pt]   [align=left] {{\footnotesize $C_{Y_1Y_2Y_0}$}};
	\draw (72.17,102.83) node [anchor=north west][inner sep=0.75pt]   [align=left] {{\footnotesize $Y_2$}};
	\draw (133.17,139.5) node [anchor=north west][inner sep=0.75pt]   [align=left] {{\footnotesize $Y_1$}};
	\draw (172.17,62.83) node [anchor=north west][inner sep=0.75pt]   [align=left] {{\footnotesize $Y_0$}};
	\end{tikzpicture},
\end{equation}
where $Y_i$'s are the Young tableaux associated to the legs, and $C_{Y_1Y_2Y_0}(q)$ is the factor associated to the vertex, which can be expressed in terms of Schur and skew-Schur functions \cite{Aganagic:2003db}. Albeit not labelled explicitly, each leg also has a direction such that the three legs attached to the same vertex all have outcoming or incoming directions. Then each leg is assigned a vector $\bm{v}_i=(v_{i1},v_{i2})$ in that direction, such that the sum of the three vectors vanishes due to charge conservation and det$(\bm{v}_i,\bm{v}_{i+1})=\pm1$ ($i\in\mathbb{Z}_3$). Now two topological vertices
\begin{equation}
\scalebox{0.7}{
	\tikzset{every picture/.style={line width=0.75pt}} 
	\begin{tikzpicture}[x=0.75pt,y=0.75pt,yscale=-1,xscale=1]
	\draw [line width=1.5]    (372.17,134.5) -- (312.17,133.83) ;
	\draw [line width=1.5]    (412.17,93.83) -- (372.17,134.5) ;
	\draw [line width=1.5]    (372.17,134.5) -- (372.17,193.17) ;
	\draw [line width=1.5]    (472.17,94.5) -- (412.17,93.83) ;
	\draw [line width=1.5]    (412.17,35.17) -- (412.17,93.83) ;
	\draw    (372.17,193.17) -- (372.17,165.83) ;
	\draw [shift={(372.17,163.83)}, rotate = 450] [color={rgb, 255:red, 0; green, 0; blue, 0 }  ][line width=0.75]    (10.93,-3.29) .. controls (6.95,-1.4) and (3.31,-0.3) .. (0,0) .. controls (3.31,0.3) and (6.95,1.4) .. (10.93,3.29)   ;
	\draw    (312.17,133.83) -- (340.17,134.14) ;
	\draw [shift={(342.17,134.17)}, rotate = 180.64] [color={rgb, 255:red, 0; green, 0; blue, 0 }  ][line width=0.75]    (10.93,-3.29) .. controls (6.95,-1.4) and (3.31,-0.3) .. (0,0) .. controls (3.31,0.3) and (6.95,1.4) .. (10.93,3.29)   ;
	\draw    (412.17,93.83) -- (440.17,94.14) ;
	\draw [shift={(442.17,94.17)}, rotate = 180.64] [color={rgb, 255:red, 0; green, 0; blue, 0 }  ][line width=0.75]    (10.93,-3.29) .. controls (6.95,-1.4) and (3.31,-0.3) .. (0,0) .. controls (3.31,0.3) and (6.95,1.4) .. (10.93,3.29)   ;
	\draw    (412.17,93.83) -- (393.57,112.74) ;
	\draw [shift={(392.17,114.17)}, rotate = 314.53] [color={rgb, 255:red, 0; green, 0; blue, 0 }  ][line width=0.75]    (10.93,-3.29) .. controls (6.95,-1.4) and (3.31,-0.3) .. (0,0) .. controls (3.31,0.3) and (6.95,1.4) .. (10.93,3.29)   ;
	\draw    (412.17,93.83) -- (412.17,66.5) ;
	\draw [shift={(412.17,64.5)}, rotate = 450] [color={rgb, 255:red, 0; green, 0; blue, 0 }  ][line width=0.75]    (10.93,-3.29) .. controls (6.95,-1.4) and (3.31,-0.3) .. (0,0) .. controls (3.31,0.3) and (6.95,1.4) .. (10.93,3.29)   ;
	\draw (312.17,135.83) node [anchor=north west][inner sep=0.75pt]   [align=left] {{\footnotesize $Y_2$}};
	\draw (375.17,171.5) node [anchor=north west][inner sep=0.75pt]   [align=left] {{\footnotesize $Y_1$}};
	\draw (458.17,95.5) node [anchor=north west][inner sep=0.75pt]   [align=left] {{\footnotesize $Y'_2$}};
	\draw (412.17,35.17) node [anchor=north west][inner sep=0.75pt]   [align=left] {{\footnotesize $Y'_1$}};
	\draw (378.17,94.5) node [anchor=north west][inner sep=0.75pt]   [align=left] {{\footnotesize $Y_0$}};
	\draw (392.17,114.17) node [anchor=north west][inner sep=0.75pt]   [align=left] {{\footnotesize $Q_0$}};
	\end{tikzpicture}}
\end{equation}
can be glued as
\begin{equation}
	\sum_{Y_0}C_{Y_1^{\text{T}}Y_2^{\text{T}}Y_0^{\text{T}}}(q)(-1)^{(n+1)|Y_0|}q^{-n\kappa(Y_0)/2}Q_0^{|Y_0|}C_{Y'_1Y'_2Y^{}_0},
\end{equation}
where $\kappa$ is related to quadratic Casimir of the representation corresponding to $|Y_0|$, namely, $\kappa(Y_0)=\sum\limits_iy_i(y_i-2i+1)$ with $y_i$ being the number of boxes in the $i^\text{th}$ row. The framing number $n$ equals det$(\bm{v}_\text{in},\bm{v}_\text{out})$, where the two vectors are chosen such that $\bm{v}_\text{in}\cdot\bm{v}_\text{out}>0$. The parameter $Q_0$ is the (exponentiated) K\"ahler parameter for the 2-cycle corresponding to the line in the dual toric diagram.

In \cite{Iqbal:2007ii}, the above is extended to a \emph{refined topological vertex} as
\begin{eqnarray}
	C_{Y_1Y_2Y_0}(q,t)&=&\left(\frac{q}{t}\right)^{(||Y_2||^2+||Y_0||^2)/2}t^{\kappa(Y_2)/2}P_{Y^\text{T}_0}\left(t^{-\rho};q,t\right)\nonumber\\
	&&\times\sum_{\eta}\left(\frac{q}{t}\right)^{(|\eta|+|Y_1|-|Y_2|)/2}s_{Y^\text{T}_1/\eta}\left(t^{-\rho}q^{-Y_0}\right)s_{Y_2/\eta}\left(t^{-Y^\text{T}_0}q^{-\rho}\right),
\end{eqnarray}
where $P_{Y^\text{T}_0}\left(t^{-\rho};q,t\right)$ is the Macdonald function and $s_{\alpha/\beta}$'s are the skew-Schur functions. The squared double slash denotes the quadratic sum of the number of boxes in each row of the Young tableau. Notice that the three Young tableaux are not cyclically symmetric and $Y_0$ corresponds to the preferred leg for gluing. One may check that when the $\Omega$-background parameters satisfy $q=t$, we would recover the unrefined topological vertex.

Define the framing factors,
\begin{equation}
	f_{Y}(q,t)=(-1)^{|Y|}q^{||Y^\text{T}||^2/2}t^{-||Y||^2/2},~\tilde{f}_{Y}(q,t)=(-1)^{|Y|}q^{(||Y^\text{T}||^2+|Y|)/2}t^{-(||Y||^2+|Y|)/2},
\end{equation}
and the edge factor, $(-Q_0)^{|Y_0|}\times[\text{framing~factor}]$. Then the topological string partition function takes the sum over all the Young tableaux of internal legs\footnote{The Young tableaux of external legs would be $\emptyset$.} as
\begin{equation}
	Z_\text{topo}=\sum_{Y_i}\prod_\text{edges}[\text{edge~factor}]\prod_\text{vertices}[\text{vertex~factor}].
\end{equation}

Again, let us contemplate the SU(2) gauge theory with 4 flavours. The dual web diagram is
\begin{equation}
\scalebox{0.7}{
	\tikzset{every picture/.style={line width=0.75pt}} 
	\begin{tikzpicture}[x=0.75pt,y=0.75pt,yscale=-1,xscale=1]
	\draw [line width=1.5]    (138.17,358.5) -- (78.17,357.83) ;
	\draw [line width=1.5]    (178.17,317.83) -- (138.17,358.5) ;
	\draw [line width=1.5]    (138.17,358.5) -- (138.17,417.17) ;
	\draw [line width=1.5]    (238.17,318.5) -- (178.17,317.83) ;
	\draw [line width=1.5]    (178.17,259.17) -- (178.17,317.83) ;
	\draw    (138.17,417.17) -- (138.17,389.83) ;
	\draw [shift={(138.17,387.83)}, rotate = 450] [color={rgb, 255:red, 0; green, 0; blue, 0 }  ][line width=0.75]    (10.93,-3.29) .. controls (6.95,-1.4) and (3.31,-0.3) .. (0,0) .. controls (3.31,0.3) and (6.95,1.4) .. (10.93,3.29)   ;
	\draw    (78.17,357.83) -- (106.17,358.14) ;
	\draw [shift={(108.17,358.17)}, rotate = 180.64] [color={rgb, 255:red, 0; green, 0; blue, 0 }  ][line width=0.75]    (10.93,-3.29) .. controls (6.95,-1.4) and (3.31,-0.3) .. (0,0) .. controls (3.31,0.3) and (6.95,1.4) .. (10.93,3.29)   ;
	\draw    (178.17,317.83) -- (236.17,318.48) ;
	\draw [shift={(238.17,318.5)}, rotate = 180.64] [color={rgb, 255:red, 0; green, 0; blue, 0 }  ][line width=0.75]    (10.93,-3.29) .. controls (6.95,-1.4) and (3.31,-0.3) .. (0,0) .. controls (3.31,0.3) and (6.95,1.4) .. (10.93,3.29)   ;
	\draw    (178.17,317.83) -- (159.57,336.74) ;
	\draw [shift={(158.17,338.17)}, rotate = 314.53] [color={rgb, 255:red, 0; green, 0; blue, 0 }  ][line width=0.75]    (10.93,-3.29) .. controls (6.95,-1.4) and (3.31,-0.3) .. (0,0) .. controls (3.31,0.3) and (6.95,1.4) .. (10.93,3.29)   ;
	\draw    (178.17,317.83) -- (178.17,261.17) ;
	\draw [shift={(178.17,259.17)}, rotate = 450] [color={rgb, 255:red, 0; green, 0; blue, 0 }  ][line width=0.75]    (10.93,-3.29) .. controls (6.95,-1.4) and (3.31,-0.3) .. (0,0) .. controls (3.31,0.3) and (6.95,1.4) .. (10.93,3.29)   ;
	\draw [line width=1.5]    (337.83,358.5) -- (397.83,357.83) ;
	\draw [line width=1.5]    (297.83,317.83) -- (337.83,358.5) ;
	\draw [line width=1.5]    (337.83,358.5) -- (337.83,417.17) ;
	\draw [line width=1.5]    (237.83,318.5) -- (297.83,317.83) ;
	\draw [line width=1.5]    (297.83,259.17) -- (297.83,317.83) ;
	\draw    (337.83,358.5) -- (337.83,385.83) ;
	\draw [shift={(337.83,387.83)}, rotate = 270] [color={rgb, 255:red, 0; green, 0; blue, 0 }  ][line width=0.75]    (10.93,-3.29) .. controls (6.95,-1.4) and (3.31,-0.3) .. (0,0) .. controls (3.31,0.3) and (6.95,1.4) .. (10.93,3.29)   ;
	\draw    (337.83,358.5) -- (365.83,358.19) ;
	\draw [shift={(367.83,358.17)}, rotate = 539.36] [color={rgb, 255:red, 0; green, 0; blue, 0 }  ][line width=0.75]    (10.93,-3.29) .. controls (6.95,-1.4) and (3.31,-0.3) .. (0,0) .. controls (3.31,0.3) and (6.95,1.4) .. (10.93,3.29)   ;
	\draw    (337.83,358.5) -- (319.24,339.59) ;
	\draw [shift={(317.83,338.17)}, rotate = 405.47] [color={rgb, 255:red, 0; green, 0; blue, 0 }  ][line width=0.75]    (10.93,-3.29) .. controls (6.95,-1.4) and (3.31,-0.3) .. (0,0) .. controls (3.31,0.3) and (6.95,1.4) .. (10.93,3.29)   ;
	\draw [line width=1.5]    (297.77,200.5) -- (237.77,199.83) ;
	\draw [line width=1.5]    (337.77,159.83) -- (297.77,200.5) ;
	\draw [line width=1.5]    (297.77,200.5) -- (297.77,259.17) ;
	\draw [line width=1.5]    (397.77,160.5) -- (337.77,159.83) ;
	\draw [line width=1.5]    (337.77,101.17) -- (337.77,159.83) ;
	\draw    (297.77,200.5) -- (297.83,257.17) ;
	\draw [shift={(297.83,259.17)}, rotate = 269.93] [color={rgb, 255:red, 0; green, 0; blue, 0 }  ][line width=0.75]    (10.93,-3.29) .. controls (6.95,-1.4) and (3.31,-0.3) .. (0,0) .. controls (3.31,0.3) and (6.95,1.4) .. (10.93,3.29)   ;
	\draw    (297.77,200.5) -- (240.17,199.86) ;
	\draw [shift={(238.17,199.83)}, rotate = 360.64] [color={rgb, 255:red, 0; green, 0; blue, 0 }  ][line width=0.75]    (10.93,-3.29) .. controls (6.95,-1.4) and (3.31,-0.3) .. (0,0) .. controls (3.31,0.3) and (6.95,1.4) .. (10.93,3.29)   ;
	\draw    (397.77,160.5) -- (369.77,160.19) ;
	\draw [shift={(367.77,160.17)}, rotate = 360.64] [color={rgb, 255:red, 0; green, 0; blue, 0 }  ][line width=0.75]    (10.93,-3.29) .. controls (6.95,-1.4) and (3.31,-0.3) .. (0,0) .. controls (3.31,0.3) and (6.95,1.4) .. (10.93,3.29)   ;
	\draw    (297.77,200.5) -- (316.36,181.59) ;
	\draw [shift={(317.77,180.17)}, rotate = 494.53] [color={rgb, 255:red, 0; green, 0; blue, 0 }  ][line width=0.75]    (10.93,-3.29) .. controls (6.95,-1.4) and (3.31,-0.3) .. (0,0) .. controls (3.31,0.3) and (6.95,1.4) .. (10.93,3.29)   ;
	\draw    (337.77,101.17) -- (337.77,128.5) ;
	\draw [shift={(337.77,130.5)}, rotate = 270] [color={rgb, 255:red, 0; green, 0; blue, 0 }  ][line width=0.75]    (10.93,-3.29) .. controls (6.95,-1.4) and (3.31,-0.3) .. (0,0) .. controls (3.31,0.3) and (6.95,1.4) .. (10.93,3.29)   ;
	\draw [line width=1.5]    (178.17,200.5) -- (238.17,199.83) ;
	\draw [line width=1.5]    (138.17,159.83) -- (178.17,200.5) ;
	\draw [line width=1.5]    (178.17,200.5) -- (178.17,259.17) ;
	\draw [line width=1.5]    (78.17,160.5) -- (138.17,159.83) ;
	\draw [line width=1.5]    (138.17,101.17) -- (138.17,159.83) ;
	\draw    (138.17,159.83) -- (110.17,160.14) ;
	\draw [shift={(108.17,160.17)}, rotate = 359.36] [color={rgb, 255:red, 0; green, 0; blue, 0 }  ][line width=0.75]    (10.93,-3.29) .. controls (6.95,-1.4) and (3.31,-0.3) .. (0,0) .. controls (3.31,0.3) and (6.95,1.4) .. (10.93,3.29)   ;
	\draw    (138.17,159.83) -- (156.76,178.74) ;
	\draw [shift={(158.17,180.17)}, rotate = 225.47] [color={rgb, 255:red, 0; green, 0; blue, 0 }  ][line width=0.75]    (10.93,-3.29) .. controls (6.95,-1.4) and (3.31,-0.3) .. (0,0) .. controls (3.31,0.3) and (6.95,1.4) .. (10.93,3.29)   ;
	\draw    (138.17,159.83) -- (138.17,132.5) ;
	\draw [shift={(138.17,130.5)}, rotate = 450] [color={rgb, 255:red, 0; green, 0; blue, 0 }  ][line width=0.75]    (10.93,-3.29) .. controls (6.95,-1.4) and (3.31,-0.3) .. (0,0) .. controls (3.31,0.3) and (6.95,1.4) .. (10.93,3.29)   ;
	\draw (226.17,322.5) node [anchor=north west][inner sep=0.75pt]   [align=left] {{\footnotesize $Q_B$}};
	\draw (180.17,240.83) node [anchor=north west][inner sep=0.75pt]   [align=left] {{\footnotesize $Q_F$}};
	\draw (141.17,321.5) node [anchor=north west][inner sep=0.75pt]   [align=left] {{\footnotesize $Q_4$}};
	\draw (330.83,321.5) node [anchor=north east][inner sep=0.75pt] [align=left] {{\footnotesize $Q_2$}};
	\draw (239.77,202.83) node [anchor=north west][inner sep=0.75pt]   [align=left] {{\footnotesize $Q_B$}};
	\draw (300.77,240.5) node [anchor=north west][inner sep=0.75pt]   [align=left] {{\footnotesize $Q_F$}};
	\draw (330.77,170.5) node [anchor=north west][inner sep=0.75pt]   [align=left] {{\footnotesize $Q_1$}};
	\draw (141.17,170.5) node [anchor=north east][inner sep=0.75pt]  [align=left] {{\footnotesize $Q_3$}};
	\end{tikzpicture}}.
\end{equation}
Following the gluing process, the partition function reads\footnote{More generally, for SU($N_c$) gauge group with $N_f=2N_c$, the partition function was given in \cite{Awata:2008ed}.}
\begin{eqnarray}
	Z_\text{topo}&=&\sum_{\lambda\rho\nu}(-Q_F)^{|\lambda_1|}\tilde{f}_{\lambda_1}(q,t)(-Q_F)^{|\lambda_2|}\tilde{f}_{\lambda_2}(t,q)(-Q_B)^{|\rho_1|}f_{\rho_1^\text{T}}(q,t)(-Q_B)^{|\rho_2|}f_{\rho_2^\text{T}}(t,q)\nonumber\\
	&&\times(-Q_1)^{|\nu_1|}(-Q_2)^{|\nu_2|}(-Q_3)^{|\nu_3|}(-Q_4)^{|\nu_4|}C_{\lambda_1^\text{T}\nu_1^\text{T}\rho_1^\text{T}}(q,t)C_{\nu_4\lambda_1\rho_2}(q,t)\nonumber\\
	&&\times C_{\lambda_2^\text{T}\nu_2^\text{T}\rho_2^\text{T}}(t,q)C_{\nu_3\lambda_2\rho_1}(t,q)C_{\nu_3^\text{T}\emptyset\emptyset}(q,t)C_{\emptyset\nu_2\emptyset}(q,t)C_{\nu_4^\text{T}\emptyset\emptyset}(t,q)C_{\emptyset\nu_1\emptyset}(t,q).
\end{eqnarray}
Recall the 4d instanton partition function \eqref{Zinst4flav}, which can be lifted to 5d as \cite{Taki:2007dh}
\begin{equation}
	Z_\text{inst,5d}=\frac{1}{Z_\text{U(1),5d}}\sum_{Y_{1,2}}\text{e}^{2\pi\text{i}\tau(|Y_1|+|Y_2|)}z_\text{vec,5d}(\vec{a},\vec{Y})z_\text{matter,5d}.
\end{equation}
It is discussed in \cite{Bao:2011rc,Bao:2013pwa,Taki:2007dh,Foda:2015ana,Foda:2017tnv} that under the parameter identification\footnote{Often $Q_i$ would be written as $\text{e}^{-R(-\mu_i-a)}$ for $i=2,4$. However, due to invariance under Weyl group symmetry, the Nekrasov partition function should not change under $\mu_{2,4}\rightarrow-\mu_{2,4}$.}
\begin{equation}
q=\text{e}^{-R\epsilon_1},~
t=\text{e}^{R\epsilon_2},~
Q_i=\text{e}^{-R(\mu_i-a)},~
Q_B=\text{e}^{2\pi\text{i}\tau}\text{e}^{\frac{1}{2}R\left(-4a+\sum\limits_i\mu_i\right)},~
Q_F=\text{e}^{-2Ra},
\end{equation}
where $R$ is the radius of the compactified dimension $S^1$, $Z_\text{topo}$ reproduces $Z_\text{inst,5d}$ up to perturbative part and U(1)/extra factor \cite{Hayashi:2016abm,Kim:2015jba,Hwang:2014uwa,Hayashi:2013qwa,Bao:2013pwa,Bergman:2013ala,Bergman:2013aca}. Notice that when $\epsilon_1=-\epsilon_2$, i.e., $Q=0$, we have the unrefinement $q=t$. In particular, under the 4d limit $R\rightarrow0$, the 4d topological A-model partition function would give the 4d instanton partition function.

\subsection{From Topological String Partition Functions to Seiberg-Witten Curves}\label{topostring2SW}
As aforementioned, the low energy effective theory for 4d $\mathcal{N}=2$ can be encoded by the prepotential $\mathcal{F}$, where $\mathcal{F}=\lim\limits_{\epsilon_{1,2}\rightarrow0}\epsilon_1\epsilon_2\log Z_\text{Nek}$ in terms of the Nekrasov partition function\footnote{The finite non-zero deformation parameters could also physically make sense for SW curves, for example in the context of topological B-models as in Appendix \ref{bmodel}.} which is in turn naturally related to topological partition functions in the A-model.

On the other hand, in the SW solution, the prepotential can be determined by the SW curve. Given such auxiliary curve $\Sigma$, it is possible to translate into the form $\lambda^2 = q(z)$, where $\lambda$ is the \emph{Seiberg-Witten differential}, and $q(z)$ is the meromorphic quadratic differential on the Gaiotto curve $\mathcal{C}$ \cite{Gaiotto:2009we,Tachikawa:2013kta}. In this subsection, we demonstrate how this translation runs for the theory with a single SU(2) factor and $N_f=4$. This theory will constitute our running example throughout the paper.

To begin, following \S9.1 of \cite{Tachikawa:2013kta}, $\Sigma$ for the SU(2) with $N_f=4$ theory in hyperelliptic form is
\begin{equation}
\frac{f}{\tilde{z}}(\tilde{x}-\tilde{\mu}_1)(\tilde{x}-\tilde{\mu}_2)+(f'\tilde{z})(\tilde{x}-\tilde{\mu}_3)(\tilde{x}-\tilde{\mu}_4)=\tilde{x}^2-u,\label{Sigma}
\end{equation}
where $f$ and $f'$ are complex numbers and $u$ parametrizes the space of supersymmetric vacua, viz, the $u$-plane. We first choose the coordinate of $\tilde{z}$ so that $f'=1$. Completing a square in $\tilde{x}$ by defining
\begin{equation}
x=\tilde{x}+\frac{\frac{f}{\tilde{z}}(\tilde{\mu}_1+\tilde{\mu}_2)+\tilde{z}(\tilde{\mu}_3+\tilde{\mu}_4)}{2\left(1-\tilde{z}-\frac{f}{\tilde{z}}\right)},
\end{equation}
we obtain $x^2=g(\tilde{z})$ where $g$ has double poles at $c_{1,2}(f)$. Now rescale $z=\tilde{z}/c_1(f)$ so that the poles are at $z=1,\zeta$, and we get
\begin{equation}
x^2=\frac{P(z)}{(z-1)^2(z-\zeta)^2}\label{SWcurve}
\end{equation}
for some quartic polynomial $P(z)$ determined by $\tilde{\mu}_i$, $\zeta$ and $u$. Then $a$ and its magnetic dual $a_D$ can be obtained by integrating the SW differential $\lambda\equiv x\text{d}z/z$ along $A$- and $B$-cycles:
\begin{equation}
	a=\oint_A\lambda,~a_D=\frac{\partial\mathcal{F}}{\partial a}=\oint_B\lambda.
\end{equation}

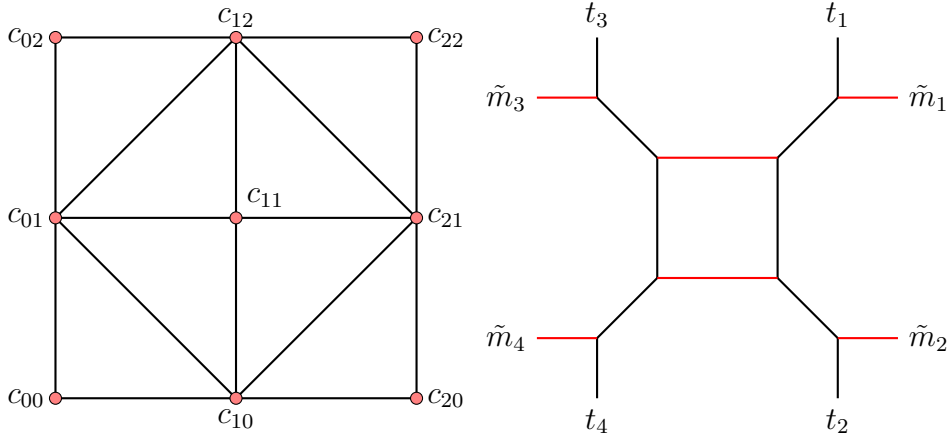
\begin{figure}
	\begin{centering}
		\begin{tikzpicture}[scale=.8]
			\draw [thick] (0,0)--(3,3)--(6,0)--(3,-3)--(0,0);
			\draw [thick] (0,0)--(6,0);
			\draw [thick] (3,3)--(3,-3);
			\draw [thick] (3,3)--(6,3)--(6,0)--(6,-3)--(3,-3);
			\draw [thick] (3,3)--(0,3)--(0,0)--(0,-3)--(3,-3);
			\draw [fill=red!50] (6,3) circle (.1cm);
			\draw [fill=red!50] (6,-3) circle (.1cm);
			\draw [fill=red!50] (0,3) circle (.1cm);
			\draw [fill=red!50] (0,-3) circle (.1cm);
			\foreach \i in {0,1} {
				\draw [fill=red!50] (3*\i,3*\i) circle (.1cm);
				\draw [fill=red!50](3*\i+3,3*\i-3) circle (.1cm);}
			\draw [fill=red!50] (3,0) circle (.1cm);
			\node [above] at (3,3){$c_{12}$};
			\node [below] at (3,-3){$c_{10}$};
			\node [left] at (0,0){$c_{01}$};
			\node [left] at (0,3) {$c_{02}$};
			\node [left] at (0,-3) {$c_{00}$};
			\node [right] at (6,3) {$c_{22}$};
			\node [right] at (6,-3) {$c_{20}$};
			\node [right] at (6,0){$c_{21}$};
			\node [above right] at (3,0){$c_{11}$};
			\draw [xshift=8cm,thick] (1,3)--(1,2)--(2,1)--(2,-1)--(1,-2)--(1,-3);
			\draw [xshift=8cm,thick] (5,3)--(5,2)--(4,1)--(4,-1)--(5,-2)--(5,-3);
			\draw [xshift=8cm,red,thick] (6,2)--(5,2);
			\draw [xshift=8cm,red,thick] (6,-2)--(5,-2);
			\draw [xshift=8cm,red,thick] (1,2)--(0,2);
			\draw [xshift=8cm,red,thick] (1,-2)--(0,-2);
			\draw [xshift=8cm,red,thick] (2,1)--(4,1);
			\draw [xshift=8cm,red,thick] (2,-1)--(4,-1);
			\node [left] at(8,2){$\tilde{m}_3$};
			\node [left] at(8,-2){$\tilde{m}_4$};
			\node [above] at (9,3){$t_3$};
			\node [below] at(9,-3){$t_4$};
			\node [right] at(14,2){$\tilde{m}_1$};
			\node [right] at(14,-2){$\tilde{m}_2$};
			\node [above] at (13,3){$t_1$};
			\node [below] at(13,-3){$t_2$};
		\end{tikzpicture} 
		\par\end{centering}
	\protect\protect\caption{The toric diagram and its dual diagram for SU(2) with 4 flavours.}\label{web}
\end{figure}

\paragraph{Construction of SW Curves from Toric Diagrams} For 5d gauge theories, the 5-brane web diagrams can be used to construct SW curves. In fact, such a web diagram is exactly the same as the dual toric diagram in the geometric engineering in \S\ref{Zinst2topostring} \cite{Leung:1997tw,Gorsky:1997mw}. The standard algorithm of constructing SW curves from toric diagrams are proposed in \cite{Aharony:1997bh} and elaborated in \cite{Kim:2014nqa}. Here, we will still focus on SU(2) with 4 flavours, where the web/dual diagram is reproduced in Figure \ref{web}, along with its toric diagram.

For each vertex $(i,j)$ in the toric diagram, we assign a non-zero number $c_{ij}$.
With these coefficients, the SW curve is given as
\begin{equation}
	\sum_{i,j} c_{ij} t^i w^j = 0.
\end{equation}
By multiplying an overall constant to this equation, we can impose 
\begin{equation}
	c_{22}  =1
\end{equation}
without the loss of generality. 
There are four boundaries, so the boundary conditions according to the toric diagram now are 
\begin{eqnarray}
	|w| & \gg & 1,\quad c_{02}w^2+c_{12}tw^2+c_{22}t^2w^2=c_{22}w^2(t-t_1 )(t-t_3 ),\nonumber\\
	|w^{-1}| & \gg & 1,\quad c_{20}t^2+c_{10}t+c_{00}=c_{20}(t-t_2 )(t-t_4  ),\nonumber\\
	|t| & \gg & 1,\quad c_{20}t^2+c_{21}t^2w+c_{22}t^2w^2=c_{22}t^2(w-\tilde{m}_1 )(w-\tilde{m}_2 ),\nonumber\\
	|t^{-1}| & \gg & 1,\quad c_{02}w^2+c_{01}w+c_{00}=c_{02}(w-\tilde{m}_3 )(w-\tilde{m}_4  ),
\end{eqnarray}
where $t$ and $w$ can be thought of as the horizontal and vertical coordinates of the diagram respectively. 
In order for the curve to satisfy all the conditions consistently, we need the compatibility condition which reads
\begin{equation}
	\tilde{m}_1 t_1^{-1} \tilde{m}_2 t_2 = \tilde{m}_3 t_3 \tilde{m}_4 t_4^{-1}.
\end{equation}
Since the SW differential is invariant under the rescaling of $t$, we can impose
\begin{equation}
	t_2 = 1
\end{equation}
for simplicity. 
Also, by rescaling $w$, we can further impose 
\begin{equation}
	\tilde{m}_1 t_1^{-1} \tilde{m}_2 t_2 =1.
\end{equation}
This condition turns out to correspond to the traceless condition of the vacuum expectation value of the SU(2) vector multiplet \cite{Bao:2011rc}. 
The instanton factor is the geometric average of $t_i $, which is
\begin{equation}
	\zeta \equiv \left(\frac{t_3 t_4}{t_1 t_2}\right)^{\frac12}.
\end{equation}
The only undetermined coefficient $c_{11}=-U'$ is interpreted as Coulomb moduli parameter. 
Defining a parameter $S\equiv\tilde{m}_1 \tilde{m}_2 \tilde{m}_3 \tilde{m}_4$, we have
\begin{equation}
	t_1 = \tilde{m}_1 \tilde{m}_2, \quad 
	t_2 = 1 , \quad
	t_3 = \tilde{m}_1 \tilde{m}_2 \left( S \right)^{-\frac12}\zeta , \quad
	t_4 = \left( S \right)^{\frac12 }  \zeta,
\end{equation}
and thus, the SW curve for 5d $\mathcal{N}=1$ SU(2) gauge theory with $N_f =4$ is
\begin{eqnarray}
	&&{\rm t}^2\left(w - \tilde{m}_1 \right)\left(w - \tilde{m}_2 \right)
	-{\rm t} \left(w^2\tilde{m}_1  \tilde{m}_2 
	\left(1 + \zeta \left( S \right)^{-\frac12}\right) + w U'   + 
	\tilde{m}_1  \tilde{m}_2  \left(1 + \zeta \left( S \right)^{\frac12 } \right)\right)\nonumber\\
	&&+ \zeta \left(S \right)^{-\frac12} 
	\left(\tilde{m}_1  \tilde{m}_2 \right)^2 
	\left( w - \tilde{m}_3 \right)
	\left( w - \tilde{m}_4 \right)=0.
\end{eqnarray}

\paragraph{The 4d limit curve} Till now, the SW curve is a 5d curve, its 4d SW curve can be obtained by taking the vanishing limit of size of compactification circle $\beta \rightarrow 0$, where we have
\begin{equation}
w=e^{-\beta v},\ \tilde{m}_i =e^{-\beta \mu_i}.
\end{equation}
We then expand the 5d Coulomb parameter $U'$ as 
\begin{equation}
U^{\prime}=\sum_{k=0}^\infty u_k\beta^k.
\end{equation}
In particular,
\begin{equation}
	u_0=-2(1+\zeta),~u_1=2(\zeta+1)(\mu_1+\mu_2).
\end{equation}
For the $N_f = 4$ curve, we then have the 4d limit
\begin{align}
& t^2
\left( v - \mu_1 \right)
\left( v - \mu_2 \right) + \zeta \left( v - \mu_3 \right)
\left( v - \mu_4 \right) +
\nonumber 
\\
& t(1+\zeta) \left( - v^2 + \frac{\zeta v}{1+\zeta}
\sum_{i=1}^4\mu_i+ U \right) = 0,\label{eq:4dcurve}
\end{align}
where
\begin{equation}
	U=-\frac{u_2}{1+\zeta}-(\mu_1+\mu_2)^2-\frac{\zeta}{4(1+\zeta)}\left(\sum_{i=1}^4\mu_i\right)^2.
\end{equation}

\paragraph{Reparametrization} Let us rewrite the curve as
\begin{align}
\frac{t}{1+\zeta} \left(v-\mu_1 \right) \left(v - \mu_2 \right) - v^2 
+  \frac{ \zeta v \sum\limits_{i=1}^4 \mu_i}{1+\zeta}+U
+\frac{\zeta}{t}\frac{1}{1+\zeta}\left( v - \mu_3 \right) \left( v - \mu_4 \right) = 0.
\end{align}
Redefining
\begin{eqnarray}
    &&f/\tilde{z}=\frac{t}{1+\zeta},~f'\tilde{z}=\frac{\zeta}{t(1+\zeta)},~\tilde{x}=v-\frac{\zeta}{2(1+\zeta)}\sum\limits_{i=1}^{4} \mu_i,~\tilde{\mu}_i=\mu_i-\frac{\zeta}{2(1+\zeta)}\sum\limits_{i=1}^4 \mu_i,\nonumber\\
	&&u=\left(\frac{\zeta}{2(1+\zeta)}\sum\limits_{i=1}^4\mu_i\right)^2+U
\end{eqnarray}
recovers the curve of form \eqref{Sigma}, which is reproduced here:
\begin{equation}
	\frac{f}{\tilde{z}}(\tilde{x}-\tilde{\mu}_1)(\tilde{x}-\tilde{\mu}_2)+(f'\tilde{z})(\tilde{x}-\tilde{\mu}_3)(\tilde{x}-\tilde{\mu}_4)=\tilde{x}^2-u.
\end{equation}

\subsection{From Seiberg-Witten Curves to Dessins d'Enfants}\label{SW2dessin}

We begin with a refresher on some preliminary definitions and key results \cite{ggd2011,10.1112/blms/28.6.561}.

\begin{definition}
	A \emph{dessin d'enfant}, or \emph{child's drawing}, is an ordered pair $(X,\mathcal{D})$, where $X$ is an oriented compact topological surface and $\mathcal{D}\subset X$ is a finite graph, such that
	\begin{enumerate}
		\item $\mathcal{D}$ is a connected bipartite graph, and
		\item $X\backslash\mathcal{D}$ is the union of finitely many topological discs that are the faces of $\mathcal{D}$.
	\end{enumerate}
\end{definition}
There is a bijection between the dessins and Belyi maps known as the \emph{Grothendieck correspondence} \cite{sketch}, where
\begin{definition}
	A \emph{Belyi map} $\beta$ is a holomorphic map from the Riemann surface X to $\mathbb{P}^1$ ramified at only 3 points, which can be taken to be $\{0,1,\infty\}\in\mathbb{P}^1$. 
\end{definition}
Recall that \emph{ramification} means that the only points $\tilde{x}\in X$ where $\frac{d}{dx}\beta(x)\big|_{\tilde{x}}=0$ are such that $\beta(\tilde{x})=0,1$ or $\infty$. In other words, the local Taylor expansion of $\beta(x)$ about the pre-images $\tilde{x}$ of $\{0,1,\infty\}$ have (at least) vanishing linear term.

\paragraph{From Belyi maps to dessins} We can associate $\beta(x)$ to a dessin via its \emph{ramification indices}: the order of vanishing of the Taylor series for $\beta(x)$ at $\tilde{x}$ is the ramification index $r_{\beta(\tilde{x})\in\{0,1,\infty\}}(i)$ at that $i^\text{th}$ point. By convention, we mark one white node for the $i^\text{th}$ pre-image of 0 with $r_0(i)$ edges emanating therefrom. Similarly, we mark one black node for the $j^\text{th}$ pre-image of 1 with $r_1(j)$ edges. We then connect the nodes with the edges, joining only black with white, such that each face is a polygon with $2r_\infty(k)$ sides. In other words, there is one pre-image of $\infty$ corresponding to each polygon of $\mathcal{D}$. Moreover, there is a cyclic ordering arising from local monodromy winding around vertices, i.e., around local covering sheets that contain a common point.

The power of dessins comes from Belyi's remarkable theorem.
\begin{theorem}
	There exists an algebraic model of $X$ (as a Riemann surface) defined over $\bar{\mathbb{Q}}$ iff there exists a Belyi map on $X$.
\end{theorem}
Thus, the existence of a dessin on $X$ is equivalent to $X$ admitting an algebraic equation over the algebraic numbers. Moreover, the Galois group Gal$(\bar{\mathbb{Q}}:\mathbb{Q})$ acts faithfully on the space of dessins.

\paragraph{Quadratic Differentials} A \emph{(holomorphic) quadratic differential} $q$ on a Riemann surface $X$ is a holomorphic section of the symmetric square of the contangent bundle. In terms of local coordinates $z$, $q=f(z)\text{d}z\otimes\text{d}z$, for some holomorphic function $f(z)$.

A curve $\gamma(t)\subset X$ can be classified by $q$ as
\begin{itemize}
	\item Horizontal trajectory: $f(\gamma(t))\dot{\gamma}(t)^2>0$;
	\item Vertical trajectory: $f(\gamma(t))\dot{\gamma}(t)^2<0$.
\end{itemize}
Locally, one can find coordinates so that horizontal tracjectories look like concentric circles while vertical trajectories look like rays emanating from a single point.

Then we can define the \emph{Strebel differential}:
\begin{definition}
	For a Riemann surface $X$ of genus $g\geq0$ with $n\geq1$ marked points $\{p_1,\dots,p_n\}$ such that $2-2g<n$, and a given $n$-tuple $a_{i=1,\dots,n}\in \mathbb{R}^+$, a Strebel differential $q=f(z)\textup{d}z^2$ is a quadratic differential such that
	\begin{itemize}
		\item $f$ is holomorphic on $X\backslash\{p_1,\dots,p_n\}$;
		\item $f$ has a second-order pole at each $p_i$;
		\item the union of all non-compact horizontal trajectories of $q$ is a closed subset of X of measure 0;
		\item every compact horizontal of $q$ is a simple loop $A_i$ centered at $p_i$ such that $a_i=\oint_{A_i}\sqrt{q}$. (Here the branch of the square root is chosen so that the integral has a positive value with respect to the positive orientation of $A_i$ that is determined by the complex structure of $X$.)
	\end{itemize}
\end{definition}
The upshot is that \cite{mp}
\begin{theorem}
	The Strebel differential is the pull-back, by a Belyi map $\beta: X\rightarrow\mathbb{P}^1$, of a quadratic differential on $\mathbb{P}^1$ with 3 punctures,
	\begin{equation}
		q=\beta^*\left(\frac{\textup{d}\zeta^2}{4\pi^2\zeta(1-\zeta)}\right)=\frac{(\textup{d}\beta)^2}{4\pi^2\beta(1-\beta)}=\frac{(\beta^\prime)^2}{4\pi^2\beta(1-\beta)}\textup{d}z^2,\label{Strebel}
	\end{equation}
	where $z$ and $\zeta$ are coordinates on $X$ and $\mathbb{P}^1$ respectively.
\end{theorem}

Recall the definition of the SW differential
\begin{equation}
\lambda=v\frac{\text{d}z}{z}.
\end{equation}
Then
\begin{equation}
q=\lambda^2=v^2\frac{\text{d}z^2}{z^2}=:\phi(z)\text{d}z^2\label{q}
\end{equation}
is the quadratic differential on $\mathcal{C}$. For our purposes, the important point to note is that the SW curve (\ref{SWcurve}) can be written in the form (\ref{q}) \cite{Tachikawa:2013kta}. This construction will prove essential in what follows.

\paragraph{SW curves and Dessins} As mentioned above, the SW curve $\Sigma$ is related to the quadratic differential $q$. Moving in the moduli space of the theory in question will alter the parameters in the SW curve, thereby altering the parameters in $q$ \cite{He:2015vua}. Following \cite{mp}, it was found in \cite{He:2015vua} that at certain isolated points in the Coulomb branch $\mathcal{U}_{g,n}$, where $g$ is the genus of the Gaiotto curve $\mathcal{C}$ with $n$ marked points, $q$ is completely fixed and becomes a Strebel differential $q=\phi(t)\text{d}t^2=\frac{\text{d}\beta^2}{4\pi^2\beta(t)(1-\beta(t))}$.

As examples for SU(2) with $N_f=4$, we will discuss 6 Strebel points in $\mathcal{U}_{g,n}\times\mathbb{R}^n$, for which the Belyi maps are presented in Table \ref{table.belyi}.
\begin{table}[h]
	\centering
	\begin{tabular}{|c|c|c|c|}
		\hline
		Graph & $\beta(t)$ & Ramification & Strebel $q$\\
		\hline
		$\Gamma(3)$ & 
		$\frac{t^3(t+6)^3(t^2-6t+36)^3}{1728(t-3)^3(t^2+3t+9)^3}$
		&
		$[3^4 | 2^6 | 3^4 ]$
		&
		$-\frac{9t(t^3+216)}{4\pi^2(t^3-27)^2}$
		\\
		\hline
		$\Gamma_0(4)\cap\Gamma(2)$ & 
		$\frac{(t^4+224t^2+256)^3}{1728t^2(t-4)^4(t+4)^4}$
		&
		$[3^4 | 2^6 | 4^2,2^2 ]$
		&
		$-\frac{4t^4+896t^2+1024}{4\pi^2t^2(t^2-16)^2}$
		\\
		\hline
		$\Gamma_1(5)$ & 
		$\frac{(t^4+248t^3+4064t^2+22312t+40336)^3}{1728(t+5)(t^3-t-31)^5}$
		&
		$[3^4 | 2^6 | 5^2,1^2 ]$
		&
		$-\frac{t^4+248t^3+4064t^2+22312t+40336}{4\pi^2(t+5)^2(t^2-t-31)^2}$
		\\
		\hline
		$\Gamma_0(6)$ & 
		$\frac{(t+7)^3(t^3+237t^2+1443t+2287)^3}{1728(t+3)^2(t+4)^3(t-5)^6}$
		&
		$[3^4 | 2^6 | 6,3,2,1 ]$
		&
		$-\frac{(t+7)(t^3+237t^2+1443t+2287)}{4\pi^2(t+5)^2(t+3)^2(t+4)^2}$
		\\
		\hline
		$\Gamma_0(8)$ & 
		$\frac{(t^4+240t^3+2144t^2+3840t+256)^3}{1728t(t+4)^2(t-4)^8}$
		&
		$[3^4 | 2^6 | 8,2,1^2 ]$
		&
		$-\frac{t^4+240t^3+2144t^2+3840t+256}{4\pi^2t^2(t^2-16)^2}$
		\\
		\hline
		$\Gamma_0(9)$ & 
		$\frac{(t+6)^3(t^3+234t^2+756t+2160)^3}{1728(t^2+3t+9)(t-3)^9}$
		&
		$[3^4 | 2^6 | 9,1^3 ]$
		&
		$-\frac{(t+6)(t^3+234t^2+756t+2160)}{4\pi^2(t^3-27)^2}$
		\\
		\hline
	\end{tabular}
	\caption{The list of the six genus-zero, torsion-free, congruence subgroups of the modular group $\Gamma$, of index 12. The corresponding Belyi maps $\beta(t)$ and their ramification indices, as well as the Strebel differentials are also shown. Note that the ramification indices for all 6 are such that there are 4 pre-images of 0 of order 3 and 6 pre-images of 1 of order 2. The pre-images of $\infty$ (aka the \emph{cusp widths}) all add to 12, as do the ramification indices for 0 and 1. This is required by the fact that all the subgroups are of index 12 within $\Gamma$.}\label{table.belyi}
\end{table}
These six Belyi maps are those found in \cite{He:2012jn,MS} to be associated to the six \emph{genus zero}, \emph{torsion-free}, \emph{congruence} subgroups of the modular group $\Gamma=\text{PSL}(2,\mathbb{Z})\cong\mathbb{Z}_2*\mathbb{Z}_3$, where $*$ denotes the free product\footnote{For the background on the congruence subgroups of $\Gamma$, see Appendix \ref{conggp}. It remains an open question whether dessins associated to other subgroups of the modular group, perhaps of higher index, arise for other $\mathcal{N}=2$ generalised quiver theories in a parallel manner.}.

\begin{figure}[h]
	\centering
	\begin{minipage}[t]{0.25\textwidth}
		\begin{center}
			\includegraphics[scale=0.15]{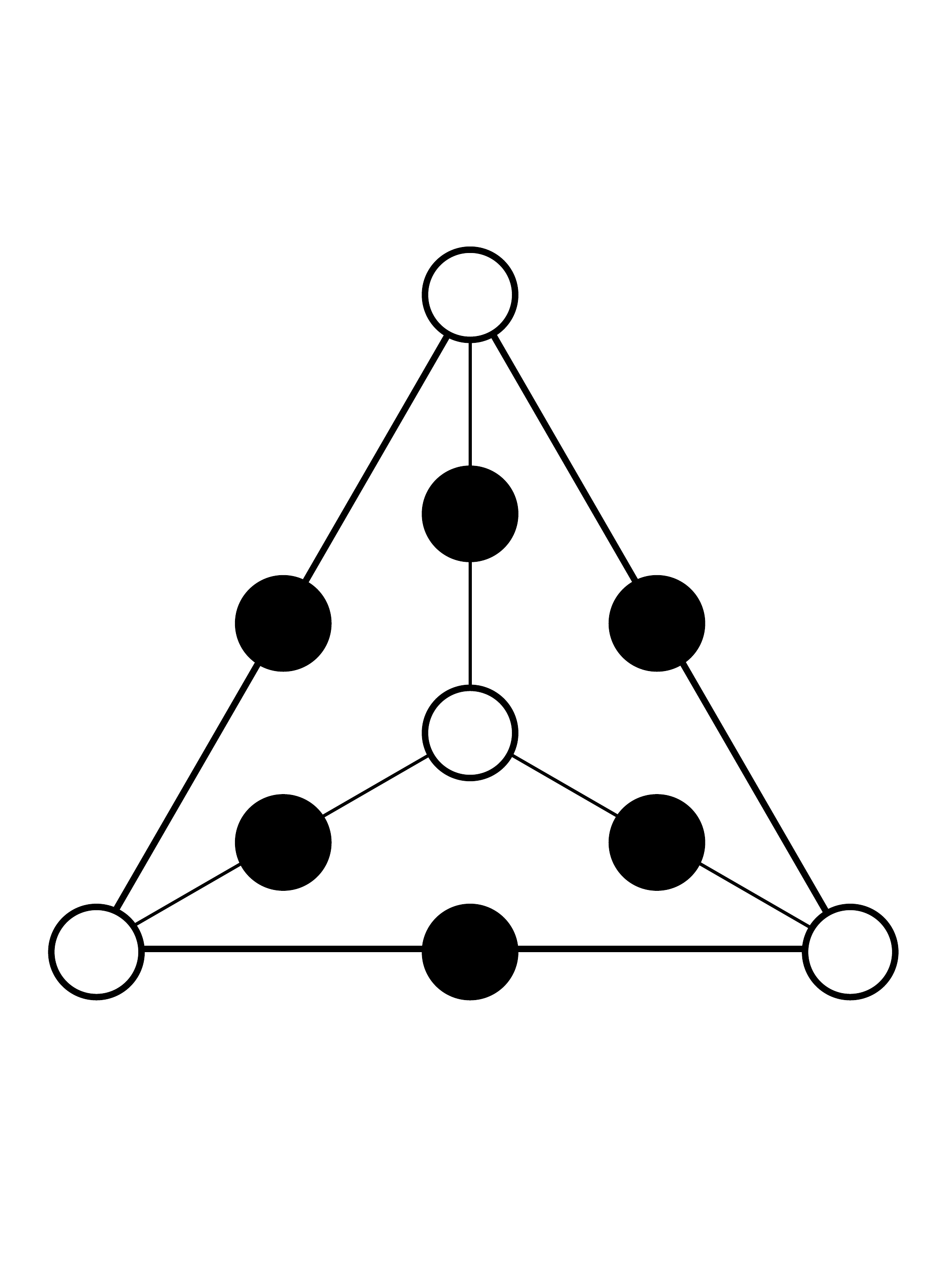}
			\par
			$\Gamma (3)$
		\end{center}
	\end{minipage}
	\begin{minipage}[t]{0.25\textwidth}
		\begin{center}
			\includegraphics[scale=0.15]{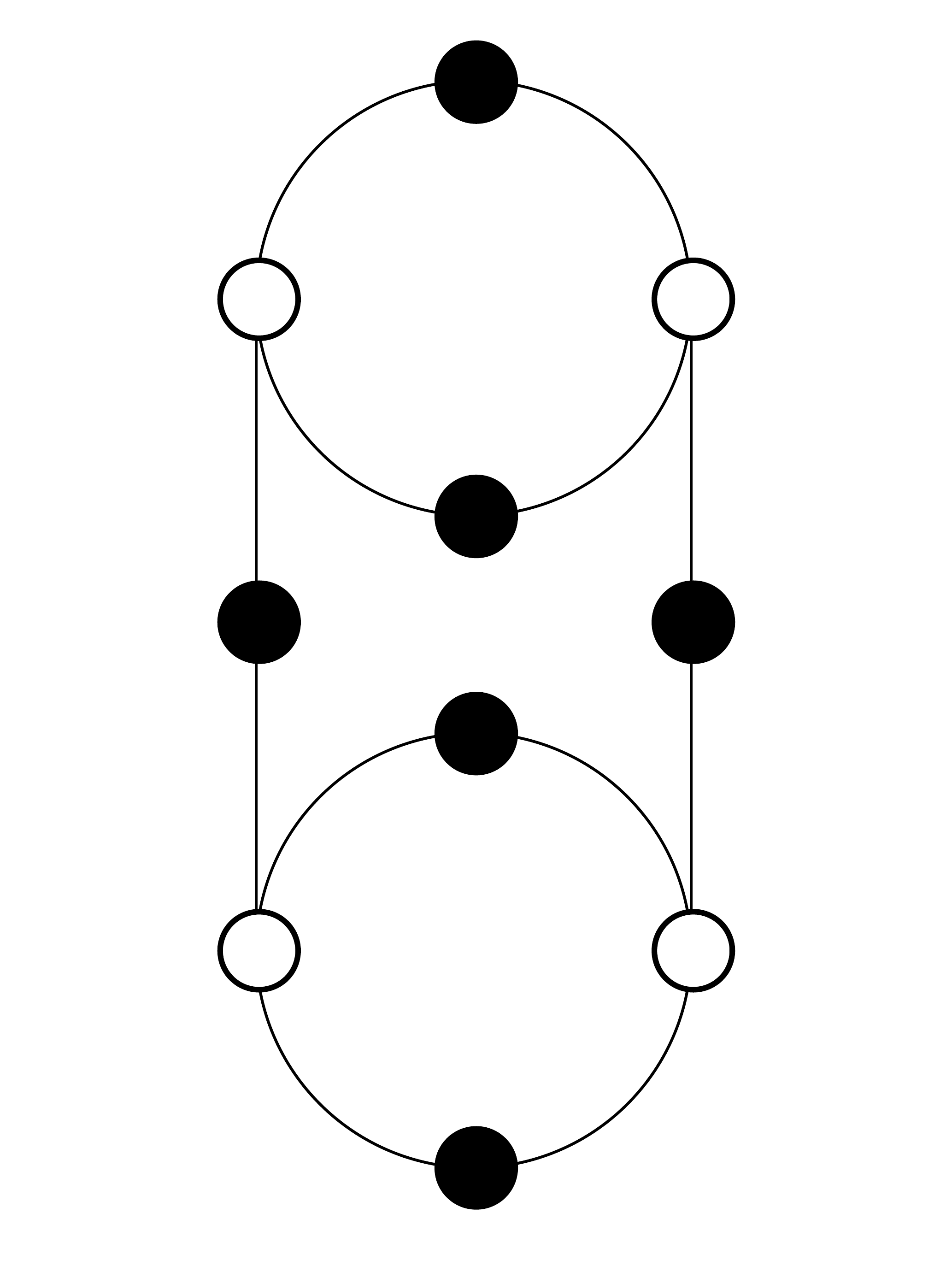}
			\par
			$\Gamma_0 (4) \cap \Gamma (2)$
		\end{center}
	\end{minipage}
	\begin{minipage}[t]{0.25\textwidth}
		\begin{center}
			\includegraphics[scale=0.15]{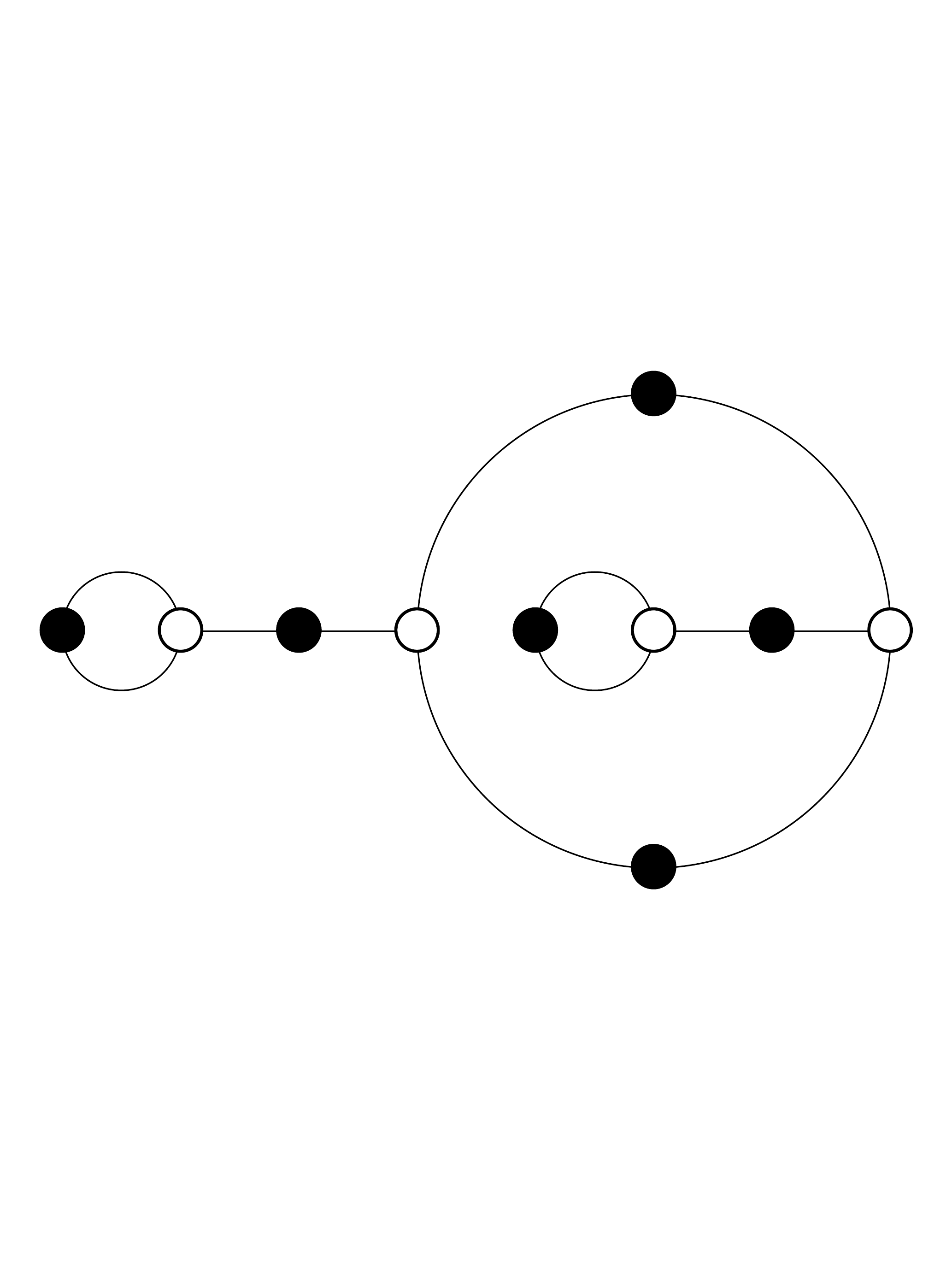}
			\par
			$\Gamma_1 (5)$
		\end{center}
	\end{minipage}
	\begin{minipage}[t]{0.25\textwidth}
		\begin{center}
			\includegraphics[scale=0.15]{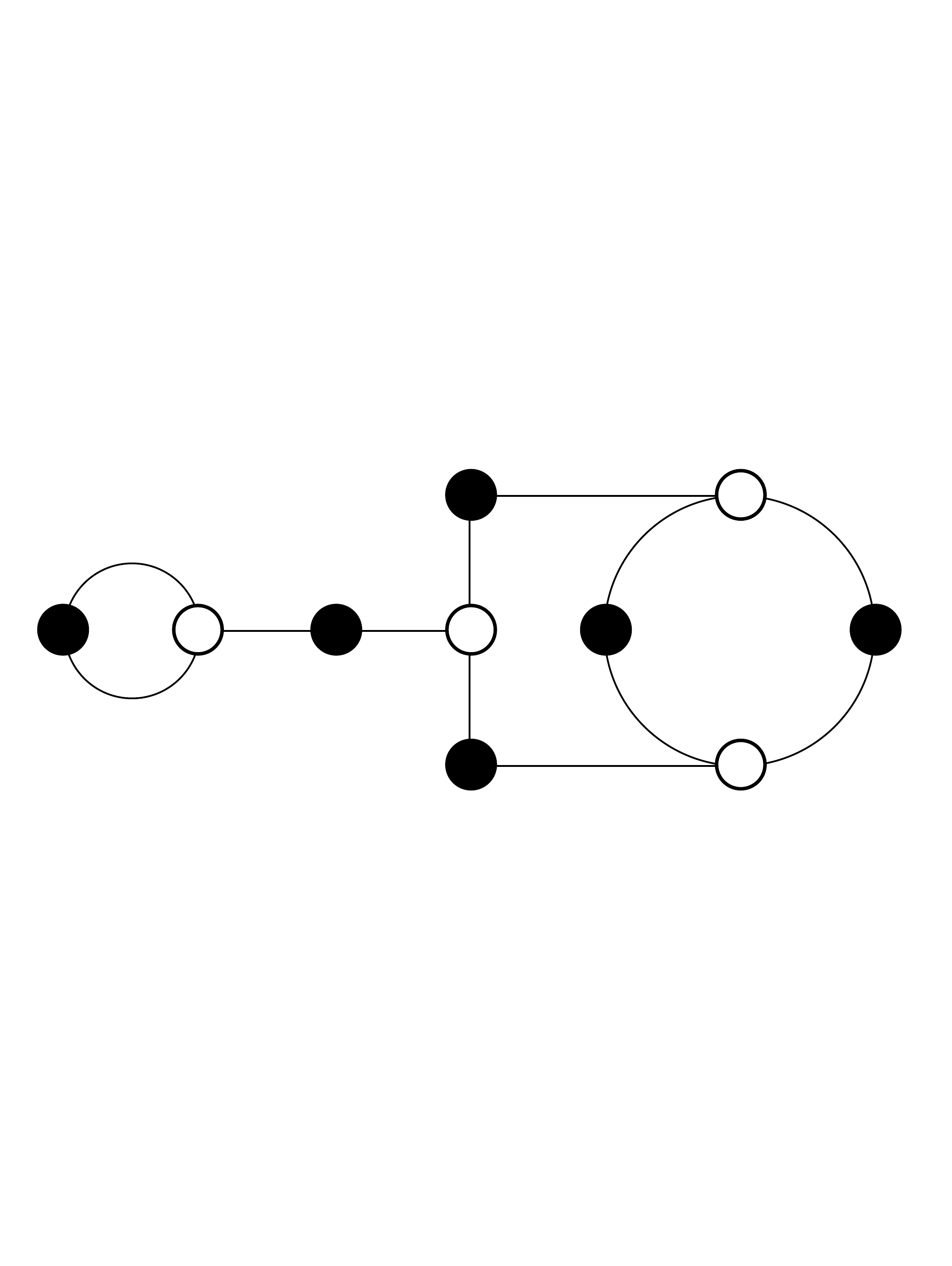}
			$\Gamma_0 (6)$
		\end{center}
	\end{minipage}
	\begin{minipage}[t]{0.25\textwidth}
		\begin{center}
			\includegraphics[scale=0.15]{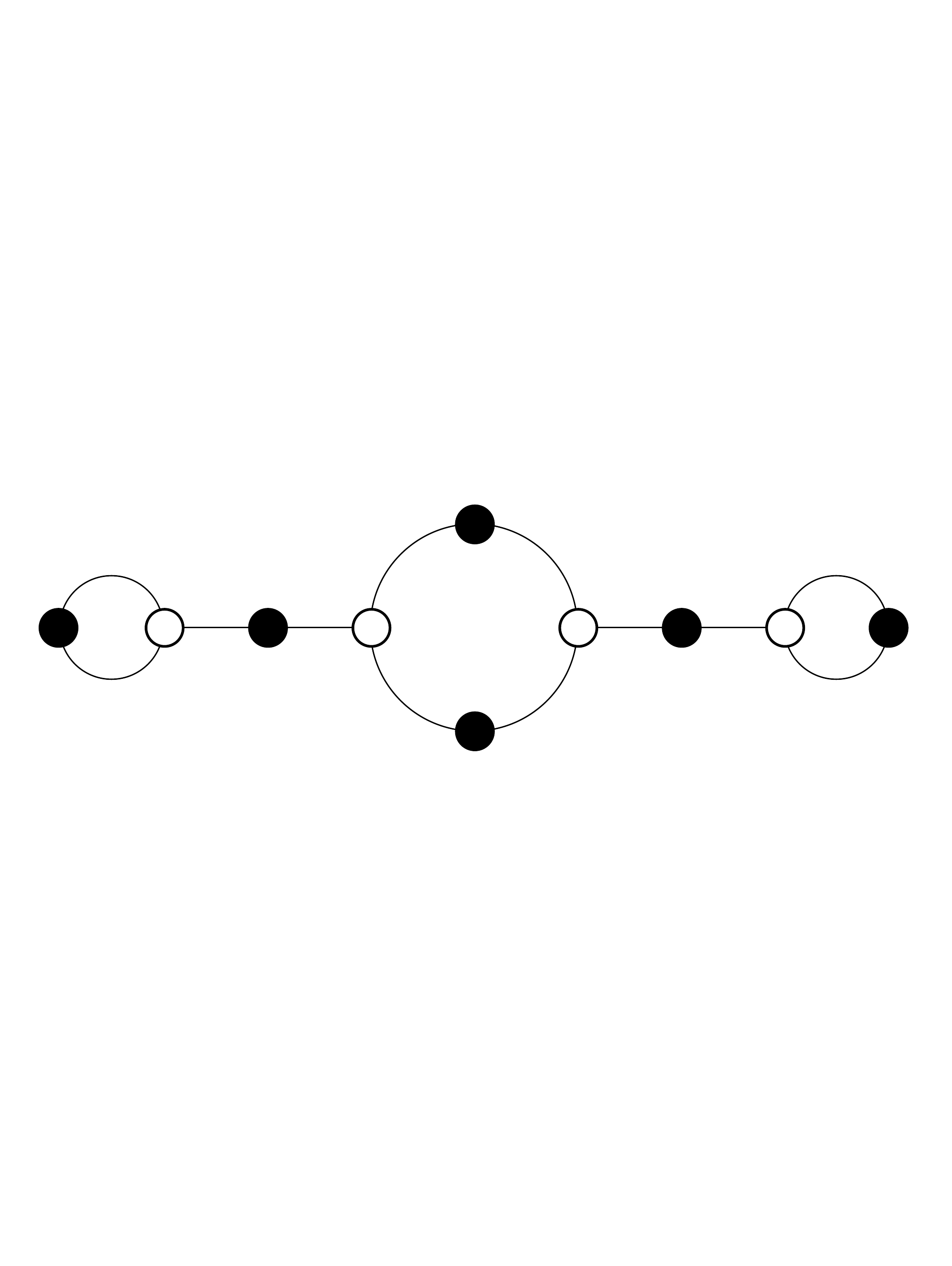}
			\par
			$\Gamma_0 (8)$
		\end{center}
	\end{minipage}
	\begin{minipage}[t]{0.25\textwidth}
		\begin{center}
			\includegraphics[scale=0.15]{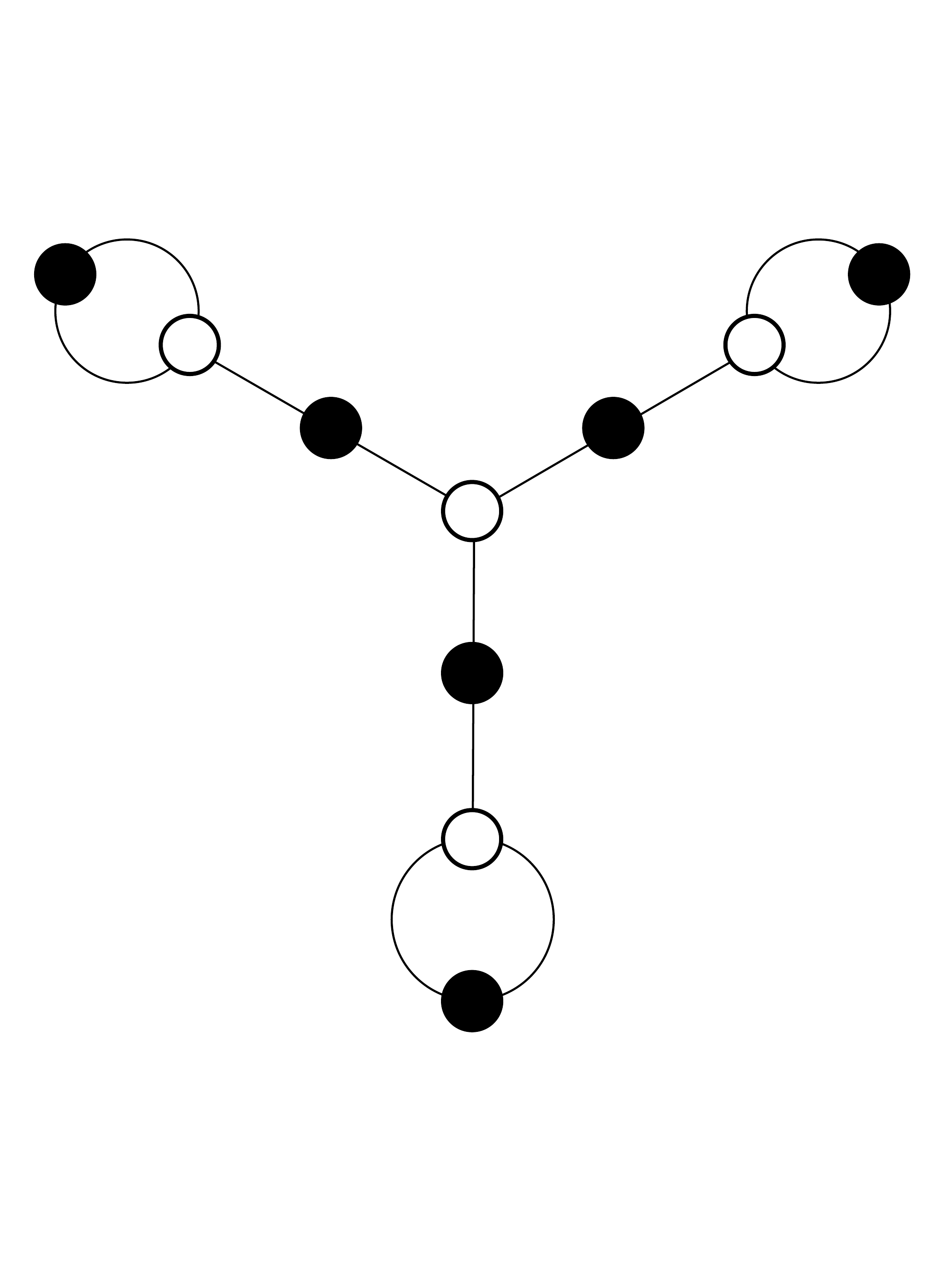}
			\par
			$\Gamma_0 (9)$
		\end{center}
	\end{minipage}
	\caption{The dessins d'enfants associated to the six Strebel points of the SU(2), $N_f=4$ theory.}\label{figure.dessins}
\end{figure}

From the Belyi maps in Table \ref{table.belyi}, we can compute the associated dessins as displayed in Figure \ref{figure.dessins}.
The dessins d'enfants associated to each Strebel point of the generalised quiver theory in question turn out to have an interpretation as so-called \emph{ribbon graphs} on the Gaiotto curve $\mathcal{C}$. For details, the readers are referred to \cite{He:2015vua,mp}.

\section{From Dessins to Conformal Blocks}\label{dessin2CB}
Let us now complete the cycle of the route map above by considering what gauge theory and CFT data we can obtain starting from these 6 dessins.

\subsection{The SU(2) with 4 Flavours}\label{SU2Nf4}
Given that all our graphs in Figure \ref{figure.dessins} are drawn on the Riemann surface 
(genus zero) with 4 marked points (one for each face), we can naturally interpret these 
as Gaiotto curves \cite{He:2012kw,He:2015vua}, and thence $\mathcal{N}=2$ gauge theories.

To begin, the Seiberg-Witten curve $\Sigma$ for the SU(2) $N_f=4$ theory in algebraic form is standard \cite{Tachikawa:2013kta}. For future convenience, we start with the SW curve of form \eqref{eq:4dcurve} and write the SW differential as \cite{Eguchi:2009gf,Kozcaz:2010af}
\begin{equation}\label{lamSW}
\lambda_\text{SW} = \frac{\sqrt{P_4(z)}}{z(z-1)(z- \zeta)} \text{d}z, 
\qquad
P_4(z) = m_0^2 \prod \limits_{i=1}^4 (z - \lambda_i) = m_0^2 \sum \limits_{i=0}^4 z^{4-i} S_i,
\end{equation}
under the substitution
\begin{eqnarray}
	&&\lambda_\text{SW}=v\text{d}z/z,~t=z,\nonumber\\
	&&\mu_1=m_2+m_0,~\mu_2=m_2-m_0,~\mu_3=m_3+m_1,~\mu_4=m_3-m_1,
\end{eqnarray}
The parameters $S_i$ are given in terms of the flavour mass and coupling parameters $m_{0,1,2,3},\zeta,U\in\mathbb{C}$ so that $S_0 = 1$ for the top coefficient and
\begin{align}
m_0^2 S_1 &= - \left( m_0^2 + m_2^2 (\zeta - 1) + m_0^2 \zeta + 2 m_2 m_3 \zeta+ (1 + \zeta) U \right),
\nonumber\\
m_0^2 S_2 &=   (m_0^2 + m_1^2 - m_3^2 + 2 m_2 m_3)\zeta+ m_2^2 (\zeta - 1) \zeta + 2 m_2 m_3  \zeta^2 + m_3^2  \zeta^ 2 + (1 + \zeta)^2 U,
\nonumber\\
m_0^2 S_3 &= - \left( (m_1^2 - m_3^2) \zeta + (m_1^2 + 2 m_2 m_3 + m_3^2)  \zeta^2 +\zeta (1 + \zeta) U \right), 
\nonumber\\
m_0^2 S_4 &= m_1^2  \zeta^2 \ .\label{Spar}
\end{align}
Now the SW curve is of the form
\begin{eqnarray}
	&&z^2(v-(m_0+m_2))(v-(m_2-m_0))+z(1+\zeta)\left(-v^2+\frac{2\zeta}{(1+\zeta)}(m_2+m_3)v+U\right)\nonumber\\
	&&+\zeta(v-(m_1+m_3))(v-(m_3-m_1))=0.
\end{eqnarray}

On the other hand, the $S$-parameters can be written in terms of the $\lambda_i$ as standard symmetric polynomials,
\begin{equation}
S_k = \sum\limits_{1 \leq j_1 \leq \ldots \leq j_k \leq 4} \lambda_{j_1} \ldots \lambda_{j_k} \ .\label{symS}
\end{equation}
Following Appendix \ref{ellipticK}, we can then write
\begin{equation}
\frac{\text{d}a(U)}{\text{d}U} = - \frac{1}{\pi \text{i}}
\frac{1+ \zeta}{m_0 \sqrt{(\lambda_2 - \lambda_3)(\lambda_4 - \lambda_1)}}
K(r),\label{dadu}
\end{equation}
where
\begin{equation}
r^2 = \frac{(\lambda_1 - \lambda_2)(\lambda_3 - \lambda_4)}{(\lambda_2 - \lambda_3)(\lambda_4 - \lambda_1)},
\end{equation}
and $K(r)$ is the elliptic integral of the first kind. The right hand side of \eqref{dadu} implicitly depends on $U$, through $\lambda_i$ and thence $S_i$, thus we only need to integrate it to obtain $a(U)$ as a function of $U$, which could be a daunting task 
analytically.

Let us nevertheless attempt at some simplifications. First, we see that the right hand side depends only on the cross-terms in the four $\lambda_i$, which we will denote as $\lambda_{(ij)(kl)}=(\lambda_i-\lambda_j)(\lambda_k-\lambda_l)$. Combining with \eqref{symS}, let us see whether these can be directly expressed in terms of $S_i$, and thence, in terms of $U$. This is a standard algebraic elimination problem and we readily find the following:
\begin{lemma}
	Consider the monic cubic polynomial,
	{\small\begin{eqnarray}
	&&x^3 + \left( -2 S_2^2 + 6 S_1 S_3 - 24 S_4 \right) x^2+\left( S_2^4-6 S_1 S_3 S_2^2+24 S_4 S_2^2+9 S_1^2 S_3^2+144 S_4^2-72 S_1 S_3 S_4 \right) x  \nonumber\\
	&&+27 S_4^2 S_1^4+4 S_3^3 S_1^3-18 S_2 S_3 S_4 S_1^3-144 S_2 S_4^2 S_1^2 + 4 S_2^3 S_4 S_1^2 + 6 S_3^2 S_4 S_1^2 - 18 S_2 S_3^3 S_1+192 S_3 S_4^2 S_1 \nonumber\\ 
	&&+ 80 S_2^2 S_3 S_4 S_1 + 27 S_3^4-256 S_4^3 + 4 S_2^3 S_3^2-S_1^2 S_2^2 S_3^2 + 128 S_2^2 S_4^2 - 16 S_2^4 S_4 - 144 S_2 S_3^2 S_4.
	\end{eqnarray}}
	Then the squares of the 3 cross-products 
	\begin{equation}
	x_1 = \lambda_{(12)(34)}^2, \ 
	x_2 = \lambda_{(23)(41)}^2, \ 
	x_3 = \lambda_{(13)(24)}^2
	\end{equation}
	are the three roots of it.
\end{lemma}

Of course, we can substitute the $S_i$ parameters in terms of the $m_i, \zeta, U$ parameters from \eqref{lamSW}, though the expression become too long to present here. Now, we have
\begin{equation}
a(U_0)-a(U) = - \frac{1+ \zeta}{m_0 \pi \text{i}} \int_U^{U_0}\frac{\text{d}U'}{\sqrt[4]{x_2(U')}} K \left( \frac{\sqrt[4]{x_1(U')}}{\sqrt[4]{x_2(U')}} \right).
\end{equation}
To determine the integral constant, we choose $U_0$ such that $a(U_0)=0$. We can find such $U_0$ by solving the discriminant of $P_4(z)$ where two branch points coincide and the $A$-cycle shrinks\footnote{Alternatively, we may also integrate from $U$ to $\infty$ as the large $U$ behaviour can be determined as in Appendix \ref{ellipticLog}.}. Hence,
\begin{equation}
	a(U) = \frac{1+ \zeta}{m_0 \pi \text{i}} \int_U^{U_0}\frac{\text{d}U'}{\sqrt[4]{x_2(U')}} K \left( \frac{\sqrt[4]{x_1(U')}}{\sqrt[4]{x_2(U')}} \right).\label{aU}
\end{equation}
In general, when we integrate from some $U$ to $U_0$, the positions of branch points and cuts might change. Therefore, this is really a sum of integrals:
\begin{equation}
	\int_U^{U_0}=\int_U^{U_1}+\int_{U_1}^{U_2}+\dots+\int_{U_{n-1}}^{U_n=U_0}\label{intsum}
\end{equation}
such that $x_i$ does not change its expression for each integral on the right hand side.

Recall the definition of the Seiberg-Witten differential from \eqref{q}, we 
have that 
\begin{equation}
	\lambda_\text{SW}^2 = \phi_\text{SW}(z) \text{d} z^2\label{qq}
\end{equation}
is a quadratic differential. This is the above mentioned meromorphic quadratic differential on $\mathcal{C}$. Moving in the moduli space of the theory in question will alter the parameters  
in the Seiberg-Witten curve, thereby altering the parameters in $q$ (cf. \cite{He:2015vua}).
Following \cite{mp}, it was found in \cite{He:2015vua} that at certain isolated points in the Coulomb branch of the moduli space $\mathcal{U}_{g, n}$ of the gauge theory in question, where $g$ is the genus of the Gaiotto curve $\mathcal{C}$ with $n$ marked points, $q$ is completely fixed, which becomes a Strebel differential.

We therefore have two forms of the Strebel differentials, $\phi_{\beta}(t)$ coming 
from the dessin and $\phi_\text{SW}(z)$ coming from the physics. Now, because dessins are 
\emph{rigid}, they have no parameters. The insight of Belyi and Grothendieck is precisely that the maps $\beta$ have parameters fixed at very special algebraic points in moduli space. Thus, $\phi_{\beta}(t)$ is of a particular form, as a rational function in $t$ with fixed algebraic coefficients.

On the other hand $\phi_\text{SW}(z)$ from the gauge theory has parameters which we saw earlier, corresponding to masses, couplings etc. Therefore, up to redefinition of the variables $(t,z)$ and identifying $\phi_\text{SW}(z)$ and $\phi_{\beta}(t)$ it is natural to ask how the special values of the parameters from the dessin perspective \emph{fix} the physical parameters in the gauge theory and if a dessin implicates any interesting physical theory.

We have now introduced all the necessary dramatis personae of our tale and our strategy is thus clear. There are also some further details that we should be careful about in the calculations. We will work through an example in detail to illustrate them in the following subsection.

\subsection{Example: $\Gamma(3)$}\label{gamma3}
Let us take the dessin for $\Gamma(3)$, whose Belyi map is
\begin{equation}
	\beta(t)=\frac{t^3(t+6)^3(t^2-6t+36)^3}{1728(t-3)^3(t^2+3t+9)^3}.
\end{equation}
We can readily get the pre-images of 0, 1 and $\infty$:
\begin{equation}
\begin{array}{| c | c | c |} \hline
& \mbox{Pre-image} & \mbox{Ramification} 
\\ 
\hline 
\hline
\beta^{-1} (0) & -6                & 3 \\
&  0                & 3 \\
&  3 - 3 i \sqrt3 & 3 \\
&  3 + 3 i \sqrt3 & 3 \\
\hline
\beta^{-1} (1) &  3 (1 - \sqrt3) & 2 \\
&  3 (1 + \sqrt3) & 2 \\
&  \left(\frac32 +\frac{3 i}{2} \right)   \left(\sqrt3 + (-2-i)\right)  & 2 \\
&  \left(-\frac32 -\frac{3 i}{2}\right)  \left(\sqrt3 + ( 2+i) \right)  & 2 \\
&       \frac12                    \left((-3+9 i)-(3-3 i) \sqrt3\right) & 2 \\
&       \frac12                    \left((-3+9 i)+(3-3 i) \sqrt3\right) & 2 \\
\hline
\beta^{-1}(\infty) & \infty & 3 \\
&            3                                                      & 3 \\
&       -\frac32 i \left(\sqrt3 - i\right)                          & 3 \\
&        \frac32 i \left(\sqrt3 + i\right)                          & 3 \\
\hline
\end{array}.
\end{equation}
We can construct the corresponding dessin as in Figure \ref{figure.dessins}. Subsequently, using \eqref{Strebel}, we see that the Strebel differential is $q=\phi_\beta(t)\text{d}t^2$, where
\begin{equation}
	\phi_\beta(t)=-\frac{9t(t^3+216)}{4\pi^2(t^3-27)^2}.\label{eq:gamma3}
\end{equation}
We have marked $\phi$ with a subscript $\beta$ to emphasize its dessin origin. On the other side, we have the Seiberg-Witten curve and quadratic differential for SU(2) with $N_f = 4$ from \eqref{lamSW} and \eqref{qq}, to be
\begin{align}
\phi_\text{SW}(z) & = \frac{P_4(z)}{ \left( z(z-1)(z- \zeta) \right)^2} \ ,  \quad \mbox{where}
\nonumber\\
P_4(z) &= z^4 m_0^2 - z^3  \left( m_0^2 + m_2^2 (\zeta - 1) + m_0^2 \zeta + 2 m_2 m_3 \zeta+ (1 + \zeta) U \right)
\nonumber\\
& + z^2 \left( (m_0^2 + m_1^2 - m_3^2 + 2 m_2 m_3)\zeta+ m_2^2 (\zeta - 1) \zeta + 2 m_2 m_3  \zeta^2 + m_3^2  \zeta^ 2 + (1 + \zeta)^2 U \right)
\nonumber\\
& - z \left( (m_1^2 - m_3^2) \zeta + (m_1^2 + 2 m_2 m_3 + m_3^2)  \zeta^2 +\zeta (1 + \zeta) U\right) + m_1^2  \zeta^2.\label{SW_curve.2}
\end{align}

Here, likewise we have marked $\phi$ with a subscript ``SW'' to emphasize its Seiberg-Witten
origin. We have also explicitly written the differential coming from the Seiberg-Witten 
side in terms of the parameters $m_{0,1,2,3},\zeta, U$.

We need to match \eqref{eq:gamma3} with \eqref{SW_curve.2}, up to an PGL$(2,\mathbb{C})$ 
transformation on the complex variable $z$. The reason for this is that we are dealing in 
this example with a quadratic differential on the \emph{sphere}. For curves of higher 
genus, such PGL$(2,\mathbb{C})$ transformations are generically not permitted, as they 
will not preserve the structure of the poles and zeros of the quadratic differential.

We can therefore write
\begin{equation}
z = \frac{a t + b}{c t + d},
\qquad
a, b, c, d \in \mathbb{C}
\end{equation}
and solve for complex coefficients $a, b, c, d$ as well as the parameters $m_{0,1,2,3},\zeta,U$ so that we have identically for all $t$ that
\begin{equation}
\phi_{\beta}(t) = \phi_\text{SW}\left(  \frac{a t + b}{c t + d} \right).
\end{equation}
There are actually continuous families of $2\times2$ matrices solving this equation for a given dessin. As the elliptic curve is the same up to an overall factor, it turns out that each continuous family would simply scale the SW differential by $\phi_\text{SW}\rightarrow k^2\phi_\text{SW}$ with $k^2\in\mathbb{R}$, where the square comes from the $\lambda_\text{SW}^2$ in the differential. Obviously, equating the numerators of $\phi_\beta$ and $\phi_\text{SW}$ as well as equating their denominators would give a solution. For future convenience, such solution will be referred to as the ``basic'' values of the parametrization. Then other parametrizations would simply follow
\begin{equation}
	\phi_\text{SW}=k^2\phi_\text{SW,basic}.
\end{equation}
There are two points we should pay attention to:
\begin{itemize}
	\item As we will try to relate this to minimal models, due to modular invariance, we can only allow primary states with pure imaginary charges \cite{francesco2012conformal}. Recall the AGT relation \eqref{agt}, which in terms of $m_i$ is
	\begin{equation}
		\frac{m_0}{\sqrt{\epsilon_1\epsilon_2}}+\frac{Q}{2}=\alpha_4,~\frac{m_1}{\sqrt{\epsilon_1\epsilon_2}}+\frac{Q}{2}=\alpha_1,~\frac{m_2}{\sqrt{\epsilon_1\epsilon_2}}=\alpha_3,~\frac{m_3}{\sqrt{\epsilon_1\epsilon_2}}=\alpha_2,~\frac{a}{\sqrt{\epsilon_1\epsilon_2}}+\frac{Q}{2}=\alpha_\text{int}.\label{agt2}
	\end{equation}
	In fact, $\epsilon_{1,2}$ are not completely free once $Q=(\epsilon_1+\epsilon_2)/\sqrt{\epsilon_1\epsilon_2}$ is chosen. Moreover, to have real conformal dimensions, $m_i$'s and $a$ should only be real or pure imaginary (depending on $\epsilon_{1,2}$). This is also the reason why $k^2$ should be real.
	\item One may easily check that an SW differential/elliptic curve would have the same $j$-invariant under $\phi\rightarrow k^2\phi$. As a result, the parameters, based on their mass dimensions or by looking at $P_4(z)$ and $a(U)$, would follow
	\begin{eqnarray}
		&&m_i\rightarrow km_i,~a\rightarrow ka,~\epsilon_i\rightarrow k\epsilon_i,~U\rightarrow k^2U;\nonumber\\
		&&\zeta\rightarrow\zeta,~\alpha_{i,\text{int}}\rightarrow \alpha_{i,\text{int}},~Q\rightarrow Q.\label{fam}
	\end{eqnarray}
    Therefore, rather than discrete parameters, we would have \emph{families} of differentials. Importantly, we can see that the coupling $\zeta$ is \emph{invariant}. Following the AGT map, the dimensionless CFT parameters, $\alpha_{i,\text{int}}$ and $Q$, are also invariant under the scaling though we still have the freedom to choose $\sqrt{\epsilon_1\epsilon_2}$.
\end{itemize}

Now expanding the above and setting all the coefficients of $t$ to vanish identically gives 
a complicated polynomial system in $(a, b, c, d, m_{0,1,2,3}, \zeta, U)$ for which one 
can find many solutions. For example, the following constitutes a solution (with $k^2=1$),
\begin{align}
m_0 = -m_1 = m_2 = -m_3 = \frac{1}{2 \sqrt{3} \pi },\quad
\zeta = \frac{1}{2}+\frac{\text{i} \sqrt{3}}{2} \ , \quad
U = \frac{1}{9 \pi ^2}
\label{parGamma3}
\end{align}
with $(a,b,c,d) = 
\left(
\frac{1+\text{i}}{\sqrt{2} 3^{3/4}} , \
\text{i} \left(\frac{\sqrt[4]{3}}{2 \sqrt{2}}-\frac{3^{3/4}}{2 \sqrt{2}}\right)+\frac{\sqrt[4]{3}}{2 \sqrt{2}}+\frac{3^{3/4}}{2 \sqrt{2}} , \
0, \
\frac{(1-\text{i}) 3^{3/4}}{\sqrt{2}}\right)$. The numerator of the SW differential takes the form
\begin{equation}
P_4(z) = 
-\frac{6 z^4-4 \text{i} \left(\sqrt{3}-3 \text{i}\right) z^3+\left(6+6 \text{i} \sqrt{3}\right) z^2-8 \text{i} \sqrt{3} z+3 \text{i} \sqrt{3}-3}{72 \pi ^2}.
\end{equation}
We now need the roots $\lambda_i$ of $P_4(z)$ as given in \eqref{lamSW}:
\begin{equation}
z^4+\left(-2-\frac{2 \text{i}}{\sqrt{3}}\right) z^3+\left(1+\text{i} \sqrt{3}\right) z^2-\frac{4 \text{i} }{\sqrt{3}}z+\frac{1}{2} \text{i} \left(\sqrt{3}+\text{i}\right)
= \prod \limits_{i=1}^4 (z - \lambda_i).
\end{equation}
The SW curve itself is genus 1 and is in fact an elliptic curve. We can recast \eqref{SW_curve.2} as
\begin{align}
	y^2 & = z^4 m_0^2 - z^3  \left( m_0^2 + m_2^2 (\zeta - 1) + m_0^2 \zeta + 2 m_2 m_3 \zeta+ (1 + \zeta) U \right)
	\nonumber\\
	& + z^2 \left( (m_0^2 + m_1^2 - m_3^2 + 2 m_2 m_3)\zeta+ m_2^2 (\zeta - 1) \zeta + 2 m_2 m_3  \zeta^2 + m_3^2  \zeta^ 2 + (1 + \zeta)^2 U \right)
	\nonumber\\
	& - z \left( (m_1^2 - m_3^2) \zeta + (m_1^2 + 2 m_2 m_3 + m_3^2)  \zeta^2 +\zeta (1 + \zeta) U\right) + m_1^2  \zeta^2,
\end{align}
where the redefinition $y^2=(z(z-1)(z-\zeta))^2\phi_\text{SW}(z)=P_4(z)$ is used. Following Appendix \ref{ellip}, as one may check, the $j$-invariant we get from the parameterization \eqref{parGamma3} agrees with the one directly from the Strebel differential \eqref{eq:gamma3}:
\begin{equation}
	j = 0.
\end{equation}
Indeed, $j=0$ corresponds to a special elliptic curve with $\mathbb{Z}/3\mathbb{Z}$-symmetry, much like the dessin for $\Gamma(3)$ itself.

In this case, we can integrate \eqref{aU} numerically to obtain $a(U)=\frac{1}{3\sqrt{3}\pi}$ \footnote{Numerical integration would often give decimals rather than precise values. For instance, here we get $a\approx0.06125877$. In some cases like here, we give exact values for $a$ with the help of minimal models. This can be done by checking multiple minimal models and finding certain $\epsilon_{1,2}$ so that $\Delta_i$ would fit into their Kac tables. Then the correct closed form of $a$ can be obtained if $\Delta_\text{int}$ also lives in these Kac tables for \emph{all these} minimal models. (To get the correct CBs under AGT map, we further need the fusion rule, but even if the corresponding $\Delta_\text{int}$ does not satisfy the fusion rule for some CB, this could still be regarded as a verfication of fine-tuning $a$ as long as $\Delta_\text{int}$, along with $\Delta_i$, belongs to the Kac table.)}. Now we can use the AGT relation \eqref{agt2} to get the parametrizations for CBs. If we take $Q=0$, we have
\begin{equation}
	\alpha_1=\alpha_2=-\alpha_3=-\alpha_4=\frac{\text{i}}{2\sqrt{3}\pi},~\alpha_\text{int}=\frac{\text{i}}{3\sqrt{3}\pi},\label{cftparamex}
\end{equation}
where we have chosen $-\epsilon_1=\epsilon_2=1$ as an example.

We can also have pure imaginary $m_i$'s and $a$ for the above example such as
\begin{equation}
	m_0 = -m_1 = m_2 = -m_3 = \frac{\text{i}}{2 \sqrt{3} \pi },~\zeta = \frac{1}{2}+\frac{\text{i} \sqrt{3}}{2},~U = -\frac{1}{9 \pi ^2},~a=\frac{\text{i}}{3\sqrt{3}\pi}.
\end{equation}
Then we can still get the same CFT parameters for $Q=0$ as in \eqref{cftparamex} with the choice $-\epsilon_1=\epsilon_2=\text{i}$.

\subsection{Matching Parameters}\label{matching}
Here, we report all parameters from the six dessins in Table \ref{param1}$\sim$\ref{param6}. Notice that we are only giving solutions coming from $(\pm)\phi_\text{SW,basic}$ with pure imaginary $m_i$'s and $a$. There is actually a family for each parametrization following \eqref{fam}.
\begin{center}
	\centering
	\renewcommand*{\arraystretch}{1.2}
	\begin{longtable}{|c|c|c|c|c|c|c|c|}
		\hline
		$\zeta=\text{e}^{2\pi\text{i}\tau}$ & $m_0$ & $m_1$ & $m_2$ & $m_3$ & $U$ & $\sum\limits_im_i$ & $a=\frac{\alpha_\text{int}}{\sqrt{\epsilon_1\epsilon_2}}+\frac{Q}{2}$ \\ \hline\hline
		& $-\frac{i}{2\sqrt{3}\pi}$ & $-\frac{i}{2\sqrt{3}\pi}$ & $\frac{i}{2\sqrt{3}\pi}$ & $-\frac{i}{2\sqrt{3}\pi}$ & $-\frac{1}{9\pi^2}$ & $\frac{i}{\sqrt{3}\pi}$ & $-\frac{i}{3\sqrt{3}\pi}$ \\ \cline{2-8}
		$\frac{1}{2}(1-i\sqrt{3})$ & $-\frac{i}{2\sqrt{3}\pi}$ & $\frac{i}{2\sqrt{3}\pi}$ & $-\frac{i}{2\sqrt{3}\pi}$ & $\frac{i}{2\sqrt{3}\pi}$ & $-\frac{1}{9\pi^2}$ & 0 & $-\frac{i}{3\sqrt{3}\pi}$ \\ \cline{2-8}
		& $\frac{i}{2\sqrt{3}\pi}$ & $-\frac{i}{2\sqrt{3}\pi}$ & $\frac{i}{2\sqrt{3}\pi}$ & $-\frac{i}{2\sqrt{3}\pi}$ & $-\frac{1}{9\pi^2}$ & 0 & $\frac{i}{3\sqrt{3}\pi}$ \\ \cline{2-8}
		& $\frac{i}{2\sqrt{3}\pi}$ & $\frac{i}{2\sqrt{3}\pi}$ & $-\frac{i}{2\sqrt{3}\pi}$ & $\frac{i}{2\sqrt{3}\pi}$ & $-\frac{1}{9\pi^2}$ & $-\frac{i}{\sqrt{3}\pi}$ & $\frac{i}{3\sqrt{3}\pi}$ \\ \cline{1-8}
		& $-\frac{i}{2\sqrt{3}\pi}$ & $-\frac{i}{2\sqrt{3}\pi}$ & $\frac{i}{2\sqrt{3}\pi}$ & $-\frac{i}{2\sqrt{3}\pi}$ & $-\frac{1}{9\pi^2}$ & $\frac{i}{\sqrt{3}\pi}$ & $\frac{i}{3\sqrt{3}\pi}$ \\ \cline{2-8}
		$\frac{1}{2}(1+i\sqrt{3})$ & $-\frac{i}{2\sqrt{3}\pi}$ & $\frac{i}{2\sqrt{3}\pi}$ & $-\frac{i}{2\sqrt{3}\pi}$ & $\frac{i}{2\sqrt{3}\pi}$ & $-\frac{1}{9\pi^2}$ & 0 & $\frac{i}{3\sqrt{3}\pi}$ \\ \cline{2-8}
		& $\frac{i}{2\sqrt{3}\pi}$ & $-\frac{i}{2\sqrt{3}\pi}$ & $\frac{i}{2\sqrt{3}\pi}$ & $-\frac{i}{2\sqrt{3}\pi}$ & $-\frac{1}{9\pi^2}$ & 0 & $-\frac{i}{3\sqrt{3}\pi}$ \\ \cline{2-8}
		& $\frac{i}{2\sqrt{3}\pi}$ & $\frac{i}{2\sqrt{3}\pi}$ & $-\frac{i}{2\sqrt{3}\pi}$ & $\frac{i}{2\sqrt{3}\pi}$ & $-\frac{1}{9\pi^2}$ & $-\frac{i}{\sqrt{3}\pi}$ & $-\frac{i}{3\sqrt{3}\pi}$ \\ \hline
		\caption{The parameters obtained from $\Gamma(3)$. Using \eqref{agt2}, we can get the values for $\alpha_i$'s.}\label{param1}
	\end{longtable}
\end{center}
\begin{center}
	\centering
	\renewcommand*{\arraystretch}{1.2}
	\begin{longtable}{|c|c|c|c|c|c|c|c|}
		\hline
		$\zeta=\text{e}^{2\pi\text{i}\tau}$ & $m_0$ & $m_1$ & $m_2$ & $m_3$ & $U$ & $\sum\limits_im_i$ & $a$ \\ \hline\hline
		& $-\frac{i}{8\pi}$ & $-\frac{i}{4\pi}$ & $-\frac{i}{4\pi}$ & $-\frac{i}{8\pi}$ & $-\frac{1}{192\pi^2}$ & $\frac{3i}{4\pi}$ & $\frac{i}{8\pi}$ \\ \cline{2-8}
		& $\frac{i}{8\pi}$ & $-\frac{i}{4\pi}$ & $-\frac{i}{4\pi}$ & $-\frac{i}{8\pi}$ & $-\frac{1}{192\pi^2}$ & $\frac{i}{2\pi}$ & $-\frac{i}{8\pi}$ \\ \cline{2-8}
		& $-\frac{i}{8\pi}$ & $-\frac{i}{4\pi}$ & $-\frac{i}{4\pi}$ & $\frac{i}{8\pi}$ & $-\frac{3}{64\pi^2}$ & $\frac{i}{2\pi}$ & $\frac{i}{8\pi}$ \\ \cline{2-8}
		& $-\frac{i}{8\pi}$ & $\frac{i}{4\pi}$ & $-\frac{i}{4\pi}$ & $-\frac{i}{8\pi}$ & $-\frac{1}{192\pi^2}$ & $\frac{i}{4\pi}$ & $\frac{i}{8\pi}$ \\ \cline{2-8}
		& $\frac{i}{8\pi}$ & $-\frac{i}{4\pi}$ & $-\frac{i}{4\pi}$ & $\frac{i}{8\pi}$ & $-\frac{3}{64\pi^2}$ & $\frac{i}{4\pi}$ & $-\frac{i}{8\pi}$ \\ \cline{2-8}
		& $-\frac{i}{8\pi}$ & $-\frac{i}{4\pi}$ & $\frac{i}{4\pi}$ & $-\frac{i}{8\pi}$ & $-\frac{3}{64\pi^2}$ & $\frac{i}{4\pi}$ & $\frac{i}{8\pi}$ \\ \cline{2-8}
		& $\frac{i}{8\pi}$ & $\frac{i}{4\pi}$ & $-\frac{i}{4\pi}$ & $-\frac{i}{8\pi}$ & $-\frac{1}{192\pi^2}$ & 0 & $-\frac{i}{8\pi}$ \\ \cline{2-8}
		$\frac{1}{2}$ & $-\frac{i}{8\pi}$ & $\frac{i}{4\pi}$ & $-\frac{i}{4\pi}$ & $\frac{i}{8\pi}$ & $-\frac{3}{64\pi^2}$ & 0 & $\frac{i}{8\pi}$ \\ \cline{2-8}
		& $\frac{i}{8\pi}$ & $-\frac{i}{4\pi}$ & $\frac{i}{4\pi}$ & $-\frac{i}{8\pi}$ & $-\frac{3}{64\pi^2}$ & 0 & $-\frac{i}{8\pi}$ \\ \cline{2-8}
		& $-\frac{i}{8\pi}$ & $-\frac{i}{4\pi}$ & $\frac{i}{4\pi}$ & $\frac{i}{8\pi}$ & $-\frac{1}{192\pi^2}$ & 0 & $\frac{i}{8\pi}$ \\ \cline{2-8}
		& $\frac{i}{8\pi}$ & $\frac{i}{4\pi}$ & $-\frac{i}{4\pi}$ & $\frac{i}{8\pi}$ & $-\frac{3}{64\pi^2}$ & $-\frac{i}{4\pi}$ & $-\frac{i}{8\pi}$ \\ \cline{2-8}
		& $-\frac{i}{8\pi}$ & $\frac{i}{4\pi}$ & $\frac{i}{4\pi}$ & $-\frac{i}{8\pi}$ & $-\frac{3}{64\pi^2}$ & $-\frac{i}{4\pi}$ & $\frac{i}{8\pi}$ \\ \cline{2-8}
		& $\frac{i}{8\pi}$ & $-\frac{i}{4\pi}$ & $\frac{i}{4\pi}$ & $\frac{i}{8\pi}$ & $-\frac{1}{192\pi^2}$ & $-\frac{i}{4\pi}$ & $-\frac{i}{8\pi}$ \\ \cline{2-8}
		& $\frac{i}{8\pi}$ & $\frac{i}{4\pi}$ & $\frac{i}{4\pi}$ & $-\frac{i}{8\pi}$ & $-\frac{3}{64\pi^2}$ & $-\frac{i}{2\pi}$ & $-\frac{i}{8\pi}$ \\ \cline{2-8}
		& $-\frac{i}{8\pi}$ & $\frac{i}{4\pi}$ & $\frac{i}{4\pi}$ & $\frac{i}{8\pi}$ & $-\frac{1}{192\pi^2}$ & $-\frac{i}{2\pi}$ & $\frac{i}{8\pi}$ \\ \cline{2-8}
		& $\frac{i}{8\pi}$ & $\frac{i}{4\pi}$ & $\frac{i}{4\pi}$ & $\frac{i}{8\pi}$ & $-\frac{1}{192\pi^2}$ & $-\frac{3i}{4\pi}$ & $\frac{i}{8\pi}$ \\ \cline{1-8}
		& $-\frac{i}{4\pi}$ & $-\frac{i}{2\pi}$ & $-\frac{i}{4\pi}$ & $-\frac{i}{2\pi}$ & $\frac{1}{6\pi^2}$ & $\frac{3i}{2\pi}$ & $\frac{i}{2\pi}$ \\ \cline{2-8}
		& $\frac{i}{4\pi}$ & $-\frac{i}{2\pi}$ & $-\frac{i}{4\pi}$ & $-\frac{i}{2\pi}$ & $\frac{1}{6\pi^2}$ & $\frac{i}{\pi}$ & $-\frac{i}{2\pi}$ \\ \cline{2-8}
		& $-\frac{i}{4\pi}$ & $-\frac{i}{2\pi}$ & $\frac{i}{4\pi}$ & $-\frac{i}{2\pi}$ & $-\frac{1}{6\pi^2}$ & $\frac{i}{\pi}$ & $\frac{i}{2\pi}$ \\ \cline{2-8}
		& $\frac{i}{4\pi}$ & $-\frac{i}{2\pi}$ & $\frac{i}{4\pi}$ & $-\frac{i}{2\pi}$ & $-\frac{1}{6\pi^2}$ & $\frac{i}{2\pi}$ & $-\frac{i}{2\pi}$ \\ \cline{2-8}
		& $-\frac{i}{4\pi}$ & $-\frac{i}{2\pi}$ & $-\frac{i}{4\pi}$ & $\frac{i}{2\pi}$ & $-\frac{1}{6\pi^2}$ & $\frac{i}{2\pi}$ & $\frac{i}{2\pi}$ \\ \cline{2-8}
		& $-\frac{i}{4\pi}$ & $\frac{i}{2\pi}$ & $-\frac{i}{4\pi}$ & $-\frac{i}{2\pi}$ & $\frac{1}{6\pi^2}$ & $\frac{i}{2\pi}$ & $\frac{i}{2\pi}$ \\ \cline{2-8}
		& $\frac{i}{4\pi}$ & $\frac{i}{2\pi}$ & $-\frac{i}{4\pi}$ & $-\frac{i}{2\pi}$ & $\frac{1}{6\pi^2}$ & 0 & $-\frac{i}{2\pi}$ \\ \cline{2-8}
		2 & $-\frac{i}{4\pi}$ & $\frac{i}{2\pi}$ & $\frac{i}{4\pi}$ & $-\frac{i}{2\pi}$ & $-\frac{1}{6\pi^2}$ & 0 & $\frac{i}{2\pi}$ \\ \cline{2-8}
		& $\frac{i}{4\pi}$ & $-\frac{i}{2\pi}$ & $-\frac{i}{4\pi}$ & $\frac{i}{2\pi}$ & $-\frac{1}{6\pi^2}$ & 0 & $-\frac{i}{2\pi}$ \\ \cline{2-8}
		& $-\frac{i}{4\pi}$ & $-\frac{i}{2\pi}$ & $\frac{i}{4\pi}$ & $\frac{i}{2\pi}$ & $\frac{1}{6\pi^2}$ & 0 & $\frac{i}{2\pi}$ \\ \cline{2-8}
		& $\frac{i}{4\pi}$ & $-\frac{i}{2\pi}$ & $\frac{i}{4\pi}$ & $\frac{i}{2\pi}$ & $\frac{1}{6\pi^2}$ & $-\frac{i}{2\pi}$ & $-\frac{i}{2\pi}$ \\ \cline{2-8}
		& $\frac{i}{4\pi}$ & $\frac{i}{2\pi}$ & $\frac{i}{4\pi}$ & $-\frac{i}{2\pi}$ & $-\frac{1}{6\pi^2}$ & $-\frac{i}{2\pi}$ & $-\frac{i}{2\pi}$ \\ \cline{2-8}
		& $-\frac{i}{4\pi}$ & $\frac{i}{2\pi}$ & $-\frac{i}{4\pi}$ & $\frac{i}{2\pi}$ & $-\frac{1}{6\pi^2}$ & $-\frac{i}{2\pi}$ & $\frac{i}{2\pi}$ \\ \cline{2-8}
		& $\frac{i}{4\pi}$ & $\frac{i}{2\pi}$ & $-\frac{i}{4\pi}$ & $\frac{i}{2\pi}$ & $-\frac{1}{6\pi^2}$ & $-\frac{i}{\pi}$ & $-\frac{i}{2\pi}$ \\ \cline{2-8}
		& $-\frac{i}{4\pi}$ & $\frac{i}{2\pi}$ & $\frac{i}{4\pi}$ & $\frac{i}{2\pi}$ & $\frac{1}{6\pi^2}$ & $-\frac{i}{\pi}$ & $\frac{i}{2\pi}$ \\ \cline{2-8}
		& $\frac{i}{4\pi}$ & $\frac{i}{2\pi}$ & $\frac{i}{4\pi}$ & $\frac{i}{2\pi}$ & $\frac{1}{6\pi^2}$ & $-\frac{3i}{2\pi}$ & $-\frac{i}{2\pi}$ \\ \cline{1-8}
		& $-\frac{i}{4\pi}$ & $-\frac{i}{4\pi}$ & $-\frac{i}{2\pi}$ & $\frac{i}{2\pi}$ &  & $\frac{i}{2\pi}$ &  \\ \cline{2-5}\cline{7-7}
		& $-\frac{i}{4\pi}$ & $-\frac{i}{4\pi}$ & $\frac{i}{2\pi}$ & $-\frac{i}{2\pi}$ &  & $\frac{i}{2\pi}$ &  \\ \cline{2-5}\cline{7-7}
		& $\frac{i}{4\pi}$ & $-\frac{i}{4\pi}$ & $-\frac{i}{2\pi}$ & $\frac{i}{2\pi}$ &  & 0 &  \\ \cline{2-5}\cline{7-7}
		$-1$ & $\frac{i}{4\pi}$ & $-\frac{i}{4\pi}$ & $\frac{i}{2\pi}$ & $-\frac{i}{2\pi}$ & Any & 0 & 0 \\ \cline{2-5}\cline{7-7}
		& $-\frac{i}{4\pi}$ & $\frac{i}{4\pi}$ & $-\frac{i}{2\pi}$ & $\frac{i}{2\pi}$ & value\footnote{Here, any complex number can be a basic value for $U$ since all the terms of $U$ in $P_4(z)$ contain $(1+\zeta)$ as well. Moreover, the integral for $a$ always vanishes.} & 0 &  \\ \cline{2-5}\cline{7-7}
		& $-\frac{i}{4\pi}$ & $\frac{i}{4\pi}$ & $\frac{i}{2\pi}$ & $-\frac{i}{2\pi}$ &  & 0 &  \\ \cline{2-5}\cline{7-7}
		& $\frac{i}{4\pi}$ & $\frac{i}{4\pi}$ & $-\frac{i}{2\pi}$ & $\frac{i}{2\pi}$ &  & $-\frac{i}{2\pi}$ &  \\ \cline{2-5}\cline{7-7}
		& $\frac{i}{4\pi}$ & $\frac{i}{4\pi}$ & $\frac{i}{2\pi}$ & $-\frac{i}{2\pi}$ &  & $-\frac{i}{2\pi}$ &  \\ \hline
		\caption{The parameters obtained from $\Gamma_0(4)\cap\Gamma(2)$. Using \eqref{agt2}, we can get the values for $\alpha_i$'s.}\label{param2}
	\end{longtable}
\end{center}
As the size of the table increases, we will give a more compact version for the remaining cases below. For each $\zeta$, there are usually $2^4=16$ possibilities. For $a$, as the sign of $a$ only depends on the sign of $m_0$ (in the following sense), ``$\pm$'' in $a$ means that $a$ has the same sign as $m_0$ while ``$\mp$'' in $a$ indicates that $m_0$ and $a$ have opposite signs.
\begin{center}
	\centering
	\renewcommand*{\arraystretch}{1.2}
	\begin{longtable}{|c|c|c|c|c|c|c|}
		\hline
		$\zeta=\text{e}^{2\pi\text{i}\tau}$ & $m_0$ & $m_1$ & $m_2$ & $m_3$ & $U$ & $a$ \\ \hline\hline
		& $\pm\frac{i\sqrt{5}}{50\pi}$ & $\pm\frac{i\sqrt{5}}{10\pi}$ & $\frac{i\sqrt{5}}{10\pi}$ & $\frac{i\sqrt{5}}{50\pi}$ & $\frac{(-607+85\sqrt{5})}{62750\pi^2}$ & $\pm\frac{i\sqrt{5}}{25\pi}$ \\ \cline{4-5}
		$\frac{1}{2}-\frac{11}{50}\sqrt{5}$ &  &  & $-\frac{i\sqrt{5}}{10\pi}$ & $-\frac{i\sqrt{5}}{50\pi}$ &  &  \\ \cline{2-7}
		& $\pm\frac{i\sqrt{5}}{50\pi}$ & $\pm\frac{i\sqrt{5}}{10\pi}$ & $\frac{i\sqrt{5}}{10\pi}$ & $-\frac{i\sqrt{5}}{50\pi}$ & $\frac{9(-69+20\sqrt{5})}{31375\pi^2}$ & $\mp\frac{i\sqrt{5}}{25\pi}$ \\ \cline{4-5}
		&  &  & $-\frac{i\sqrt{5}}{10\pi}$ & $\frac{i\sqrt{5}}{50\pi}$ &  &  \\ \hline
		& $\pm\frac{i\sqrt{5}}{50\pi}$ & $\pm\frac{i\sqrt{5}}{10\pi}$ & $\frac{i\sqrt{5}}{10\pi}$ & $\frac{i\sqrt{5}}{50\pi}$ & $\frac{(-607-85\sqrt{5})}{62750\pi^2}$ & $\mp\frac{i\sqrt{5}}{25\pi}$ \\ \cline{4-5}
		$\frac{1}{2}+\frac{11}{50}\sqrt{5}$ &  &  & $-\frac{i\sqrt{5}}{10\pi}$ & $-\frac{i\sqrt{5}}{50\pi}$ &  &  \\ \cline{2-7}
		& $\pm\frac{i\sqrt{5}}{50\pi}$ & $\pm\frac{i\sqrt{5}}{10\pi}$ & $\frac{i\sqrt{5}}{10\pi}$ & $-\frac{i\sqrt{5}}{50\pi}$ & $\frac{9(-69-20\sqrt{5})}{31375\pi^2}$ & $\mp\frac{i\sqrt{5}}{25\pi}$ \\ \cline{4-5}
		&  &  & $-\frac{i\sqrt{5}}{10\pi}$ & $\frac{i\sqrt{5}}{50\pi}$ &  &  \\ \hline
		& $\pm\frac{i(5\sqrt{5}+11)}{4\pi}$ & $\pm\frac{i5(5\sqrt{5}+11)}{4\pi}$ & $\frac{i(5\sqrt{5}+11)}{4\pi}$ & $\frac{i5(5\sqrt{5}+11)}{4\pi}$ & $100.010534$ & $\pm\frac{i2(5\sqrt{5}+11)}{4\pi}$ \\ \cline{4-5}
		$\frac{125}{2}+\frac{55}{2}\sqrt{5}$ &  &  & $-\frac{i(5\sqrt{5}+11)}{4\pi}$ & $-\frac{i5(5\sqrt{5}+11)}{4\pi}$ &  &  \\ \cline{2-7}
		& $\pm\frac{i(5\sqrt{5}+11)}{4\pi}$ & $\pm\frac{i5(5\sqrt{5}+11)}{4\pi}$ & $\frac{i(5\sqrt{5}+11)}{4\pi}$ & $-\frac{i5(5\sqrt{5}+11)}{4\pi}$ & $38.200625$ & $\pm\frac{i2(5\sqrt{5}+11)}{4\pi}$ \\ \cline{4-5}
		&  &  & $-\frac{i(5\sqrt{5}+11)}{4\pi}$ & $\frac{i5(5\sqrt{5}+11)}{4\pi}$ &  &  \\ \hline
		& $\pm\frac{i\sqrt{5}}{50\pi}$ & $\pm\frac{i\sqrt{5}}{50\pi}$ & $\frac{i\sqrt{5}}{10\pi}$ & $\frac{i\sqrt{5}}{10\pi}$ & $-0.000843$ & $\pm\frac{i\sqrt{5}}{25\pi}$ \\ \cline{4-5}
		$-\frac{123}{2}+\frac{55}{2}\sqrt{5}$ &  &  & $-\frac{i\sqrt{5}}{10\pi}$ & $-\frac{i\sqrt{5}}{10\pi}$ &  &  \\ \cline{2-7}
		& $\pm\frac{i\sqrt{5}}{50\pi}$ & $\pm\frac{i\sqrt{5}}{50\pi}$ & $\frac{i\sqrt{5}}{10\pi}$ & $-\frac{i\sqrt{5}}{10\pi}$ & $-0.000674$ & $\mp\frac{i\sqrt{5}}{25\pi}$ \\ \cline{4-5}
		&  &  & $-\frac{i\sqrt{5}}{10\pi}$ & $\frac{i\sqrt{5}}{10\pi}$ &  &  \\ \hline
		& $\pm\frac{i\sqrt{5}}{50\pi}$ & $\pm\frac{i\sqrt{5}}{10\pi}$ & $\frac{i\sqrt{5}}{50\pi}$ & $\frac{i\sqrt{5}}{10\pi}$ & $-0.001278$ & $\mp\frac{i\sqrt{5}}{25\pi}$ \\ \cline{4-5}
		$\frac{125}{2}-\frac{55}{2}\sqrt{5}$ &  &  & $-\frac{i\sqrt{5}}{50\pi}$ & $-\frac{i\sqrt{5}}{10\pi}$ &  &  \\ \cline{2-7}
		& $\pm\frac{i\sqrt{5}}{50\pi}$ & $\pm\frac{i\sqrt{5}}{10\pi}$ & $\frac{i\sqrt{5}}{50\pi}$ & $-\frac{i\sqrt{5}}{10\pi}$ & $-0.003346$ & $\pm\frac{i\sqrt{5}}{25\pi}$ \\ \cline{4-5}
		&  &  & $-\frac{i\sqrt{5}}{50\pi}$ & $\frac{i\sqrt{5}}{10\pi}$ &  &  \\ \hline
		& $\pm\frac{i(5\sqrt{5}+11)}{4\pi}$ & $\pm\frac{i(5\sqrt{5}+11)}{4\pi}$ & $\frac{i5(5\sqrt{5}+11)}{4\pi}$ & $\frac{i5(5\sqrt{5}+11)}{4\pi}$ & $303.899917$ & $\pm\frac{i2(5\sqrt{5}+11)}{4\pi}$ \\ \cline{4-5}
		$-\frac{123}{2}-\frac{55}{2}\sqrt{5}$ &  &  & $-\frac{i5(5\sqrt{5}+11)}{4\pi}$ & $-\frac{i5(5\sqrt{5}+11)}{4\pi}$ &  &  \\ \cline{2-7}
		& $\pm\frac{i(5\sqrt{5}+11)}{4\pi}$ & $\pm\frac{i(5\sqrt{5}+11)}{4\pi}$ & $\frac{i5(5\sqrt{5}+11)}{4\pi}$ & $-\frac{i5(5\sqrt{5}+11)}{4\pi}$ & $-10.195921$ & $\pm\frac{i2(5\sqrt{5}+11)}{4\pi}$ \\ \cline{4-5}
		&  &  & $-\frac{i5(5\sqrt{5}+11)}{4\pi}$ & $\frac{i5(5\sqrt{5}+11)}{4\pi}$ &  &  \\ \hline
    	\caption{The parameters obtained from $\Gamma_1(5)$. Using \eqref{agt2}, we can get the values for $\alpha_i$'s and $Q$.}\label{param3}
    \end{longtable}
\end{center}
\begin{center}
	\centering
	\renewcommand*{\arraystretch}{1.2}
	\begin{longtable}{|c|c|c|c|c|c|c|}
		\hline
		$\zeta=\text{e}^{2\pi\text{i}\tau}$ & $m_0$ & $m_1$ & $m_2$ & $m_3$ & $U$ & $a$ \\ \hline\hline
		& $\pm\frac{i}{4\pi}$ & $\pm\frac{i\sqrt{109}}{2\pi}$ & $\frac{2i}{\pi}$ & $\frac{27i}{4\pi}$ & $\frac{595}{48\pi^2}$ & $\pm0.30258i$ \\ \cline{4-5}
		$\frac{1}{2}$ &  &  & $-\frac{2i}{\pi}$ & $-\frac{27i}{4\pi}$ &  &  \\ \cline{2-7}
		& $\pm\frac{i}{4\pi}$ & $\pm\frac{i\sqrt{109}}{2\pi}$ & $\frac{2i}{\pi}$ & $-\frac{27i}{4\pi}$ & $-\frac{269}{48\pi^2}$ & $\pm0.30258i$ \\ \cline{4-5}
		&  &  & $-\frac{2i}{\pi}$ & $\frac{27i}{4\pi}$ &  &  \\ \hline
		& $\pm\frac{i}{4\pi}$ & $\pm\frac{2i}{\pi}$ & $\frac{i\sqrt{109}}{2\pi}$ & $\frac{27i}{4\pi}$ & $\frac{-665+108\sqrt{109}}{48\pi^2}$ & $\pm0.741431i$ \\ \cline{4-5}
		$\frac{1}{2}$ &  & &  $-\frac{i\sqrt{109}}{2\pi}$ & $-\frac{2i}{\pi}$ &  &  \\ \cline{2-7}
		& $\pm\frac{i}{4\pi}$ & $\pm\frac{2i}{\pi}$ & $\frac{i\sqrt{109}}{2\pi}$ & $-\frac{27i}{4\pi}$ & $\frac{-665-108\sqrt{109}}{48\pi^2}$ & $\pm0.741431i$ \\ \cline{4-5}
		&  &  & $-\frac{i\sqrt{109}}{2\pi}$ & $\frac{27i}{4\pi}$ &  &  \\ \hline
		& $\pm\frac{i}{2\pi}$ & $\pm\frac{i\sqrt{109}}{\pi}$ & $\frac{27i}{2\pi}$ & $\frac{4i}{\pi}$ & $\frac{455}{3\pi^2}$ & $\pm0.6051525i$ \\ \cline{4-5}
		2 &  &  & $-\frac{27i}{2\pi}$ & $-\frac{4i}{\pi}$ &  &  \\ \cline{2-7}
		& $\pm\frac{i}{2\pi}$ & $\pm\frac{i\sqrt{109}}{\pi}$ & $\frac{27i}{2\pi}$ & $-\frac{4i}{\pi}$ & $\frac{23}{3\pi^2}$ & $\pm0.6051525i$ \\ \cline{4-5}
		&  &  & $-\frac{27i}{2\pi}$ & $\frac{4i}{\pi}$ &  &  \\ \hline
		& $\pm\frac{i}{2\pi}$ & $\pm\frac{4i}{\pi}$ & $\frac{27i}{2\pi}$ & $\frac{i\sqrt{109}}{\pi}$ & $\frac{125+54\sqrt{109}}{3\pi^2}$ & $\pm1.4828632i$ \\ \cline{4-5}
		2 &  &  & $-\frac{27i}{2\pi}$ & $-\frac{i\sqrt{109}}{\pi}$ &  &  \\ \cline{2-7}
		& $\pm\frac{i}{2\pi}$ & $\pm\frac{4i}{\pi}$ & $\frac{27i}{2\pi}$ & $-\frac{i\sqrt{109}}{\pi}$ & $\frac{125-54\sqrt{109}}{3\pi^2}$ & $\pm1.4828632i$ \\ \cline{4-5}
		&  &  & $-\frac{27i}{2\pi}$ & $\frac{i\sqrt{109}}{\pi}$ &  &  \\ \hline
		\caption{The parameters obtained from $\Gamma_0(6)$. Using \eqref{agt2}, we can get the values for $\alpha_i$'s and $Q$.}\label{param4}
	\end{longtable}
\end{center}
\begin{center}
	\centering
	\renewcommand*{\arraystretch}{1.2}
	\begin{longtable}{|c|c|c|c|c|c|c|}
		\hline
		$\zeta=\text{e}^{2\pi\text{i}\tau}$ & $m_0$ & $m_1$ & $m_2$ & $m_3$ & $U$ & $a$ \\ \hline\hline
		& $\pm\frac{i}{16\pi}$ & $\pm\frac{i}{8\pi}$ & $\frac{i}{2\pi}$ & $\frac{i}{16\pi}$ & $\frac{11}{768\pi^2}$ & $\pm0.0528623$ \\ \cline{4-5}
		$\frac{1}{2}$ &  &  & $-\frac{i}{2\pi}$ & $-\frac{i}{16\pi}$ &  &  \\ \cline{2-7}
		& $\pm\frac{i}{16\pi}$ & $\pm\frac{i}{8\pi}$ & $\frac{i}{2\pi}$ & $-\frac{i}{16\pi}$ & $-\frac{7}{256\pi^2}$ & $\pm0.0528623$ \\ \cline{4-5}
		&  &  & $-\frac{i}{2\pi}$ & $\frac{i}{16\pi}$ &  &  \\ \hline
		& $\pm\frac{i}{8\pi}$ & $\pm\frac{i}{4\pi}$ & $\frac{i}{8\pi}$ & $-\frac{i}{\pi}$ & $\frac{7}{48\pi^2}$ & $\pm0.1057$ \\ \cline{4-5}
		2 &  &  & $-\frac{i}{8\pi}$ & $\frac{i}{\pi}$ &  &  \\ \cline{2-7}
		& $\pm\frac{i}{8\pi}$ & $\pm\frac{i}{4\pi}$ & $\frac{i}{8\pi}$ & $\frac{i}{\pi}$ & $\frac{23}{48\pi^2}$ & $\pm0.1057$ \\ \cline{4-5}
		&  &  & $-\frac{i}{8\pi}$ & $-\frac{i}{\pi}$ &  &  \\ \hline
		\caption{The parameters obtained from $\Gamma_0(8)$. Using \eqref{agt2}, we can get the values for $\alpha_i$'s and $Q$.}\label{param5}
	\end{longtable}
\end{center}
\begin{center}
	\centering
	\renewcommand*{\arraystretch}{1.2}
	\begin{longtable}{|c|c|c|c|c|c|c|}
		\hline
		$\zeta=\text{e}^{2\pi\text{i}\tau}$ & $m_0$ & $m_1$ & $m_2$ & $m_3$ & $U$ & $a$ \\ \hline\hline
		& $\pm\frac{i}{6\sqrt{3}\pi}$ & $\pm\frac{i}{6\sqrt{3}\pi}$ & $\frac{i}{6\sqrt{3}\pi}$ & $\frac{i\sqrt{3}}{2\pi}$ & $-\frac{i(33i+25\sqrt{3})}{162\pi^2}$ & $\pm(-0.1402495+0.0315441i)$ \\ \cline{4-5}
		$\frac{1-i\sqrt{3}}{2}$ &  &  & $-\frac{i}{6\sqrt{3}\pi}$ & $-\frac{i\sqrt{3}}{2\pi}$ &  &  \\ \cline{2-7}
		& $\pm\frac{i}{6\sqrt{3}\pi}$ & $\pm\frac{i}{6\sqrt{3}\pi}$ & $\frac{i}{6\sqrt{3}\pi}$ & $-\frac{i\sqrt{3}}{2\pi}$ & $-\frac{i(3i+8\sqrt{3})}{81\pi^2}$ & $\pm(-0.0887502+0.0362071i)$ \\ \cline{4-5}
		&  &  & $-\frac{i}{6\sqrt{3}\pi}$ & $\frac{i\sqrt{3}}{2\pi}$ &  &  \\ \cline{1-7}
		& $\pm\frac{i}{6\sqrt{3}\pi}$ & $\pm\frac{i}{6\sqrt{3}\pi}$ & $\frac{i}{6\sqrt{3}\pi}$ & $\frac{i\sqrt{3}}{2\pi}$ & $\frac{i(-33i+25\sqrt{3})}{162\pi^2}$ & $\pm(-0.1402495-0.0315441i)$ \\ \cline{4-5}
		$\frac{1+i\sqrt{3}}{2}$ &  &  & $-\frac{i}{6\sqrt{3}\pi}$ & $-\frac{i\sqrt{3}}{2\pi}$ &  &  \\ \cline{2-7}
		& $\pm\frac{i}{6\sqrt{3}\pi}$ & $\pm\frac{i}{6\sqrt{3}\pi}$ & $\frac{i}{6\sqrt{3}\pi}$ & $-\frac{i\sqrt{3}}{2\pi}$ & $\frac{i(-3i+8\sqrt{3})}{81\pi^2}$ & $\pm(-0.0887502-0.0362071i)$ \\ \cline{4-5}
		&  &  & $-\frac{i}{6\sqrt{3}\pi}$ & $\frac{i\sqrt{3}}{2\pi}$ &  &  \\ \hline
		\caption{The parameters obtained from $\Gamma_0(9)$. Using \eqref{agt2}, we can get the values for $\alpha_i$'s and $Q$.}\label{param6}
	\end{longtable}
\end{center}

Based on the above calculations, there are some remarks we can make:
\begin{itemize}
	\item One may check that the elliptic curves parametrized by these $m_i$, $\zeta$ and $U$ have the same $j$-invariants as in Table \ref{table.j.invariant} for the six Belyi maps.
\begin{table}[h]
	\centering
	\renewcommand*{\arraystretch}{1.2}
	\begin{tabular}{|c|c|}
		\hline
		$\Gamma(3)$ & $ 0 $\\
		\hline
		$\Gamma_0(4)\cap\Gamma(2)$ & $ \frac{35152}{9} $\\
		\hline
		$\Gamma_1(5)$ & $\frac{131072}{9}$\\
		\hline
		$\Gamma_0(6)$ & $-3072$\\
		\hline
		$\Gamma_0(8)$ & $\frac{21952}{9}$\\
		\hline
		$\Gamma_0(9)$ & $ 0 $\\
		\hline
	\end{tabular}
	\par  
	\vspace{0.3cm}
	\caption{The $j$-invariants that correspond to the six index-12 Belyi maps.}
	\label{table.j.invariant}
\end{table}
Moreover, there are two cases with $\zeta=(1\pm i\sqrt{3})/2$, which are the cusp points for the fundamental diagram of SL(2,$\mathbb{Z}$). They are exactly the dessins whose Belyi maps have $j$-invariant 0.
\item It is obvious that for each dessin, the parametrizations for different $\zeta$'s are related by triality
\begin{equation}
	\zeta~\leftrightarrow~\zeta'=\frac{1}{\zeta}~\leftrightarrow~\zeta''=1-\zeta.
\end{equation}
This is explicitly listed in Table \ref{triality}. Modular invariance of the curve also leads to the following transformations of mass parameters:
\begin{eqnarray}
	&&\zeta\leftrightarrow\frac{1}{\zeta}:~(m_0,m_1,m_2,m_3)\leftrightarrow\frac{1}{|\zeta|}(m_0,m_1,m_3,m_2);\nonumber\\
	&&\zeta\leftrightarrow1-\zeta:~~(m_0,m_1,m_2,m_3)\leftrightarrow(m_0,m_2,m_1,m_3).
\end{eqnarray}
\begin{table}[h]
	\centering
	\renewcommand*{\arraystretch}{1.2}
	\begin{tabular}{|c|c|c|c|}
		\hline
		Dessin & $\zeta$ & $\zeta'$ &$\zeta''$  \\ \hline\hline
		$\Gamma(3)$ & $\frac{1}{2}(1\pm i\sqrt{3})$ & $\frac{1}{2}(1\mp i\sqrt{3})$ & $\frac{1}{2}(1\mp i\sqrt{3})$ \\ \hline
		$\Gamma_0(4)\cap\Gamma(2)$ & 2 & $\frac{1}{2}$ & $-1$ \\ \hline
		$\Gamma_1(5)$ & $\frac{1}{2}-\frac{11}{50}\sqrt{5}$ & $\frac{125}{2}+\frac{55}{2}\sqrt{5}$ & $\frac{1}{2}+\frac{11}{50}\sqrt{5}$ \\ \cline{2-4} 
		& $-\frac{123}{2}+\frac{55}{2}\sqrt{5}$ & $-\frac{123}{2}-\frac{55}{2}\sqrt{5}$ & $\frac{125}{2}-\frac{55}{2}\sqrt{5}$ \\ \hline
		$\Gamma_0(6)$ & 2 & $\frac{1}{2}$ & - \\ \hline
		$\Gamma_0(8)$ & 2 & $\frac{1}{2}$ & - \\ \hline
		$\Gamma_0(9)$ & $\frac{1}{2}(1\pm i\sqrt{3})$ & - & $\frac{1}{2}(1\mp i\sqrt{3})$ \\ \hline
	\end{tabular}
    \caption{The parametrizations for each case are related by triality. The hyphens indicate that such $\zeta$ either gives no solution to mass parameters ($\Gamma_0(6)$ and $\Gamma_0(8)$) or does not satisfy the transformations of masses ($\Gamma_0(9)$).}\label{triality}
\end{table}
In particular, the two rows for $\Gamma_1(5)$ are also related by triality: $1-\left(\frac{125}{2}+\frac{55}{2}\sqrt{5}\right)=-\frac{123}{2}-\frac{55}{2}\sqrt{5}$.
\end{itemize}

\subsection{Minimal Models and $\Gamma(3)$}\label{Gamma3mm}
As an example, let us match the parametrizations for $\Gamma(3)$ obtained above to 4-point CBs in minimal models. In fact, as we will see, such CB first appears for the tetracritical Ising model when $p'=6$ and $p=5$, that is, $c=4/5$. As usual, we can write the 4-point CB as
\begin{equation}
	\tikzset{every picture/.style={line width=0.75pt}} 
	\begin{tikzpicture}[x=0.75pt,y=0.75pt,yscale=-1,xscale=1]
		\draw [line width=1.5]    (179.9,119.77) -- (90.17,119.83) ;
		\draw [line width=1.5]    (120.17,120.5) -- (120.23,90.77) ;
		\draw [line width=1.5]    (150.17,119.5) -- (149.9,90.43) ;
		\draw (81,122.4) node [anchor=north west][inner sep=0.75pt]  [font=\footnotesize]  {$\alpha _{1}$};
		\draw (170,122.4) node [anchor=north west][inner sep=0.75pt]  [font=\footnotesize]  {$\alpha _{4}$};
		\draw (102,81.4) node [anchor=north west][inner sep=0.75pt]  [font=\footnotesize]  {$\alpha _{2}$};
		\draw (152.9,80.83) node [anchor=north west][inner sep=0.75pt]  [font=\footnotesize]  {$\alpha _{3}$};
		\draw (129,123.4) node [anchor=north west][inner sep=0.75pt]  [font=\footnotesize]  {$\alpha _{\text{int}}$};
	\end{tikzpicture}.\label{CB4pt}
\end{equation}
Then the intermediate field $\phi_{k,l}$ should satisfy the fusion rule
\begin{equation}
	\phi_{r,s}\times\phi_{m,n}=\sum_{\substack{k=|m-r|+1\\k-m+r-1\in2\mathbb{Z}}}^{\min(m+r-1,2p-1-m-r)}\sum_{\substack{l=|n-s|+1\\l-n+s-1\in2\mathbb{Z}}}^{\min(n+s-1,2p'-1-n-s)}\phi_{k,l}\label{fusion},
\end{equation}
where the entire conformal family of a primary is implicit in the above abuse of notation. Let $\phi_{r_i,s_i}$ correspond to $\alpha_{1,4}$ and $\phi_{m_i,n_i}$ correspond to $\alpha_{2,3}$ ($i=1,2$). Then the fusion rule for the 4-point CB is
\begin{equation}
	\phi_{k,l}\in\phi_{r_1,s_1}\times\phi_{m_1,n_1},~\phi_{k,l}\in\phi_{r_2,s_2}\times\phi_{m_2,n_2}
\end{equation}
with constraints on $k,l$ indicated in \eqref{fusion}.

Before we insert the specific values of the parametrizations, we can make some simplifications:
\begin{itemize}
	\item Recall that the mass parameters are real or pure imaginary. If we have some parametrization with $m_i\in\mathbb{R}$, without loss of generality we can choose $\epsilon_1<0<\epsilon_2$. Then since $\frac{\epsilon_1+\epsilon_2}{\sqrt{\epsilon_1\epsilon_2}}=Q=\text{i}\left(\sqrt{\frac{p'}{p}}-\sqrt{\frac{p}{p'}}\right)$, we have $\sqrt{\epsilon_1\epsilon_2}=-\text{i}\epsilon_2\sqrt{\frac{p}{p'}}$. Likewise, for some parametrization with $m_i\in\text{i}\mathbb{R}$, without loss of generality we can choose $\epsilon_1/\text{i}<0<\epsilon_2/\text{i}$. Such two cases related by $m_i\rightarrow\text{i}m_i$ should give the same $\epsilon_{1,2}$ up to a factor of $\text{i}$.
	\item If we make the choice in the above point for some specific $m_i$, then $m_i\rightarrow-m_i$ should give the same CFT parameters with $\epsilon_{1,2}\rightarrow\text{i}\epsilon_{1,2}$. If we only have $m_0\rightarrow-m_0$ or $m_1\rightarrow-m_1$, then we should always get the same parametrization even without changing $\epsilon_{1,2}$ since the corresponding conformal dimension is $\frac{Q^2}{4}-\frac{m_{0,1}^2}{\epsilon_1\epsilon_2}$.
	\item Swapping $m_2\leftrightarrow m_3$ and swapping $m_0\leftrightarrow m_1$ simultaneously should give the same CFT parameters (for same $\epsilon_{1,2}$) due to the AGT map. This simply corresponds to read the CB \eqref{CB4pt} from the left or from the right.
\end{itemize}

In light of these points, it suffices to only contemplate one parametrization\footnote{Since $\Delta_1=\Delta_4$, when considering $\zeta\leftrightarrow1/\zeta$, it is equivalent to swapping both $m_2\leftrightarrow m_3$ and $m_0\leftrightarrow m_1$. Therefore, $\zeta=(1\pm i\sqrt{3})/2$ should give the same parametrizations. Even if $|\zeta|\neq1$, as long as $\Delta_1=\Delta_4$, swapping $2\leftrightarrow3$ always gives same CFT parameters as the extra factor of $1/|\zeta|$ can be absorbed into $\sqrt{\epsilon_1\epsilon_2}$.}, say $m_0=-m_1=m_2=-m_3=-\frac{1}{2\sqrt{3}}$, for $\Gamma(3)$. When $p'=6,p=5$,  we find that there is only one possibility for $\Delta_1$ and $\Delta_4$, that is,
\begin{equation}
	\Delta_1=\Delta_4=\frac{1}{15}.
\end{equation}
There are two possible solutions for the remaining mass parameters (and deformation parameters):
\begin{eqnarray}
	&&\epsilon_2=\frac{2}{\sqrt{3}\pi},~\Delta_2=\frac{1}{40},~\Delta_3=\frac{1}{8};\label{soln1}\\
	&&\epsilon_2=-\frac{2}{\sqrt{3}\pi},~\Delta_2=\frac{1}{8},~\Delta_3=\frac{1}{40}\label{soln2}.
\end{eqnarray}
Moreover, for the intermediate channel,
\begin{equation}
	a=-\frac{1}{3\sqrt{3}\pi},~\Delta_\text{int}=\frac{1}{40}.
\end{equation}
Hence, the intermediate channel $(k,l)$ obtained from $\Gamma(3)$ corresponds to $(2,2)$ or $(3,4)$ (and another $(k,l)$ satifying the fusion rule but not from the dessin is $(2,4)$ or $(3,2)$). It is not hard to see that the above two solutions both give the 8 CBs in Table \ref{CB8}.
\begin{table}[h]
	\centering
	\begin{tabular}{c|cccccccc}
		$\Delta_1$ & (2,3) & (2,3) & (2,3) & (2,3) & (3,3) & (3,3) & (3,3) & (3,3) \\
		$\Delta_2$ & (2,2) & (2,2) & (3,4) & (3,4) & (2,2) & (2,2) & (3,4) & (3,4) \\
		$\Delta_3$ & (1,2) & (4,4) & (1,2) & (4,4) & (1,2) & (4,4) & (1,2) & (4,4) \\
		$\Delta_4$ & (3,3) & (2,3) & (2,3) & (3,3) & (2,3) & (3,3) & (3,3) & (2,3) \\
		$\Delta_\text{int}$ & (3,4) & (3,4) & (2,2) & (2,2) & (2,2) & (2,2) & (3,4) & (3,4)
	\end{tabular}
    \caption{There are 8 possible combinations. Each column gives a CB. In the leftmost column, $\Delta_i$'s follow the nomenclature correpsonding to \eqref{soln1}. For \eqref{soln2}, it just swaps $2\leftrightarrow3$ (and $\Delta_1=\Delta_4$). Therefore, it essentially gives the same CBs. In other words, the two solutions just correspond to reading the 4-point CB \eqref{CB4pt} from left or from right.}\label{CB8}
\end{table}

In fact, this corresponds to not only a CB in the tetracritical Ising model, but also CBs in many other minimal models. In Figure \ref{kac}, we give the Kac tables for a few examples.
\begin{figure}[h]
	\centering
	\begin{subfigure}{6cm}
		\centering
			\begin{tabular}{c|ccccc}
				4& 3 & $\frac{13}{8}$ & $\frac{2}{3}$ & \textcolor{cyan}{$\frac{1}{8}$} & 0 \\
				3& $\frac{7}{5}$ & $\frac{21}{40}$ & \textcolor{cyan}{$\frac{1}{15}$} & \textcolor{cyan}{$\frac{1}{40}$} & $\frac{2}{5}$ \\
				2& $\frac{2}{5}$ & \textcolor{cyan}{$\frac{1}{40}$} & \textcolor{cyan}{$\frac{1}{15}$} & $\frac{21}{40}$ & $\frac{7}{5}$ \\
				1& 0 & \textcolor{cyan}{$\frac{1}{8}$} & $\frac{2}{3}$ & $\frac{13}{8}$ & 3 \\ \hline
				& 1 & 2 & 3 & 4 & 5
			\end{tabular}
		\caption{}
	\end{subfigure}
    \begin{subfigure}{6cm}
    	\centering
    	\begin{tabular}{c|cccccc}
    		4& $\frac{15}{4}$ & $\frac{16}{7}$ & $\frac{33}{28}$ & \textcolor{cyan}{$\frac{3}{7}$} & $\frac{1}{28}$ & 0 \\
    		3& $\frac{9}{5}$ & $\frac{117}{140}$ & \textcolor{cyan}{$\frac{8}{35}$} & $-\frac{3}{140}$ & \textcolor{cyan}{$\frac{3}{35}$} & $\frac{11}{20}$ \\
    		2& $\frac{11}{20}$ & \textcolor{cyan}{$\frac{3}{35}$} & $-\frac{3}{140}$ & \textcolor{cyan}{$\frac{8}{35}$} & $\frac{117}{140}$ & $\frac{9}{5}$ \\
    		1& 0 & $\frac{1}{28}$ & \textcolor{cyan}{$\frac{3}{7}$} & $\frac{33}{28}$ & $\frac{16}{7}$ & $\frac{15}{4}$ \\ \hline
    		& 1 & 2 & 3 & 4 & 5 & 6
    	\end{tabular}
    	\caption{}
    \end{subfigure}
	\begin{subfigure}{6cm}
		\centering
		\begin{tabular}{c|cccccc}
			5& 5 & $\frac{22}{7}$ & $\frac{12}{7}$ & $\frac{5}{7}$ & $\frac{1}{7}$ & 0 \\
			4& $\frac{23}{8}$ & $\frac{85}{56}$ & $\frac{33}{56}$ & \textcolor{cyan}{$\frac{5}{56}$} & \textcolor{cyan}{$\frac{1}{56}$} & $\frac{3}{8}$ \\
			3& $\frac{4}{3}$ & $\frac{10}{21}$ & \textcolor{cyan}{$\frac{1}{21}$} & \textcolor{cyan}{$\frac{1}{21}$} & $\frac{10}{21}$ & $\frac{4}{3}$ \\
			2& $\frac{3}{8}$ & \textcolor{cyan}{$\frac{1}{56}$} & \textcolor{cyan}{$\frac{5}{56}$} & $\frac{33}{56}$ & $\frac{85}{56}$ & $\frac{23}{8}$ \\
			1& 0 & $\frac{1}{7}$ & $\frac{5}{7}$ & $\frac{12}{7}$ & $\frac{22}{7}$ & 5 \\ \hline
			& 1 & 2 & 3 & 4 & 5 & 6
		\end{tabular}
		\caption{}
	\end{subfigure}
    \begin{subfigure}{6cm}
    	\centering
    	\begin{tabular}{c|ccccccc}
    		4& $\frac{9}{2}$ & $\frac{95}{32}$ & $\frac{7}{4}$ & \textcolor{cyan}{$\frac{27}{32}$} & $\frac{1}{4}$ & $-\frac{1}{32}$ & 0 \\
    		3& $\frac{11}{5}$ & $\frac{187}{160}$ & \textcolor{cyan}{$\frac{9}{20}$} & $\frac{7}{160}$ & $-\frac{1}{20}$ & \textcolor{cyan}{$\frac{27}{160}$} & $\frac{7}{10}$ \\
    		2& $\frac{7}{10}$ & \textcolor{cyan}{$\frac{27}{160}$} & $-\frac{1}{20}$ & $\frac{7}{160}$ & \textcolor{cyan}{$\frac{9}{20}$} & $\frac{187}{160}$ & $\frac{11}{5}$ \\
    		1& 0 & $-\frac{1}{32}$ & $\frac{1}{4}$ & \textcolor{cyan}{$\frac{27}{32}$} & $\frac{7}{4}$ & $\frac{95}{32}$ & $\frac{9}{2}$ \\ \hline
    		& 1 & 2 & 3 & 4 & 5 & 6 & 7
    	\end{tabular}
    	\caption{}
    \end{subfigure}
	\begin{subfigure}{6cm}
		\centering
		\begin{tabular}{c|ccccccc}
			6& $\frac{15}{2}$ & $\frac{165}{32}$ & $\frac{13}{4}$ & $\frac{57}{32}$ & $\frac{3}{4}$ & $\frac{5}{32}$ & 0 \\
			5& $\frac{34}{7}$ & $\frac{675}{224}$ & \textcolor{green}{$\frac{45}{28}$} & \textcolor{green}{$\frac{143}{224}$} & $\frac{3}{28}$ & \textcolor{cyan}{$\frac{3}{224}$} & $\frac{5}{14}$ \\
			4& $\frac{39}{14}$ & \textcolor{green}{$\frac{323}{224}$} & $\frac{15}{28}$ & \textcolor{cyan}{$\frac{15}{224}$} & \textcolor{cyan}{$\frac{1}{28}$} & $\frac{99}{224}$ & \textcolor{green}{$\frac{9}{7}$} \\
			3& \textcolor{green}{$\frac{9}{7}$} & $\frac{99}{224}$ & \textcolor{cyan}{$\frac{1}{28}$} & \textcolor{cyan}{$\frac{15}{224}$} & $\frac{15}{28}$ & \textcolor{green}{$\frac{323}{224}$} & $\frac{39}{14}$ \\
			2& $\frac{5}{14}$ & \textcolor{cyan}{$\frac{3}{224}$} & $\frac{3}{28}$ & \textcolor{green}{$\frac{143}{224}$} & \textcolor{green}{$\frac{45}{28}$} & $\frac{675}{224}$ & $\frac{34}{7}$\\
			1& 0 & $\frac{5}{32}$ & $\frac{3}{4}$ & $\frac{57}{32}$ & $\frac{13}{4}$ & $\frac{165}{32}$ & $\frac{15}{2}$ \\ \hline
			& 1 & 2 & 3 & 4 & 5 & 6 & 7
		\end{tabular}
		\caption{}
	\end{subfigure}
	\caption{Here we list the first five possible examples of CBs that $\Gamma(3)$ corresponds to: (a) $p'=6,p=5$, (b) $p'=7,p=5$, (c) $p'=7,p=6$, (d) $p'=8,p=5$, (e) $p'=8,p=7$. Those appeared in the CBs are in cyan in the Kac tables. For (e), we also have another combination of CBs in green.}\label{kac}
\end{figure}

By looking at these examples, one might see some patterns of the minimal models and the positions of conformal dimensions in cyan appeared in the Kac tables. Now, we are going to show
\begin{proposition}
	The dessin $\Gamma(3)$ gives rise to the charges/momenta of the states in 4-point conformal blocks, where the corresponding weights of the primaries satisfy the conditions in Table \ref{conditions}, in minimal models.
\end{proposition}
\begin{table}
	\centering
	\begin{tabular}{c|c}
		Cases & Conditions \\ \hline\hline
		All & $(r_1,s_1)=(p-r_2,p'-s_2)\in3(\mathbb{Z},\mathbb{Z})$ \\ \hline
		$(r_1,s_1),(r_1+1,s_1+1),(r_1-1,s_1-1),(r_1,s_1)$ &  \\ \cline{1-1}
		$(r_1,s_1),(r_1+1,s_1+1),(r_2+1,s_2+1),(r_2,s_2)$ & $r_1\leq\frac{3(p-1)}{4},s_1\leq\frac{3(p'-1)}{4},$ \\ \cline{1-1}
		$(r_1,s_1),(r_1-1,s_1-1),(r_2-1,s_2-1),(r_2,s_2)$ & $k=\frac{2}{3}r_1,l=\frac{2}{3}s_1$ \\ \cline{1-1}
		$(r_2,s_2),(r_2+1,s_2+1),(r_2-1,s_2-1),(r_2,s_2)$ &  \\ \hline
		$(r_2,s_2),(r_1+1,s_1+1),(r_1-1,s_1-1),(r_2,s_2)$ & $\left(\frac{p+1}{2}\leq r_1\leq p-2\right.$ \\ \cline{1-1}
		$(r_2,s_2),(r_1+1,s_1+1),(r_2+1,s_2+1),(r_1,s_1)$ & $~\left.\text{or}~\frac{p+1}{2}\leq r_1\leq\frac{3(p-1)}{4}~\text{or}~p=2r_1\right)$ \\ \cline{1-1}
		$(r_2,s_2),(r_1-1,s_1-1),(r_2-1,s_2-1),(r_1,s_1)$ & and (similar relations\footnote{Note that the relation with $p=2r_1$ and $p'=2s_1$ is automatically ruled out as $p'-p=1$.} with $p\rightarrow p',r_1\rightarrow s_1$) \\ \cline{1-1}
		$(r_1,s_1),(r_2+1,s_2+1),(r_2-1,s_2-1),(r_1,s_1)$ & and $k=p-\frac{2}{3}r_1,l=p'-\frac{2}{3}s_1$ \\ \hline
	\end{tabular}
    \caption{The possible CBs of minimal models that $\Gamma(3)$ corresponds to.}\label{conditions}
\end{table}

Following the specific values for $m_i$ and $a$, we can define $M_0:=\frac{m_0}{\sqrt{\epsilon_1\epsilon_2}}$ so that
\begin{equation}
	\alpha_1=-M_0+\frac{Q}{2},~\alpha_2=-M_0,~\alpha_3=M_0,~\alpha_4=M_0+\frac{Q}{2},~\alpha_\text{int}=\frac{2M_0}{3}+\frac{Q}{2}.\label{alphaM0}
\end{equation}

There are two possible choices for $\Delta_1$ in the Kac table. For future convenience, let us denote them as $\Delta_{r_1,s_1}$ and $\Delta_{r_2,s_2}$. Then
\begin{equation}
	\frac{(p'r_i-ps_i)^2-(p'-p)^2}{4p'p}=\frac{Q^2}{4}-M_0^2=-\frac{(p'-p)^2}{4p'p}-M_0^2.
\end{equation}
Therefore,
\begin{equation}
	M_0^2=-\frac{(p'r_i-ps_i)^2}{4p'p}.
\end{equation}
It is also immediate from \eqref{alphaM0} that $\Delta_1=\Delta_4$. Hence, we can denote $\Delta_{2~\text{or}~3}$ as $\Delta_{m_i,n_i}$ without specifying whether $(m_{1,2},n_{1,2})$ corresponds to $\Delta_2$ or $\Delta_3$. We can plug this into $\Delta_{m_i,n_i}=\Delta_3=-M_0^2+QM_0$ and get
\begin{equation}
	(p'm_i-p'n_i+xn_2)^2-x^2=(p'r_j-p's_j+xs_j)^2-2x(p'r_j-p's_j+xs_j),\label{mnrsa}
\end{equation}
where $x:=p'-p$ is some positive integer. Its expansion gives
\begin{equation}
	p'^2(m_i-n_i)^2+2p'(m_i-n_i)xn_i+x^2n_i^2-x^2=p'^2(r_j-s_j)^2+2p'(r_j-s_j)(s_j-1)x+x^2s_j^2-2x^2s_j.\label{mnrsb}
\end{equation}
Since this is for general $p'$, by comparing coefficients at different orders of $p'$,  we have
\begin{equation}
	m_i-n_i=\pm (r_j-s_j),~n_i=\pm(s_j-1),~n_i^2-1=s_j^2-2s_j,
\end{equation}
where $\pm$ can be seen from the symmetry of $p'^2$ and $p'$ terms in \eqref{mnrsb}. Due to a similar symmetry for $(m_i,n_i)\leftrightarrow(p-m_i,p'-n_i)$, it is possible to replace $(m_i,n_i)$ with $(p-m_i,p'-n_i)$ or $(r_j,s_j)$ with $(p-r_j,p'-s_j)$ in \eqref{mnrsa}. It turns out that they also give the same set of equations. The third equation is actually redundant, and hence we have
\begin{equation}
	m_i-n_i=\pm(r_j-s_j),~n_i=\pm(s_j-1).
\end{equation}
Strictly speaking, in \eqref{mnrsa}, we should really have $|p'r_j-p's_j+xs_j|$ on the right hand side. Taking this into account, we would obtain another set of solutions with $-1$ replaced by $+1$. Therefore,
\begin{eqnarray}
	&&m_i=r_j-1,~n_i=s_j-1,\\
	&\text{or}~&m_i=r_j+1,~n_i=s_j+1.
\end{eqnarray}
As we also have similar relations for $\Delta_2$ and we have seen that $\Delta_{m_1,n_1}\neq\Delta_{m_2,n_2}$ for $Q\neq0$, we learn that
\begin{equation}
	(m_i,n_i)=(r_j,s_j)\pm(1,1),~(m_1,n_1)\neq(m_2,n_2),~(m_1,n_1)\neq(p-m_2,p'-n_2).
\end{equation}
For the intermediate channel, using $\frac{a}{\sqrt{\epsilon_1\epsilon_2}}=\frac{2M_0}{3}$, we have
\begin{equation}
	(p'k-p'l+l)^2=\frac{4}{9}(p'r_j-p's_j+s_j)^2,
\end{equation}
so likewise,
\begin{eqnarray}
	&&k=\frac{2}{3}r_1,~l=\frac{2}{3}s_1,\label{kl1}\\
	&\text{or}&k=p-\frac{2}{3}r_1,~l=p'-\frac{2}{3}s_1,\label{kl2}
\end{eqnarray}
where without loss of generality we have chosen $j=1$ for convenience. As $k,l$ are integers, we must have $r_1,s_1\in3\mathbb{Z}$ (or in other words, $(p-r_2),(p'-s_2)\in3\mathbb{Z}$). As $p=p'-1$, it is straightforward to see that $k,l\in2\mathbb{Z}$ for \eqref{kl1} while $(k,l)\in(2\mathbb{Z},2\mathbb{Z}+1)$ or $(k,l)\in(2\mathbb{Z}+1,2\mathbb{Z})$ for \eqref{kl2}.

We also need to take the fusion rule into account. In general, there are $2^2\times\binom{4}{2}=24$ possible choices of external legs, where $2^2$ is the number of choices of $\Delta_1$ and $\Delta_4$ and $\binom{4}{2}$ corresponds to the choices of $\Delta_2\neq\Delta_3$. Therefore, we can discuss these possibilities case by case. Here, we will provide the details for three representative cases as examples\footnote{Below we will use the correpsonding $r$'s for external legs to denote each case.}.

\paragraph{Example 1: $r_1,r_1+1,r_1-1,r_1$} In such case, the fusion rule gives
\begin{eqnarray}
	&&2\leq k\leq\min(2r_1,2p-2r_1-2);\nonumber\\
	&&2\leq k\leq\min(2r_1-2,2p-2r_1).
\end{eqnarray}
Putting them together, we have
\begin{equation}
	2\leq k\leq\min(2r_1-2,2p-2r_1-2).
\end{equation}
Therefore,
\begin{equation}
	r_1\geq2,~p-r_1\geq2.
\end{equation}
In fact, we can omit $r_1\geq2$ as we already have $r_1\in3\mathbb{Z}$. Furthermore, we also require $k-(r_1+1)+r_1-1\in2\mathbb{Z}$, that is, $k\in2\mathbb{Z}$. We can write similar conditions for $l$. In particular, $l$ should also be even, so $(k,l)$ should obey \eqref{kl1}. Therefore, we also need to plug \eqref{kl1} into the above inequality. This gives
\begin{equation}
	r_1\leq\frac{3(p-1)}{4},~s_1\leq\frac{3(p'-1)}{4}.
\end{equation}
Comparing $p-2$ with $3(p-1)/4$, we find that $p-2\leq 3(p-1)/4$ only when $p\leq5$ (with equality at $p=5$). However, for $p\leq4$, we cannot have $p-r_1\geq2$ as $r_1\in3\mathbb{Z}$. Hence, $r_1\leq\min(p-2,3(p-1)/4)=3(p-1)/4$ and likewise for $s_1$. In all, the conditions for this case are
\begin{equation}
	r_1\leq\frac{3(p-1)}{4},~s_1\leq\frac{3(p'-1)}{4},~k=\frac{2}{3}r_1,~l=\frac{2}{3}s_1.
\end{equation}

\paragraph{Example 2: $r_2,r_1+1,r_1-1,r_2$} In such case, it is not hard to see that $k$ and $l$ should satisfy \eqref{kl2}. Besides, the fusion rule gives
\begin{eqnarray}
	&&|p-2r_1-1|+1\leq k\leq p-2;\nonumber\\
	&&|p-2r_1+1|+1\leq k\leq p-2.
\end{eqnarray}
Putting them together, we have
\begin{equation}
	\max(|p-2r_1-1|+1,|p-2r_1+1|+1)\leq k\leq p-2.
\end{equation}
Since $p-2r_1-1<p-2r_1+1$, there are three possibilities:
\begin{enumerate}
	\item $p-2r_1-1\geq0$: If
	\begin{equation}
		p\geq2r_1+1,\label{ineq1}
	\end{equation}
	then
	\begin{equation}
		p-2r_1+2\leq k\leq p-2.\label{ineq2}
	\end{equation}
	Plugging $k=p-\frac{2}{3}r_1$ into \eqref{ineq2}, one may check that \eqref{ineq1} and \eqref{ineq2} are indeed consistent (they give the conditions $r_1\geq2/3$ and $r_1\geq3$ which are automatic as $r_1\in3\mathbb{Z}$).
	\item $p-2r_1+1\leq0$: If
	\begin{equation}
		p\leq2r_1-1,
	\end{equation}
	then
	\begin{equation}
		2r_1-p+2\leq k\leq p-2.
	\end{equation}
	For this inequality to hold, we need $p\geq r_1+2$. Plugging $k=p-\frac{2}{3}r_1$ into the inequalities, we need $r_1\leq\frac{3(p-1)}{4}$. Following the above same reasoning, it suffices to keep $r_1\leq\frac{3(p-1)}{4}$.
	\item $p-2r_1=0$: If
	\begin{equation}
		p=2r_1,
	\end{equation}
	then
	\begin{equation}
		2\leq k\leq p-2.
	\end{equation}
	Plugging $k=p-\frac{2}{3}r_1$ into the inequalities, one may check that these inequalities are indeed consistent (they give the conditions $r_1\geq2/3$ and $r_1\geq3$ which are automatic as $r_1\in3\mathbb{Z}$).
\end{enumerate}
The disussion for $p',l,s_1$ is the same.

\paragraph{Example 3: $r_1,r_1+1,r_1-1,r_2$} In such case, the fusion rule gives
\begin{eqnarray}
	&&2\leq k\leq\min(2r_1,2p-2r_1-2);\nonumber\\
	&&2\leq k\leq\min(r_1+r_2-2,2p-r_1-r_2)=\min(p-2,p)=p-2.
\end{eqnarray}
Putting them together, we have
\begin{equation}
	2\leq k\leq\min(2r_1-2,2p-2r_1-2,p-2).
\end{equation}
Therefore,
\begin{equation}
	r_1\geq2,~p\geq4,~p-r_1\geq2,
\end{equation}
where we can omit the first two conditions as we already have $r_1\in3\mathbb{Z}$. Furthermore, we also require $k-(r_1+1)+r_1-1\in2\mathbb{Z}$, that is, $k\in2\mathbb{Z}$. We can write the similar conditions for $l$. In particular, $l$ should also be even. However, we also have $k-(r_1-1)+r_2-1\in2\mathbb{Z}$, that is, $k-r_1-r_2=k-p=k-p'+1\in2\mathbb{Z}$. Likewise, $l-p'\in2\mathbb{Z}$. This means that $k,l$ cannot be even at the same time (i.e., they should satisfy \eqref{kl2}). Hence, we reach an contradiction and this case is not possible.

In fact, we can still reduce the number of cases to be checked. Since $r_1=p-r_2$, we have $r_1\pm1=p-(r_2\mp1)$. Therefore, we can rule out the cases where we choose $r_1\pm1,r_2\mp1$ from the $\binom{4}{2}$ possibilities as $\Delta_2\neq\Delta_3$. Hence, there are 16 cases (including the above three examples) overall. Moreover, just like in Example 3, we see that it fails to satisfy the fusion rule due to the parity of $k,l$. This can also be used to reduce the number of possible cases. One may check that
\begin{eqnarray}
	&&r=r_i,m=r_j\pm1,i=j~\Rightarrow~(k,l)\in2(\mathbb{Z},\mathbb{Z});\nonumber\\
	&&r=r_i,m=r_j\pm1,i\neq j~\Rightarrow~(k,l)\in(2\mathbb{Z},2\mathbb{Z}+1)~\text{or}~(2\mathbb{Z}+1,2\mathbb{Z}).
\end{eqnarray}
This further reduces the number of possible cases (including the first two examples) to 8. Although there are 8 distinct cases, there are only two conditions as in Example 1 and 2. This is because for the combination $r_i,r_i\pm1,r_j\pm1,r_j$, we always have
\begin{equation}
	2\leq k\leq\min(2r_1-2,2p-2r_1-2),
\end{equation}
and for the combination $r_i,r_{j\neq i}\pm1,r_\iota,r_{\kappa\neq\iota}\pm1$, we always have
\begin{equation}
	\max(|p-2r_1-1|+1,|p-2r_1+1|+1)\leq k\leq p-2.
\end{equation}

This completes the proof, and the above conditions are summarized in Table \ref{conditions}. We can also see why the tetracritical Ising model is the one with smallest $p'$ for $\Gamma(3)$. One way is to compute $p'=3,4,5$ (with possible $p$) case by case, and none of them would give parametrizations from $\Gamma(3)$. Alternatively, it is straightforward to use the above conditions as well. Likewise, we can deduce that the smallest possible $p$ is 5. Moreover, this also tells us why we cannot have $r_1=6$ or $s_1=6$ for $p'=6,7$ and why $s_1=6$ is not allowed for $(p',p)=(8,5)$ as in Figure \ref{kac} etc.

If a minimal model has CBs corresponding to $\Gamma(3)$, then $(r_1,s_1)=(3,3)$ (and hence $(r_2,s_2)=(p-3,p'-3)$) must be one solution. It is not hard to find that $(k,l)$ is $(2,2)$ or $(p-2,p'-2)$, and either $\Delta_2$ or $\Delta_3$ corresponds to $(2,2)$ or $(p-2,p'-2)$ for all the eight cases. Therefore, we may use this to solve $M_0$ and $\epsilon_{1,2}$. Suppose $\Delta_\text{int}=\Delta_3$, then
\begin{equation}
	\frac{Q^2}{4}-\frac{4}{9}M_0^2=-M_0^2+QM_0.
\end{equation}
Hence, $M_0=\frac{3Q}{10}$ or $\frac{3Q}{2}$ with $Q=\frac{i}{\sqrt{p'(p'-1)}}$. If we consider $\Delta_\text{int}=\Delta_2$ (which we have seen that this would give no new CBs), then we have the opposite values, that is, $M_0=-\frac{3Q}{10}$ or $-\frac{3Q}{2}$. Using $M_0=\frac{m_0}{\sqrt{\epsilon_1\epsilon_2}}$ and $\sqrt{\epsilon_1\epsilon_2}Q=\epsilon_1+\epsilon_2$, we may also solve $\epsilon_{1,2}$.

\subsection{Minimal Models and $\Gamma_0(4)\cap\Gamma(2)$}\label{onemoreex}
Let us now discuss one more example, $\Gamma_0(4)\cap\Gamma(2)$. We first focus on the cases when $\zeta=1/2$. In terms of the simplifications we can make as above, there are only two cases we need to consider. Again, we set $M_0=\frac{m_0}{\sqrt{\epsilon_1\epsilon_2}}$. In particular, one can find that the two cases only differ by $\Delta_3$. However, after some calculations, the fusion rule would always lead to $p',p\in2\mathbb{Z}$, which is impossible for coprime $p'$ and $p$.

Next, for $\zeta=2$, it is very similar to $\zeta=1/2$ but with a swap of $m_2,m_3$ and an overall rescaling. We also have two distinct cases. For $(+,+,-,-)$ \footnote{Here, it is still sufficient to choose two representatives for the two distinct cases. As different parametrizations of the masses would only differ by signs of $m_i$'s, we will only use their signs to denote $(m_0,m_1,m_2,m_3)$. This should be clear from the tables in \S\ref{matching}.}, using the same method yields the CBs in minimal models with conditions in Table \ref{cond1}.
\begin{center}
	\centering
	\renewcommand*{\arraystretch}{1.2}
	\begin{longtable}{c|c}
		Cases & Conditions \\ \hline\hline
		$2r_0,2r_0\pm1,r_0\pm1,r_0$ &  \\ \cline{1-1}
		$2r_0,2r_0\pm1,p-(r_0\pm1),p-r_0$ & $r_0\leq\frac{p-1\pm1}{3},$ \\ \cline{1-1}
		$p-2r_0,p-(2r_0\pm1),r_0\pm1,r_0$ & $k=2r_0$ \\ \cline{1-1}
		$p-2r_0,p-(2r_0\pm1),p-(r_0\pm1),p-r_0$ &  \\ \hline
		$p-2r_0,2r_0\pm1,p-(r_0\pm1),r_0$ & $\left(r_0<\frac{p\mp1}{4}\right.$ or \\ \cline{1-1}
		$p-2r_0,2r_0\pm1,r_0\pm1,p-r_0$ & $\frac{p\mp1}{4}\leq r_0\leq\frac{p-1/2\mp1/2}{3}$ or $\left.r_0=\frac{p-1}{2}\right)$ \\ \cline{1-1}
		$2r_0,p-(2r_0\pm1),p-(r_0\pm1),r_0$ & and $k=p-2r_0$ \\ \cline{1-1}
		$2r_0,p-(2r_0\pm1),r_0\pm1,p-r_0$ &  \\ \hline
		\caption{One set of possible CBs in minimal models that $\Gamma_0(4)\cap\Gamma(2)$ corresponds to. There are similar relations for $s_0,l,p'$ by a simple substitution of the corresponding letters, where we have set $\alpha_4=\alpha_{r_0,s_0}$.}\label{cond1}
	\end{longtable}
\end{center}
Likewise, the other case with $(-,+,-,+)$ gives the conditions in Table \ref{cond2}.
\begin{center}
	\centering
	\renewcommand*{\arraystretch}{1.2}
	\begin{longtable}{c|c}
		Cases & Conditions \\ \hline\hline
		$2r_0,2r_0\pm1,r_0\mp1,r_0$ &  \\ \cline{1-1}
		$2r_0,2r_0\pm1,p-(r_0\mp1),p-r_0$ & $r_0\leq\frac{p-1\pm1}{3},$ \\ \cline{1-1}
		$p-2r_0,p-(2r_0\pm1),r_0\mp1,r_0$ & $k=2r_0$ \\ \cline{1-1}
		$p-2r_0,p-(2r_0\pm1),p-(r_0\mp1),p-r_0$ &  \\ \hline
		$p-2r_0,2r_0\pm1,p-(r_0\mp1),r_0$ & $\left(r_0<\frac{p\mp1}{4}\right.$ or \\ \cline{1-1}
		$p-2r_0,2r_0\pm1,r_0\mp1,p-r_0$ & $\frac{p\mp1}{4}\leq r_0\leq\frac{p-1/2\mp1/2}{3}$ or $\left.r_0=\frac{p-1}{2}\right)$ \\ \cline{1-1}
		$2r_0,p-(2r_0\pm1),p-(r_0\mp1),r_0$ & and $k=p-2r_0$ \\ \cline{1-1}
		$2r_0,p-(2r_0\pm1),r_0\mp1,p-r_0$ &  \\ \hline
		\caption{The other set of possible CBs in minimal models that $\Gamma_0(4)\cap\Gamma(2)$ corresponds to. There are similar relations for $s_0,l,p'$ by a simple substitution of the corresponding letters, where we have set $\alpha_4=\alpha_{r_0,s_0}$.}\label{cond2}
	\end{longtable}
\end{center}

\begin{figure}[h]
	\centering
	\begin{subfigure}{6cm}
		\centering
		\begin{tabular}{c|cccc}
			3& $\frac{3}{2}$ & $\frac{3}{5}$ & \textcolor{cyan}{$\frac{1}{10}$} & \textcolor{cyan}{0} \\
			2& $\frac{7}{16}$ & \textcolor{cyan}{$\frac{3}{80}$} & \textcolor{cyan}{$\frac{3}{80}$} & $\frac{7}{16}$ \\
			1& \textcolor{cyan}{0} & \textcolor{cyan}{$\frac{1}{10}$} & $\frac{3}{5}$ & $\frac{3}{2}$ \\ \hline
			& 1 & 2 & 3 & 4
		\end{tabular}
	\tikzset{every picture/.style={line width=0.75pt}} 
	\begin{tikzpicture}[x=0.75pt,y=0.75pt,yscale=-1,xscale=1]
		\draw [line width=1.5]    (179.9,119.77) -- (90.17,119.83) ;
		\draw [line width=1.5]    (120.17,120.5) -- (120.23,90.77) ;
		\draw [line width=1.5]    (150.17,119.5) -- (149.9,90.43) ;
		\draw [color={rgb, 255:red, 155; green, 155; blue, 155 }  ,draw opacity=1 ][line width=1.5]    (299.9,119.77) -- (210.17,119.83) ;
		\draw [color={rgb, 255:red, 155; green, 155; blue, 155 }  ,draw opacity=1 ][line width=1.5]    (240.17,120.5) -- (240.23,90.77) ;
		\draw [color={rgb, 255:red, 155; green, 155; blue, 155 }  ,draw opacity=1 ][line width=1.5]    (270.17,119.5) -- (269.9,90.43) ;
		\draw (81,122.4) node [anchor=north west][inner sep=0.75pt]  [font=\footnotesize]  {$\frac{3}{80}$};
		\draw (170,122.4) node [anchor=north west][inner sep=0.75pt]  [font=\footnotesize]  {0};
		\draw (102,81.4) node [anchor=north west][inner sep=0.75pt]  [font=\footnotesize]  {$\frac{1}{10}$};
		\draw (152.9,80.83) node [anchor=north west][inner sep=0.75pt]  [font=\footnotesize]  {$\frac{3}{80}$};
		\draw (129,123.4) node [anchor=north west][inner sep=0.75pt]  [font=\footnotesize]  {$\frac{3}{80}$};
		\draw (201,122.4) node [anchor=north west][inner sep=0.75pt]  [font=\footnotesize,color={rgb, 255:red, 155; green, 155; blue, 155 }  ,opacity=1 ]  {$\frac{3}{80}$};
		\draw (290,122.4) node [anchor=north west][inner sep=0.75pt]  [font=\footnotesize,color={rgb, 255:red, 155; green, 155; blue, 155 }  ,opacity=1 ]  {0};
		\draw (222,81.4) node [anchor=north west][inner sep=0.75pt]  [font=\footnotesize,color={rgb, 255:red, 155; green, 155; blue, 155 }  ,opacity=1 ]  {0};
		\draw (272.9,80.83) node [anchor=north west][inner sep=0.75pt]  [font=\footnotesize,color={rgb, 255:red, 155; green, 155; blue, 155 }  ,opacity=1 ]  {$\frac{3}{80}$};
		\draw (249,123.4) node [anchor=north west][inner sep=0.75pt]  [font=\footnotesize,color={rgb, 255:red, 155; green, 155; blue, 155 }  ,opacity=1 ]  {$\frac{3}{80}$};
		\end{tikzpicture}
		\caption{}
	\end{subfigure}
	\begin{subfigure}{6cm}
		\centering
		\begin{minipage}[c]{0.49\textwidth}
			\begin{tabular}{c|ccc}
				2& $\frac{1}{2}$ & \textcolor{cyan}{$\frac{1}{16}$} & \textcolor{cyan}{0} \\
				1& \textcolor{cyan}{0} & \textcolor{cyan}{$\frac{1}{16}$} & $\frac{1}{2}$ \\ \hline
				& 1 & 2 & 3
			\end{tabular}
		\end{minipage}
	    \begin{minipage}[c]{0.49\textwidth}
	    \tikzset{every picture/.style={line width=0.75pt}} 
	    \begin{tikzpicture}[x=0.75pt,y=0.75pt,yscale=-1,xscale=1]
	    	\draw [line width=1.5]    (179.9,119.77) -- (90.17,119.83) ;
	    	\draw [line width=1.5]    (120.17,120.5) -- (120.23,90.77) ;
	    	\draw [line width=1.5]    (150.17,119.5) -- (149.9,90.43) ;
	    	\draw (81,122.4) node [anchor=north west][inner sep=0.75pt]  [font=\footnotesize]  {$\frac{1}{16}$};
	    	\draw (170,122.4) node [anchor=north west][inner sep=0.75pt]  [font=\footnotesize]  {0};
	    	\draw (102,81.4) node [anchor=north west][inner sep=0.75pt]  [font=\footnotesize]  {0};
	    	\draw (152.9,80.83) node [anchor=north west][inner sep=0.75pt]  [font=\footnotesize]  {$\frac{1}{16}$};
	    	\draw (129,123.4) node [anchor=north west][inner sep=0.75pt]  [font=\footnotesize]  {$\frac{1}{16}$};
	    \end{tikzpicture}
        \end{minipage}
		\caption{}
	\end{subfigure}
	\caption{Here we list the first possible examples of CBs that $\Gamma_0(4)\cap\Gamma(2)$ corresponds to: (a) The first CB for $(+,+,-,-)$. For reference, the one in grey is the CB from $(-,+,-,+)$ for this minimal model. (b) The first CB for $(-,+,-,+)$.}\label{kaccb}
\end{figure}
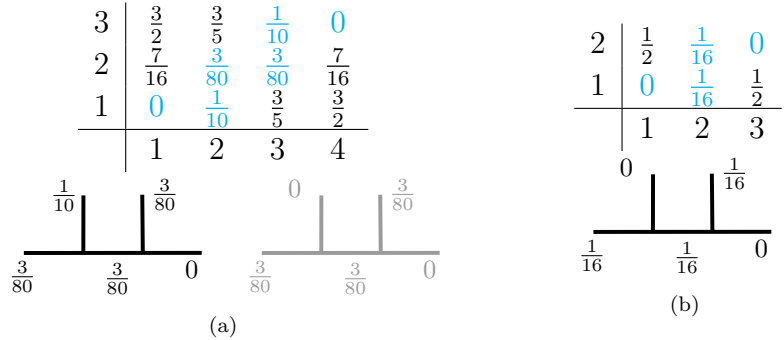

It is not hard to see that for $(+,+,-,-)$, the first CB appears in the minimal model with $p'=5,p=4$, viz, the tricritical Ising model. For $(-,+,-,+)$, the first CB appears in the minimal model with $p'=4,p=3$, viz, the (critical) Ising model. The Kac tables and corresponding CBs are shown in Figure \ref{kaccb}.

Finally, let us consider $\zeta=-1$. Since $a$ always vanishes, $\Delta_\text{int}=\frac{Q^2}{4}=-\frac{(p'-p)^2}{4p'p}$. Hence, $p'k-pl=0$, that is, $p'/p=l/k$. However, as $\gcd(p',p)=1$ and $k<p,l<p'$, this is impossible.

\begin{figure}[h]
	\centering
	\begin{minipage}[c]{1\textwidth}
		\centering
		\begin{tabular}{c|ccccc}
			4& 3 & $\frac{13}{8}$ & $\frac{2}{3}$ & \textcolor{cyan}{$\frac{1}{8}$} & \textcolor{green}{0} \\
			3& $\frac{7}{5}$ & $\frac{21}{40}$ & \textcolor{orange}{$\frac{1}{15}$} & \textcolor{orange}{$\frac{1}{40}$} & $\frac{2}{5}$ \\
			2& $\frac{2}{5}$ & \textcolor{orange}{$\frac{1}{40}$} & \textcolor{orange}{$\frac{1}{15}$} & $\frac{21}{40}$ & $\frac{7}{5}$ \\
			1& \textcolor{green}{0} & \textcolor{cyan}{$\frac{1}{8}$} & $\frac{2}{3}$ & $\frac{13}{8}$ & 3 \\ \hline
			& 1 & 2 & 3 & 4 & 5
		\end{tabular}
	\end{minipage}
	\begin{minipage}{1\textwidth}
		\centering
		\includegraphics[width=9cm]{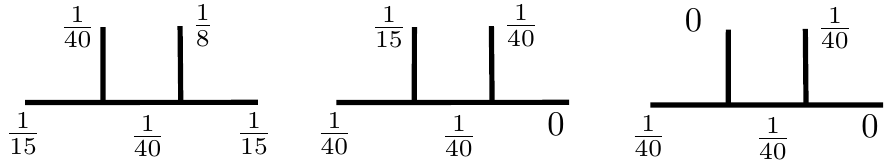}
	\end{minipage}
	\caption{The CBs from $\Gamma(3)$ (cyan) and $\Gamma_0(4)\cap\Gamma(2)$ (green) in the tetracritical Ising model. The ones in orange appear for both of the dessins. The three CBs, from left to right, come from $\Gamma(3)$, $(+,+,-,-)$ and $(-,+,-,+)$ in $\Gamma_0(4)\cap\Gamma(2)$ respectively.}\label{twodessins}
\end{figure}

Now that we have found two dessins that corresponds to CBs in minimal models, we can consider their CBs in the same minimal model. Such example would first appear when $p'=6,p=5$ as in Figure \ref{twodessins}.

\subsection{Minimal Models and General Dessins}\label{general}
Following the above steps, we can derive the results for any dessin in general.
\begin{proposition}
	Suppose for a dessin, we have the gauge theory parameters with relation
	\begin{equation}
		m_1=\pm/\mp k_1m_0,~m_2=\pm/\mp k_2m_0,~m_3=\pm/\mp k_3m_0,~a=\pm/\mp k_\textup{int}m_0,
	\end{equation}
	where $k_{i,\textup{int}}>0$. Then the dessin corresponds to the states of 4-point CBs satisfying conditions in Table \ref{gencond} in minimal models.
\end{proposition}\label{prop}
\begin{center}
	\centering
	\renewcommand*{\arraystretch}{1.2}
	\begin{longtable}{c|c}
		Cases & Conditions \\ \hline\hline
		$k_1r_0,k_2r_0\pm1,k_3r_0\pm/\mp1,r_0$ & $\max(|\blacktriangle\pm1|,|\vartriangle\pm1|)+1$ \\ \cline{1-1}
		$k_1r_0,k_2r_0\pm1,p-(k_3r_0\pm/\mp1),p-r_0$ & $\leq k_\text{int}r_0\leq\frac{1}{2}\min(\bigstar-3,4p-3-\bigstar),$ \\ \cline{1-1}
		$p-k_1r_0,p-(k_2r_0\pm1),k_3r_0\pm/\mp1,r_0$ & and $k=k_\text{int}r_0$ \\ \cline{1-1}
		$p-k_1r_0,p-(k_2r_0\pm1),p-(k_3r_0\pm/\mp1),p-r_0$ &  \\ \hline
		$p-k_1r_0,k_2r_0\pm1,p-(k_3r_0\pm/\mp1),r_0$ & $\max(|(k_1+k_2)r_2\pm1|,|(k_3+1)r_2\pm1|)+1$ \\ \cline{1-1}
		$p-k_1r_0,k_2r_0\pm1,k_3r_0\pm/\mp1,p-r_0$ & $\leq k_\text{int}r_0\leq\min(p-2-|\blacktriangle|,p-2-|\vartriangle|),$ \\ \cline{1-1}
		$k_1r_0,p-(k_2r_0\pm1),p-(k_3r_0\pm/\mp1),r_0$ & and $k=p-k_\text{int}r_0$ \\ \cline{1-1}
		$k_1r_0,p-(k_2r_0\pm1),k_3r_0\mp1,p-r_0$ &  \\ \hline
		\caption{The set of possible CBs in minimal models that a general dessin corresponds to. There are similar relations for $s_0,l,p'$ by a simple substitution of the corresponding letters, where we have set $\alpha_4=\alpha_{r_0,s_0}$ and $\blacktriangle=(k_2-k_1)r_0,\vartriangle=(1-k_3)r_0,\bigstar=\left(-|k_1+k_2-k_3-1|+\sum\limits_ik_i\right)r_0$. In particular, $k_{i,\text{int}}r_0\in\mathbb{N}^*$ is a necessary condition.}\label{gencond}
	\end{longtable}
\end{center}
In fact, we may further make the following conjecture.
\begin{conjecture}
	For a dessin satisfying the conditions in Proposition \ref{prop}, it corresponds to a family of 4-point CBs whose states follow Table \ref{gencond}. 
\end{conjecture}
So far, we have already discussed how a dessin can reproduce the charges/momenta of the states in a 4-point CB of a minimal model. However, as $\zeta$ is fixed for each dessin and we are only obtaining $\zeta$ by relating the Strebel and SW differentials rather than describing it as a concrete mathematical object in the language of dessins, further study on whether/how dessins could fully recover the CBs and the spectra is required.

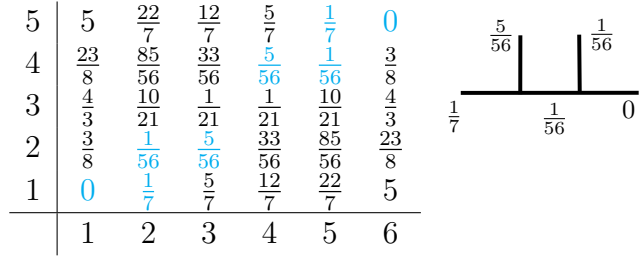
\begin{figure}[h]
		\centering
			\begin{tabular}{c|cccccc}
				5& 5 & $\frac{22}{7}$ & $\frac{12}{7}$ & $\frac{5}{7}$ & \textcolor{cyan}{$\frac{1}{7}$} & \textcolor{cyan}{0} \\
				4& $\frac{23}{8}$ & $\frac{85}{56}$ & $\frac{33}{56}$ & \textcolor{cyan}{$\frac{5}{56}$} & \textcolor{cyan}{$\frac{1}{56}$} & $\frac{3}{8}$ \\
				3& $\frac{4}{3}$ & $\frac{10}{21}$ & $\frac{1}{21}$ & $\frac{1}{21}$ & $\frac{10}{21}$ & $\frac{4}{3}$ \\
				2& $\frac{3}{8}$ & \textcolor{cyan}{$\frac{1}{56}$} & \textcolor{cyan}{$\frac{5}{56}$} & $\frac{33}{56}$ & $\frac{85}{56}$ & $\frac{23}{8}$ \\
				1& \textcolor{cyan}{0} & \textcolor{cyan}{$\frac{1}{7}$} & $\frac{5}{7}$ & $\frac{12}{7}$ & $\frac{22}{7}$ & 5 \\ \hline
				& 1 & 2 & 3 & 4 & 5 & 6
			\end{tabular}
			\tikzset{every picture/.style={line width=0.75pt}} 
			\begin{tikzpicture}[x=0.75pt,y=0.75pt,yscale=-1,xscale=1]
				\draw [line width=1.5]    (179.9,119.77) -- (90.17,119.83) ;
				\draw [line width=1.5]    (120.17,120.5) -- (120.23,90.77) ;
				\draw [line width=1.5]    (150.17,119.5) -- (149.9,90.43) ;
				\draw (81,122.4) node [anchor=north west][inner sep=0.75pt]  [font=\footnotesize]  {$\frac{1}{7}$};
				\draw (170,122.4) node [anchor=north west][inner sep=0.75pt]  [font=\footnotesize]  {0};
				\draw (102,81.4) node [anchor=north west][inner sep=0.75pt]  [font=\footnotesize]  {$\frac{5}{56}$};
				\draw (152.9,80.83) node [anchor=north west][inner sep=0.75pt]  [font=\footnotesize]  {$\frac{1}{56}$};
				\draw (129,123.4) node [anchor=north west][inner sep=0.75pt]  [font=\footnotesize]  {$\frac{1}{56}$};
			\end{tikzpicture}
	\caption{The CB on the right has conformal dimensions coloured cyan in the Kac table.}\label{exfirstcb}
\end{figure}

With the conditions in Table \ref{gencond}, we can check what CBs in minimal models we can obtain from a dessin. For instance, when $\zeta=\frac{1}{2}+\frac{11}{50}\sqrt{5}$ for $\Gamma_1(5)$, we have $k_2=1,k_1=k_3=5,k_\text{int}=2$. It is not hard to find that the first CB it corresponds to appears when $p'=7,p=6$ as in Figure \ref{exfirstcb}.

\paragraph{Examples not giving minimal models} From Proposition \ref{prop}, it is straightforward to see that there could be dessins that do not correspond to CBs in minimal models. Besides the inequalities in Table \ref{gencond}, a necessary condition is that $k_{i,\text{int}}r_0$ and $k_{i,\text{int}}s_0$ should be positive integers. Let us verify this with some examples.

For $\Gamma_0(6)$, there are two big classes of parametrizations. If $m_2$ or $m_3$ has the factor $\sqrt{109}$, then we cannot get the rational conformal dimensions for all the external legs. If instead $m_1$ has the factor $\sqrt{109}$, all the conformal dimensions can be rational since $\Delta_1=\frac{Q^2}{4}-M_0^2$. However, if we now express $M_0$ in terms of the labels $(r_2,s_2)$ for $\Delta_4$ and insert this into $\Delta_1$, we find that
\begin{equation}
	(p'r_1-ps_1)^2=4\times109(p'r_2-ps_2)^2,
\end{equation}
where 109 is not a square number, and hence no integer solutions (except when $0=0$ which is excluded for minimal models). Therefore, it is not possible to get CBs in minimal models for $\Gamma_0(6)$.

For $\Gamma_0(8)$, $m_i$ and $a$ are non-zero and cannot simultaneously be real/pure imaginary as in Table \ref{param5}. Without loss of generality, suppose $\frac{m_i}{\sqrt{\epsilon_1\epsilon_2}}$ is pure imaginary and then $\frac{a}{\sqrt{\epsilon_1\epsilon_2}}$ is real. This yields
\begin{equation}
	\Delta_\text{int}=\frac{Q^4}{4}-\frac{a^2}{\epsilon_1\epsilon_2}<\frac{Q^2}{4}.
\end{equation}
Therefore,
\begin{equation}
	\frac{(p'k-pl)^2-(p'-p)^2}{4p'p}<\frac{Q^2}{4}=-\frac{(p'-p)^2}{4p'p}.
\end{equation}
In other words,
\begin{equation}
	(p'k-pl)^2<0.
\end{equation}
Hence, it is not possible to get CBs in minimal models for $\Gamma_0(8)$.

For $\Gamma_0(9)$, since the $a$'s are not real or pure imaginary, it should not give CBs in minimal models.

\section*{Acknowledgement}
OF wishes to thank R. Santachiara for useful discussions on the topic of this note, as well as related topics. The research of JB is supported by the CSC scholarship. OF is supported by the Australian Reasearch Council. YHH would like to thank STFC for grant ST/J00037X/1. EH would like to thank STFC for the PhD studentship. YX is supported by NSFC grant No. 20191301017. FY is supported by the NSFC grant No. 11950410490, by Fundamental Research Funds for the Central Universities A0920502051904-48, by Start-up research grant A1920502051907-2-046, in part by NSFC grant No. 11501470 and No. 11671328, and by Recruiting Foreign Experts Program No. T2018050 granted by SAFEA.

\appendix

\section{The B-model and Omega Deformations}\label{bmodel}
When mapping gauge theory/SW geometry parameters to CFT parameters, we need to include a factor of $\frac{1}{\sqrt{\epsilon_1\epsilon_2}}$, which would lead to divergence under the flat space limit $\epsilon_{1,2}\rightarrow0$. Here, we discuss a way in terms of topological B-model so that the SW geometry is still physically meaningful when $\epsilon_{1,2}$ are non-zero.

Recall that we have related $\mathcal{N}=2$ gauge theories to A-model topological strings. The mirror in B-model is defined by the equation
\begin{equation}
	vw+f(x,y)=0,
\end{equation}
which is a CY$_3$ that can be considered as fibration of $uv=c$ for some constant $c$ over the Riemann surface $f(x,y)$. In particular, $f(x,y)=0$ can be identified as the SW curve $\Sigma$. Denote the multiplicity of a BPS state in this 5d theory as $N^\beta$, where $\beta$ is essentially the charge of the BPS state\footnote{More precisely, we should also include the indices denoting the $\text{SU}(2)_L\times\text{SU}(2)_R$ spin representations, but for our purpose here, it suffices to label it with the topological data $\beta$ only. For more details, see for example \cite{Huang:2013yta}.}. Mathematically, the BPS configuration can be defined by a (complex) one-dimensional sheaf $\mathcal{F}$ (plus certain section in $H^0(\mathcal{F})$) such that
\begin{equation}
	\beta=\text{ch}_2(\mathcal{F}),~n=\chi(\mathcal{F}),
\end{equation}
where $\beta\in H_2(\mathcal{M},\mathbb{Z})$ and $n\in\mathbb{Z}$.

The topological string amplitude then has the expansion
\begin{equation}
	F(\epsilon_1,\epsilon_2,t)=\log(Z)=\sum_{n,g=0}^\infty(\epsilon_1+\epsilon_2)^{2n}(\epsilon_1\epsilon_2)^{g-1}F^{(n,g)}(t),
\end{equation}
where $Z$ is known as the (refined) Pandharipande-Thomas (PT) partition function, and $g$ stands for the genus while $t$ denotes the K\"ahler parameter measuring the volume of a curve in $\beta$, which can be identified as the Coulomb parameter $a$ as we are focusing on SU(2) gauge group in this paper \cite{Huang:2013yta,Huang:2010kf,Huang:2011qx}. In particular, when $n=g=0$, $F^{(0,0)}$ is the prepotential $\mathcal{F}$. In the limit $\epsilon_{1,2}\rightarrow0$, the PT partition function is naturally identified as the Nekrasov partition function at leading order:
\begin{equation}
	\log(Z)=(\epsilon_1\epsilon_2)^{-1}F^{(0,0)}.
\end{equation}
Moreover, $F^{(0,1)}$ and $F^{(1,0)}$ can also be determined using the metric on $\mathcal{M}$ and the discriminant of $\Sigma$ as in Equation (3.22) and (3.23) in \cite{Huang:2011qx}. Then $F^{(n,g)}$ with higher $(g+n)$ can be deduced from the \emph{(generalized) holomorphic anomaly equation} \cite{Huang:2010kf,Huang:2011qx,Huang:2013yta}
\begin{equation}
	\bar{\partial}_{\bar{i}}F^{(n,g)}=\frac{1}{2}\bar{C}_{\bar{i}}^{jk}\left(\text{D}_j\text{D}_kF^{(n,g-1)}+\sum_{m,h}{\vphantom{\sum}}'\text{D}_jF^{(m,h)}\text{D}_kF^{(n-m,g-h)}\right),~g+n>1,
\end{equation}
where the three-point coupling $\bar{C}_{\bar{i}}^{jk}$ is given in \cite{Huang:2010kf,Huang:2011qx}, and D$_i$ is the covariant derivative. The prime in the sum indicates the omission of $(m,h)=(0,0),(n,g)$. We also require the first term on the right hand side to vanish if $g=0$.

Therefore, the non-zero $\epsilon_{1,2}$ would also make sense for the SW theory physically as the prepotential generates the topological string amplitudes. Hence, we could avoid the divergence when mapping the gauge theory parameters to CFT parameters as in \S\ref{dessin2CB}.

\section{Brane Configurations}\label{brane}

\subsection{The Type IIA Brane Configuration}\label{IIA}
A type IIA configuration of parallel NS/D5-branes joined by D4-branes can be represented in M theory as a single M5-brane with a more complicated world history.

Before we write the rule for finding the Seiberg-Witten curve, we need to find out whether we have a U($N$) or an SU($N$) gauge theory. This is discussed in \cite{Witten:1997sc}, and goes as follows.

First, consider D5-branes and D4-branes in type IIA superstring theory. The world-volume of a D5-brane is described as follows. D5-branes are located at $x^7=x^8=x^9=0$ and, in a semi-classical approximation, at fixed values of $x^6$. The world-volume of D5-branes are parameterised by values of $x^0, x^1, \cdots, x^5$. In addition, D4-branes are parameterised by $x^0, x^1, x^2, x^3$ and $x^6$. D4-branes have their $x^6$-coordinate finite so that they terminate on D5-branes. We need to introduce a complex variable $v = x^4 + i x^5$. Classically, every D4-brane is located at a definite value of $v$. Since a D4-brane ending on a D5-brane creates a \textit{dimple} in the D5-brane, the value $x^6$ is the value measured at $v = \infty$, far from the disturbance created by the D4-brane. By minimizing the volume of the D5-brane, at large $v$, we obtain
\begin{equation}
x^6=k\mathrm{ln}|v|+\mathrm{const}.
\end{equation}
This is not well-defined for large $v$. Nevertheless, with D4-branes attached to the left and to the right of the D5-brane, we have
\begin{equation}
x^6=k\sum_{i=1}^{q_L}\mathrm{ln}|v-a_i|-k\sum_{j=1}^{q_R}\mathrm{ln}|v-b_j|+\mathrm{const},\label{x_6}
\end{equation}
where $a_i$ and $b_j$ are the $v$-values, or $x^6$-coordinates of D4-branes ending on the left and right respectively. Now $x^6$ is well-defined for large $v$ if and only if $q_L = q_R$, that is, if the forces on both sides are balanced. For infrared divergence, we need to consider the motion of the D4-branes, whose movement causes the D5-brane to move. The motion of a D5-brane contributes to the kinetic energy of the D4-brane. The D5-brane kinetic energy is given by $\int\text{d}^4x\text{d}^2v\sum\limits_{\mu=0}^3\partial_{\mu}x^6\partial^{\mu}x^6$. Therefore, with $x^6$ in (\ref{x_6}), we have
\begin{equation}
	k^2\int\text{d}^4x\text{d}^2v\left|\text{Re}\left(\sum_i\frac{\partial_\mu a_i}{v-a_i}-\sum_j\frac{\partial_\mu b_j}{v-b_j}\right)\right|^2.
\end{equation}
This integral converges if and only if
\begin{equation}
	\partial_\mu\left(\sum_ia_i-\sum_jb_j\right)=0,
\end{equation}
so that
\begin{equation}
\sum_i a_i - \sum_j b_j = q_{\alpha},\label{bare_mass}
\end{equation}
where $q_{\alpha}$ is characteristic of $\alpha$-th plane. From the D4-brane point of view, (\ref{bare_mass}) means the U(1) part of U($k$) for $k$ D4-branes between two D5-branes are frozen. This is because $\sum\limits_i a_i$ is the scalar part of U(1) vector multiplet in one factor U($k_{\alpha}$) and $\sum\limits_j b_j$ is the scalar part of the U(1) vector multiplet in the factor U($k_{\alpha + 1}$). Since, following (\ref{bare_mass}), the difference is fixed by supersymmetry, the entire U(1) vector multiplet is missing, and we have SU($N$).

\subsection{The M-theory Brane Configuration}\label{Mthy}
The world-volume of the M5-brane is such that,
\begin{enumerate}
	\item It has arbitrary values in the first $\mathbb{M}^4$ coordinates $x^0, \cdots, x^3$, and is located at $x^7 = x^8 = x^9 = 0$;	
	\item In the remaining four coordinates, which parametrize a 4-manifold 
	$Q \cong \mathbb{R}^3 \times S^1$, D5-brane worldvolume spans 
	a 2d surface $\Sigma$;
	\item The $\mathcal{N} = 2$ supersymmetry means we give $Q$ the complex structure 
	in which $v = x^4 + ix^5$ and $s = x^6 + ix^{10}$ are holomorphic, then $\Sigma$ 
	is a complex Riemann surface in $Q$. This makes $\mathbb{M}^4 \times \Sigma$ 
	a supersymmetric cycle in the sense of \cite{Becker:1995kb} and so it ensures spacetime
	supersymmetry.
\end{enumerate}
When projected to type IIA brane diagrams, $\Sigma$ has different components described locally by saying that $s$ is constant (the D5-branes) or that $v$ is constant (the D4-branes). In type IIA, different components can meet and singularity appears in there. However, in going to M theory, singularities disappear. Hence, for generic values of parameters, $\Sigma$ will be a smooth Riemann surface in $Q$.

\section{Congruence Subgroups of the Modular Group}\label{conggp}
In this appendix, we very briefly recall some essential details regarding the modular group
$\Gamma\equiv\mathrm{\Gamma\left(1\right)=PSL}\left(2,\mathbb{Z}\right)=\mathrm{SL}\left(2,\mathbb{Z}\right)/\left\{ \pm I\right\} $, the group of linear fractional transformations $\mathbb{Z} \ni z\rightarrow\frac{az+b}{cz+d}$, with $a,b,c,d\in\mathbb{Z}$ and $ad-bc=1$. It is generated by the transformations $T$ and $S$ defined by
\begin{equation}
T(z)=z+1\quad,\quad S(z)=-1/z \ .
\end{equation}
The presentation of $\Gamma$ is $\left\langle S,T|S^{2}=\left(ST\right)^{3}=I\right\rangle$.

The most important subgroups of $\Gamma$ are the \emph{congruence} subgroups, defined by having the entries in the generating matrices $S$ and $T$ obeying some modular arithmetic. Of particular note are the following:
\begin{itemize}
	\item Principal congruence subgroups:
	\[
	\Gamma\left(m\right):=\left\{ A\in\mathrm{SL}(2;\mathbb{Z})\;;\; A_{ij}\equiv\pm I_{ij}\;\mathrm{mod}\; m\right\} /\left\{ \pm I\right\} ;
	\]
	
	\item Congruence subgroups of level $m$: subgroups of $\Gamma$ containing
	$\Gamma\left(m\right)$ but not any $\Gamma\left(n\right)$ for $n<m$;
	\item Unipotent matrices:
	\[
	\Gamma_{1}\left(m\right):=\left\{ A\in\mathrm{SL}(2;\mathbb{Z})\;;\; A_{ij}\equiv\pm\begin{pmatrix}1 & b\\
	0 & 1
	\end{pmatrix}_{ij}\;\mathrm{mod}\; m\right\} /\left\{ \pm I\right\} ;
	\]
	
	\item Upper triangular matrices:
	\[
	\Gamma_{0}\left(m\right):=\left\{ \begin{pmatrix}a & b\\
	c & d
	\end{pmatrix}\in\Gamma\;;\; c\equiv0\;\mathrm{mod}\; m\right\} /\left\{ \pm I\right\} .
	\]
\end{itemize}
In \cite{MS,He:2012kw}, attention is drawn to the conjugacy classes of a particular family of subgroups of $\Gamma$: the so-called\emph{ genus zero, torsion-free} congruence subgroups:
\begin{itemize}
	\item \emph{Torsion-free} means that the subgroup contains no element of finite order other than the identity. 
	\item To explain \emph{genus zero}, first recall that the modular group acts on the upper half-plane $\mathcal{H}:=\left\{\tau\in\mathbb{C}\;,\;\mathrm{Im}\left(\tau\right)>0\right\} $ by linear fractional transformations $z\rightarrow\frac{az+b}{cz+d}$. Then $\mathcal{H}$ gives rise to a compactification $\mathcal{H}^{*}$ when adjoining \emph{cusps}, which are points on $\mathbb{R}\sqcup\infty$ fixed under some parabolic element (i.e. an element $A\in\Gamma$ not equal to the identity and for which $\mathrm{Tr}\left(A\right)=2$). The quotient $\mathcal{H}^{*}/\Gamma$ is a compact Riemann surface of genus 0, i.e. a sphere. It turns out that with the addition of appropriate cusp points, the extended upper half plane $\mathcal{H}^{*}$ factored by various congruence subgroups will also be compact Riemann surfaces, possibly of higher genus. Such a Riemann surface, as a complex algebraic variety, is called a \emph{modular curve}. The genus of a subgroup of the modular group is the genus of the modular curve produced in this way. 
\end{itemize}
The genus zero torsion-free congruence subgroups of the modular group are very rare: there are only 33 of them, with index $I\in\left\{6,12,24,36,48,60\right\}$, as detailed in \cite{MS}.

\section{Elliptic Curves and $j$-Invariants}\label{ellip}
Given the Weierstrass function $\wp$
\begin{equation}
\wp \left( z | \ \omega_1,\omega_2 \right) = \frac{1}{z^2} + \sum_{n^2 + m^2 \neq 0} 
\left(
\frac{1}{ \left( z + m \omega_1 + n \omega_2 \right)^2} - 
\frac{1}{ \left(     m \omega_1 + n \omega_2 \right)^2}
\right),
\label{wp}
\end{equation}
where $\omega_1$ and $\omega_2$ are complex-valued vectors that span the lattice $\Lambda = \{m \omega_1 + n \omega_2:$ $m,n \in \mathbb{Z} \}$, and we can write $\wp \left( z | \  \omega_1, \omega_2 \right) = \wp \left( z | \  \Lambda \right)$. The embedding of a torus, as an elliptic curve over $\mathbb{C}$ in the complex projective plane, follows from
\begin{equation}
\left( \wp'(z) \right)^2 = 4 \left( \wp(z) \right)^3 - g_2 \wp(z) - g_3,
\label{w.ec}
\end{equation}
where $\wp'(z)$ is the derivative of $\wp(z)$ with respect to $z$. Naturally defined 
on a torus $\mathbb{C}/\Lambda$, $\wp$ is doubly-periodic with respect to lattice $\Lambda$. This torus can be embedded in the complex projective plane by $z \mapsto [1:\wp(z):\wp'(z)]$. Close to the origin, $\wp (z)$ can be expanded as
\begin{equation}
\wp \left( z | \  \Lambda \right) = 
\frac{1}{z^2} + g_2 \frac{z^2}{20} + g_3 \frac{z^4}{28} + \mathcal{O} \left( z^6 \right),
\label{laurent.wp}
\end{equation}
where
\begin{eqnarray}
g_2 &=& 60 
\sum_{(m,n) \neq (0,0)}
\left(  
\frac{1}{m \omega_1 + n \omega_2}
\right)^4,
\nonumber\\
g_3 &=& 140 
\sum_{(m,n) \neq (0,0)} 
\left(  
\frac{1}{m \omega_1 + n \omega_2}
\right)^6.
\end{eqnarray}
The summed terms in $g_2$ and $g_3$ are the first two Eisenstein series respectively. The Eisenstein series $G_{2k}$ with weight $2k$ are modular forms of weight $2k$, that is, they transform as $G_{2k}(\tau) \mapsto \left( c \tau + d \right)^{2k} G_{2k} (\tau)$ under SL$(2,\mathbb{Z})$ with $\tau = \omega_1/\omega_2$ in upper half-plane $\mathcal{H}$. If two 
lattices are related by a multiplication by a non-zero complex number $c$, then the corresponding curves are isomorphic. The $j$-invariants are defined as
\begin{equation}
j (\tau) = 1728 \frac{g_2^3}{g_2^3 - 27 g_3^2}.
\label{def:j_inv}
\end{equation}
This definition shows that the $j$-invariant is a weight-zero modular form. From the above discussion, we can see that each isomorphism class of elliptic curves over $\mathbb{C}$ has the same $j$-invariant.

As the SW curves and Strebel differentials we have are of quartic form, $y^2=az^4+bz^3+cz^2+dz+q^2$, we can make the substitution (for $q\neq0$)
\begin{equation}
	z=\frac{2q(X+c)-d^2/(2q)}{Y},~y=-q+\frac{1}{2q}\frac{2q(X+c)-d^2/(2q)}{Y}\left(\frac{2q(X+c)-d^2/2q}{Y}-d\right)
\end{equation}
so that the elliptic curve can be expressed in the standard Weierstrass form
\begin{equation}
	Y^2+a_1XY+a_3Y=X^3+a_2X^2+a_4X+a_6,
\end{equation}
where
\begin{equation}
	a_1=\frac{d}{q},~a_2=c-\frac{d^2}{4q^2},a_3=2bq,~a_4=-4aq^2,a_6=ad^2-4acq^2.
\end{equation}
Using \texttt{SAGE} \cite{sagemath}, we can compute its $j$-invariant
{\scriptsize\begin{equation}
	j=-\frac{((a_1^2 + 4a_2)^2 - 24a_1a_3 - 48a_4)^3}{(a_2a_3^2 - a_1a_3a_4 + a_1^2a_6 - a4^2 + 4a_2a_6)(a_1^2 + 4a_2)^2 + 8(a_1a_3 + 2a_4)^3 - 9(a_1^2 + 4a_2)(a_1a_3 + 2a_4)(a_3^2 + 4a_6) + 27(a_3^2 + 4a_6)^2}.
\end{equation}}
If $q=0$ such as the Strebel differential for $\Gamma(3)$, we can replace $z$ and $y$ with $1/z$ and $y/z^2$ respectively to obtain a quartic form with a non-vanishing constant term \cite{Connell1999}.

\section{Elliptic Functions and Coulomb Moduli}\label{elliptica}

\subsection{The Elliptic Integral of First Kind}\label{ellipticK}
We first give a quick review on deriving \eqref{dadu}. From \eqref{lamSW}, we have
\begin{equation}
	\frac{\text{d}\lambda_\text{SW}}{\text{d}U}=-\frac{1+\zeta}{2\sqrt{P_4(z)}}.
\end{equation}
Then
\begin{equation}
	\frac{\text{d}a}{\text{d}U}=\frac{1}{2\pi i}\oint_A\frac{\text{d}\lambda_\text{SW}}{\text{d}U}=-\frac{2}{2\pi i}\int_{\lambda_1}^{\lambda_2}\frac{1+\zeta}{2\sqrt{P_4(z)}}\text{d}z.
\end{equation}
Therefore, the integral boils down to solving
\begin{equation}
	\int_{\lambda_1}^{\lambda_2}\frac{\text{d}z}{\sqrt{(z-\lambda_1)(z-\lambda_2)(z-\lambda_3)(z-\lambda_4)}}.
\end{equation}
First, we make a PSL(2,$\mathbb{Z}$) transformation, $z=\frac{At+B}{Ct+D}$, such that $\lambda_{1,2,3}$ are mapped to $0,1,\infty$ respectively. Then $A=(\lambda_1-\lambda_2)\lambda_3,B=(\lambda_2-\lambda_3)\lambda_1,C=\lambda_1-\lambda_2,D=\lambda_2-\lambda_3$ gives a solution, and $\lambda_4=\frac{(\lambda_2-\lambda_3)(\lambda_4-\lambda_1)}{(\lambda_1-\lambda_3)(\lambda_3-\lambda_4)}=:t_0$. After substitution of variables and some algebra, the integral becomes
\begin{equation}
	\frac{1}{\sqrt{(\lambda_2-\lambda_3)(\lambda_4-\lambda_1)}}\int_0^1\frac{\text{d}t}{\sqrt{t(1-t)(1-t_0^{-1}t)}}.
\end{equation}
In particular,
\begin{equation}
	\int_0^1\frac{\text{d}t}{\sqrt{t(1-t)(1-t_0^{-1}t)}}=\pi~{}_2F_1\left(\frac{1}{2},\frac{1}{2};1,t_0^{-1}\right)=2K(t_0^{-1/2}),
\end{equation}
where $K$ is the elliptic integral of the first kind
\begin{equation}
	K(x)=\int_0^1\frac{\text{d}t}{\sqrt{(1-t^2)(1-x^2t^2)}}.
\end{equation}
Hence,
\begin{equation}
	\int_{\lambda_1}^{\lambda_2}\frac{\text{d}z}{\sqrt{(z-\lambda_1)(z-\lambda_2)(z-\lambda_3)(z-\lambda_4)}}=\frac{2K\left(\sqrt{\frac{(\lambda_1-\lambda_2)(\lambda_3-\lambda_4)}{(\lambda_2-\lambda_3)(\lambda_4-\lambda_1)}}\right)}{\sqrt{(\lambda_2-\lambda_3)(\lambda_4-\lambda_1)}}.
\end{equation}

\subsection{The Elliptic Logarithm}\label{ellipticLog}
Here we present an alternative way to obtain $a$ from $\text{d}a/\text{d}U$ by integrating from $U$ to $\infty$. Therefore, we need to determine $a$ when $U\rightarrow\infty$. At large $U$, we have
\begin{equation}
	\frac{P_4(z)}{U}\bigg|_{U\rightarrow\infty}=(-1-\zeta)z^3+(1+\zeta)^2z^2+(-\zeta(1+\zeta))z,
\end{equation}
which yields
\begin{equation}
	P_4(z)|_{U\rightarrow\infty}=-(1+\zeta)Uz^3+(1+\zeta)^2Uz^2-\zeta(1+\zeta)Uz.
\end{equation}
If $|\zeta|\leq1$, then
\begin{eqnarray}
	\frac{\text{d}a}{\text{d}U}\bigg|_{U\rightarrow\infty}&=&-\frac{1+\zeta}{2\pi i}\int_{\lambda_1}^{\lambda_2}\left((-(1+\zeta)U)(z^3-(1+\zeta)z^2+\zeta z)\right)^{-1/2}\text{d}z\nonumber\\
	&=&\frac{\sqrt{1+\zeta}}{2\pi\sqrt{U}}\int_1^\infty(z^3-(1+\zeta)z^2+\zeta z)^{-1/2}\text{d}z\nonumber\\
	&=&-\frac{\sqrt{1+\zeta}}{\pi\sqrt{U}}\text{EL}(1,0;-1-\zeta,\zeta),
\end{eqnarray}
where EL is the elliptic logarithm defined as
\begin{equation}
	\text{EL}(x,y;a,b)=\frac{1}{2}\int_\infty^x\frac{\text{d}t}{\sqrt{t^3+at^2+bt}},~~y=\sqrt{x^3+ax^2+bx}.
\end{equation}
Therefore,
\begin{equation}
	a(U)|_{U\rightarrow\infty}=-\frac{2}{\pi}\sqrt{1+\zeta}\text{EL}(1,0;-1-\zeta,\zeta)\sqrt{U}.
\end{equation}
However, notice that the above steps are not rigorous. We need to be careful about the branches of square roots. Taking this into account, when $1+\zeta<0$, there should be an minus extra sign\footnote{In practice, we usually choose a large cutoff (which can be either positive or negative) for $U$ instead of $\infty$ when performing numerical integrals. Therefore, the branches of square roots with $U$ inside are also important if we take $U$ to a large negative number.}, that is,
\begin{equation}
	a(U)|_{U\rightarrow\infty,\zeta<-1}=\frac{2}{\pi}\sqrt{1+\zeta}\text{EL}(1,0;-1-\zeta,\zeta)\sqrt{U}.
\end{equation}
Henceforth, we will not repeat this point below. As a sanity check, we can see what would happen at weak coupling. When $\zeta=0$, $\text{EL}(1,0;-1,0)=\pi/2$. We learn that\footnote{We may also have an extra minus sign for all the expressions of $a(U)|_{U\rightarrow\infty}$ here. Mathematically, this should correspond to choosing a different branch in the redefinition of square root. Physically, this is due to the action of Weyl group of the gauge symmetry.}
\begin{equation}
	a(U)|_{U\rightarrow\infty,\zeta\rightarrow0}=-\sqrt{U},
\end{equation}
which is the familiar behaviour in the (semi)classical limit.

If $|\zeta|>1$, then we can just replace $\int_1^\infty$ with $\int_\zeta^\infty$, and hence
\begin{equation}
	a(U)|_{U\rightarrow\infty}=-\frac{2}{\pi}\sqrt{1+\zeta}\text{EL}(\zeta,0;-1-\zeta,\zeta)\sqrt{U}.
\end{equation}
Likewise, we can also write down a similar expression for $a_D$ at large $U$,
\begin{equation}
	a_D=-\frac{2}{\pi}\sqrt{1+\zeta}\text{EL}(0,0;-1-\zeta,\zeta)\sqrt{U}.
\end{equation}
If we take $\zeta=0$, then $\text{EL}(0,0;-1,0)$ goes to $\infty$. This is expected as the monopoles are heavy for weak coupling.

Therefore, the integral for $a(U)$ can be written as
\begin{equation}
	a(U) = a(\infty) + \frac{1+ \zeta}{m_0 \pi \text{i}} \int_U^\infty\frac{\text{d}U'}{\sqrt[4]{x_2(U')}} K \left( \frac{\sqrt[4]{x_1(U')}}{\sqrt[4]{x_2(U')}} \right),
\end{equation}
where $\int_U^\infty$, just like \eqref{intsum}, could still be a sum of integrals due to the non-trivial monodromy.

\addcontentsline{toc}{section}{References}
\bibliographystyle{utphys}
\bibliography{references}

\providecommand{\href}[2]{#2}\begingroup\raggedright\begin{thebibliography}{10}

\bibitem{Alday:2009aq}
L.~F. Alday, D.~Gaiotto, and Y.~Tachikawa, ``{Liouville Correlation Functions
  from Four-dimensional Gauge Theories},''
  \href{http://dx.doi.org/10.1007/s11005-010-0369-5}{{\em Lett. Math. Phys.}
  {\bfseries 91} (2010) 167--197},
  \href{http://arxiv.org/abs/0906.3219}{{\ttfamily arXiv:0906.3219 [hep-th]}}.

\bibitem{Seiberg:1994rs}
N.~Seiberg and E.~Witten, ``{Electric - magnetic duality, monopole
  condensation, and confinement in N=2 supersymmetric Yang-Mills theory},''
  \href{http://dx.doi.org/10.1016/0550-3213(94)90124-4}{{\em Nucl. Phys. B}
  {\bfseries 426} (1994) 19--52},
  \href{http://arxiv.org/abs/hep-th/9407087}{{\ttfamily arXiv:hep-th/9407087}}.
  [Erratum: Nucl.Phys.B 430, 485--486 (1994)].

\bibitem{Seiberg:1994aj}
N.~Seiberg and E.~Witten, ``{Monopoles, duality and chiral symmetry breaking in
  N=2 supersymmetric QCD},''
  \href{http://dx.doi.org/10.1016/0550-3213(94)90214-3}{{\em Nucl. Phys. B}
  {\bfseries 431} (1994) 484--550},
  \href{http://arxiv.org/abs/hep-th/9408099}{{\ttfamily arXiv:hep-th/9408099}}.

\bibitem{Nekrasov:2002qd}
N.~A. Nekrasov, ``{Seiberg-Witten prepotential from instanton counting},''
  \href{http://dx.doi.org/10.4310/ATMP.2003.v7.n5.a4}{{\em Adv. Theor. Math.
  Phys.} {\bfseries 7} no.~5, (2003) 831--864},
  \href{http://arxiv.org/abs/hep-th/0206161}{{\ttfamily arXiv:hep-th/0206161}}.

\bibitem{Gaiotto:2009we}
D.~Gaiotto, ``{N=2 dualities},''
  \href{http://dx.doi.org/10.1007/JHEP08(2012)034}{{\em JHEP} {\bfseries 08}
  (2012) 034}, \href{http://arxiv.org/abs/0904.2715}{{\ttfamily arXiv:0904.2715
  [hep-th]}}.

\bibitem{He:2012kw}
Y.-H. He and J.~McKay, ``{N=2 Gauge Theories: Congruence Subgroups, Coset
  Graphs and Modular Surfaces},''
  \href{http://dx.doi.org/10.1063/1.4772976}{{\em J. Math. Phys.} {\bfseries
  54} (2013) 012301}, \href{http://arxiv.org/abs/1201.3633}{{\ttfamily
  arXiv:1201.3633 [hep-th]}}.

\bibitem{He:2012jn}
Y.-H. He, J.~McKay, and J.~Read, ``{Modular Subgroups, Dessins d'Enfants and
  Elliptic K3 Surfaces},''
  \href{http://dx.doi.org/10.1112/S1461157013000119}{{\em J. Comput. Math.}
  {\bfseries 16} (2013) 271--318},
  \href{http://arxiv.org/abs/1211.1931}{{\ttfamily arXiv:1211.1931 [math.AG]}}.

\bibitem{He:2013lha}
Y.-H. He and J.~McKay, ``{Eta Products, BPS States and K3 Surfaces},''
  \href{http://dx.doi.org/10.1007/JHEP01(2014)113}{{\em JHEP} {\bfseries 01}
  (2014) 113}, \href{http://arxiv.org/abs/1308.5233}{{\ttfamily arXiv:1308.5233
  [hep-th]}}.

\bibitem{He:2015vua}
Y.-H. He and J.~Read, ``{Dessins d'enfants in $ \mathcal{N}=2 $ generalised
  quiver theories},'' \href{http://dx.doi.org/10.1007/JHEP08(2015)085}{{\em
  JHEP} {\bfseries 08} (2015) 085},
  \href{http://arxiv.org/abs/1503.06418}{{\ttfamily arXiv:1503.06418
  [hep-th]}}.

\bibitem{He:2015yoa}
Y.-H. He and J.~McKay, ``{Sporadic and Exceptional},''
  \href{http://arxiv.org/abs/1505.06742}{{\ttfamily arXiv:1505.06742
  [math.AG]}}.

\bibitem{Ashok:2006br}
S.~K. Ashok, F.~Cachazo, and E.~Dell'Aquila, ``{Children's drawings from
  Seiberg-Witten curves},''
  \href{http://dx.doi.org/10.4310/CNTP.2007.v1.n2.a1}{{\em Commun. Num. Theor.
  Phys.} {\bfseries 1} (2007) 237--305},
  \href{http://arxiv.org/abs/hep-th/0611082}{{\ttfamily arXiv:hep-th/0611082}}.

\bibitem{He:2020eva}
Y.-H. He, E.~Hirst, and T.~Peterken, ``{Machine-Learning Dessins d'Enfants:
  Explorations via Modular and Seiberg-Witten Curves},''
  \href{http://arxiv.org/abs/2004.05218}{{\ttfamily arXiv:2004.05218
  [hep-th]}}.

\bibitem{Eguchi:2009gf}
T.~Eguchi and K.~Maruyoshi, ``{Penner Type Matrix Model and Seiberg-Witten
  Theory},'' \href{http://dx.doi.org/10.1007/JHEP02(2010)022}{{\em JHEP}
  {\bfseries 02} (2010) 022}, \href{http://arxiv.org/abs/0911.4797}{{\ttfamily
  arXiv:0911.4797 [hep-th]}}.

\bibitem{Kozcaz:2010af}
C.~Kozcaz, S.~Pasquetti, and N.~Wyllard, ``{A \& B model approaches to surface
  operators and Toda theories},''
  \href{http://dx.doi.org/10.1007/JHEP08(2010)042}{{\em JHEP} {\bfseries 08}
  (2010) 042}, \href{http://arxiv.org/abs/1004.2025}{{\ttfamily arXiv:1004.2025
  [hep-th]}}.

\bibitem{Bershtein:2014qma}
M.~Bershtein and O.~Foda, ``{AGT, Burge pairs and minimal models},''
  \href{http://dx.doi.org/10.1007/JHEP06(2014)177}{{\em JHEP} {\bfseries 06}
  (2014) 177}, \href{http://arxiv.org/abs/1404.7075}{{\ttfamily arXiv:1404.7075
  [hep-th]}}.

\bibitem{Alkalaev:2014sma}
K.~B. Alkalaev and V.~A. Belavin, ``{Conformal blocks of $W_N$ minimal models
  and AGT correspondence},''
  \href{http://dx.doi.org/10.1007/JHEP07(2014)024}{{\em JHEP} {\bfseries 07}
  (2014) 024}, \href{http://arxiv.org/abs/1404.7094}{{\ttfamily arXiv:1404.7094
  [hep-th]}}.

\bibitem{Degiovanni:1994gr}
P.~Degiovanni, ``{Moore and Seiberg equations, topological theories and Galois
  theory},'' {\em Helv. Phys. Acta} {\bfseries 67} (1994) 799--883.

\bibitem{Poland:2018epd}
D.~Poland, S.~Rychkov, and A.~Vichi, ``{The Conformal Bootstrap: Theory,
  Numerical Techniques, and Applications},''
  \href{http://dx.doi.org/10.1103/RevModPhys.91.015002}{{\em Rev. Mod. Phys.}
  {\bfseries 91} (2019) 015002},
  \href{http://arxiv.org/abs/1805.04405}{{\ttfamily arXiv:1805.04405
  [hep-th]}}.

\bibitem{BELAVIN1984333}
``Infinite conformal symmetry in two-dimensional quantum field theory,''
  \href{http://dx.doi.org/https://doi.org/10.1016/0550-3213(84)90052-X}{{\em
  Nuclear Physics B} {\bfseries 241} no.~2, (1984) 333 -- 380}.

\bibitem{Nekrasov:2003rj}
N.~Nekrasov and A.~Okounkov, {\em {Seiberg-Witten theory and random
  partitions}}, vol.~244,
  \href{http://dx.doi.org/10.1007/0-8176-4467-9\_15}{pp.~525--596}.
\newblock 2006.
\newblock \href{http://arxiv.org/abs/hep-th/0306238}{{\ttfamily
  arXiv:hep-th/0306238}}.

\bibitem{rodger2013pedagogical}
R.~J. Rodger, ``A pedagogical introduction to the agt conjecture,'' Master's
  thesis, 2013.

\bibitem{Antoniadis:1993ze}
I.~Antoniadis, E.~Gava, K.~Narain, and T.~Taylor, ``{Topological amplitudes in
  string theory},'' \href{http://dx.doi.org/10.1016/0550-3213(94)90617-3}{{\em
  Nucl. Phys. B} {\bfseries 413} (1994) 162--184},
  \href{http://arxiv.org/abs/hep-th/9307158}{{\ttfamily arXiv:hep-th/9307158}}.

\bibitem{Bershadsky:1993cx}
M.~Bershadsky, S.~Cecotti, H.~Ooguri, and C.~Vafa, ``{Kodaira-Spencer theory of
  gravity and exact results for quantum string amplitudes},''
  \href{http://dx.doi.org/10.1007/BF02099774}{{\em Commun. Math. Phys.}
  {\bfseries 165} (1994) 311--428},
  \href{http://arxiv.org/abs/hep-th/9309140}{{\ttfamily arXiv:hep-th/9309140}}.

\bibitem{Aganagic:2003db}
M.~Aganagic, A.~Klemm, M.~Marino, and C.~Vafa, ``{The Topological vertex},''
  \href{http://dx.doi.org/10.1007/s00220-004-1162-z}{{\em Commun. Math. Phys.}
  {\bfseries 254} (2005) 425--478},
  \href{http://arxiv.org/abs/hep-th/0305132}{{\ttfamily arXiv:hep-th/0305132}}.

\bibitem{Iqbal:2007ii}
A.~Iqbal, C.~Kozcaz, and C.~Vafa, ``{The Refined topological vertex},''
  \href{http://dx.doi.org/10.1088/1126-6708/2009/10/069}{{\em JHEP} {\bfseries
  10} (2009) 069}, \href{http://arxiv.org/abs/hep-th/0701156}{{\ttfamily
  arXiv:hep-th/0701156}}.

\bibitem{Awata:2005fa}
H.~Awata and H.~Kanno, ``{Instanton counting, Macdonald functions and the
  moduli space of D-branes},''
  \href{http://dx.doi.org/10.1088/1126-6708/2005/05/039}{{\em JHEP} {\bfseries
  05} (2005) 039}, \href{http://arxiv.org/abs/hep-th/0502061}{{\ttfamily
  arXiv:hep-th/0502061}}.

\bibitem{Awata:2008ed}
H.~Awata and H.~Kanno, ``{Refined BPS state counting from Nekrasov's formula
  and Macdonald functions},''
  \href{http://dx.doi.org/10.1142/S0217751X09043006}{{\em Int. J. Mod. Phys. A}
  {\bfseries 24} (2009) 2253--2306},
  \href{http://arxiv.org/abs/0805.0191}{{\ttfamily arXiv:0805.0191 [hep-th]}}.

\bibitem{Taki:2007dh}
M.~Taki, ``{Refined Topological Vertex and Instanton Counting},''
  \href{http://dx.doi.org/10.1088/1126-6708/2008/03/048}{{\em JHEP} {\bfseries
  03} (2008) 048}, \href{http://arxiv.org/abs/0710.1776}{{\ttfamily
  arXiv:0710.1776 [hep-th]}}.

\bibitem{Bao:2011rc}
L.~Bao, E.~Pomoni, M.~Taki, and F.~Yagi, ``{M5-Branes, Toric Diagrams and Gauge
  Theory Duality},'' \href{http://dx.doi.org/10.1007/JHEP04(2012)105}{{\em
  JHEP} {\bfseries 04} (2012) 105},
  \href{http://arxiv.org/abs/1112.5228}{{\ttfamily arXiv:1112.5228 [hep-th]}}.

\bibitem{Bao:2013pwa}
L.~Bao, V.~Mitev, E.~Pomoni, M.~Taki, and F.~Yagi, ``{Non-Lagrangian Theories
  from Brane Junctions},''
  \href{http://dx.doi.org/10.1007/JHEP01(2014)175}{{\em JHEP} {\bfseries 01}
  (2014) 175}, \href{http://arxiv.org/abs/1310.3841}{{\ttfamily arXiv:1310.3841
  [hep-th]}}.

\bibitem{Foda:2015ana}
O.~Foda and J.-F. Wu, ``{From topological strings to minimal models},''
  \href{http://dx.doi.org/10.1007/JHEP07(2015)136}{{\em JHEP} {\bfseries 07}
  (2015) 136}, \href{http://arxiv.org/abs/1504.01925}{{\ttfamily
  arXiv:1504.01925 [hep-th]}}.

\bibitem{Foda:2017tnv}
O.~Foda and J.-F. Wu, ``{A Macdonald refined topological vertex},''
  \href{http://dx.doi.org/10.1088/1751-8121/aa7605}{{\em J. Phys. A} {\bfseries
  50} no.~29, (2017) 294003}, \href{http://arxiv.org/abs/1701.08541}{{\ttfamily
  arXiv:1701.08541 [hep-th]}}.

\bibitem{Hayashi:2016abm}
H.~Hayashi, S.-S. Kim, K.~Lee, and F.~Yagi, ``{Equivalence of several
  descriptions for 6d SCFT},''
  \href{http://dx.doi.org/10.1007/JHEP01(2017)093}{{\em JHEP} {\bfseries 01}
  (2017) 093}, \href{http://arxiv.org/abs/1607.07786}{{\ttfamily
  arXiv:1607.07786 [hep-th]}}.

\bibitem{Kim:2015jba}
S.-S. Kim, M.~Taki, and F.~Yagi, ``{Tao Probing the End of the World},''
  \href{http://dx.doi.org/10.1093/ptep/ptv108}{{\em PTEP} {\bfseries 2015}
  no.~8, (2015) 083B02}, \href{http://arxiv.org/abs/1504.03672}{{\ttfamily
  arXiv:1504.03672 [hep-th]}}.

\bibitem{Hwang:2014uwa}
C.~Hwang, J.~Kim, S.~Kim, and J.~Park, ``{General instanton counting and 5d
  SCFT},'' \href{http://dx.doi.org/10.1007/JHEP07(2015)063}{{\em JHEP}
  {\bfseries 07} (2015) 063}, \href{http://arxiv.org/abs/1406.6793}{{\ttfamily
  arXiv:1406.6793 [hep-th]}}. [Addendum: JHEP 04, 094 (2016)].

\bibitem{Hayashi:2013qwa}
H.~Hayashi, H.-C. Kim, and T.~Nishinaka, ``{Topological strings and 5d $T_N$
  partition functions},'' \href{http://dx.doi.org/10.1007/JHEP06(2014)014}{{\em
  JHEP} {\bfseries 06} (2014) 014},
  \href{http://arxiv.org/abs/1310.3854}{{\ttfamily arXiv:1310.3854 [hep-th]}}.

\bibitem{Bergman:2013ala}
O.~Bergman, D.~Rodriguez-G\'omez, and G.~Zafrir, ``{Discrete $\theta$ and the
  5d superconformal index},''
  \href{http://dx.doi.org/10.1007/JHEP01(2014)079}{{\em JHEP} {\bfseries 01}
  (2014) 079}, \href{http://arxiv.org/abs/1310.2150}{{\ttfamily arXiv:1310.2150
  [hep-th]}}.

\bibitem{Bergman:2013aca}
O.~Bergman, D.~Rodriguez-G\'omez, and G.~Zafrir, ``{5-Brane Webs, Symmetry
  Enhancement, and Duality in 5d Supersymmetric Gauge Theory},''
  \href{http://dx.doi.org/10.1007/JHEP03(2014)112}{{\em JHEP} {\bfseries 03}
  (2014) 112}, \href{http://arxiv.org/abs/1311.4199}{{\ttfamily arXiv:1311.4199
  [hep-th]}}.

\bibitem{Tachikawa:2013kta}
Y.~Tachikawa, \href{http://dx.doi.org/10.1007/978-3-319-08822-8}{{\em {N=2
  supersymmetric dynamics for pedestrians}}}, vol.~890.
\newblock 2014.
\newblock \href{http://arxiv.org/abs/1312.2684}{{\ttfamily arXiv:1312.2684
  [hep-th]}}.

\bibitem{Leung:1997tw}
N.~C. Leung and C.~Vafa, ``{Branes and toric geometry},''
  \href{http://dx.doi.org/10.4310/ATMP.1998.v2.n1.a4}{{\em Adv. Theor. Math.
  Phys.} {\bfseries 2} (1998) 91--118},
  \href{http://arxiv.org/abs/hep-th/9711013}{{\ttfamily arXiv:hep-th/9711013}}.

\bibitem{Gorsky:1997mw}
A.~Gorsky, S.~Gukov, and A.~Mironov, ``{SUSY field theories, integrable systems
  and their stringy / brane origin. 2.},''
  \href{http://dx.doi.org/10.1016/S0550-3213(98)00106-0}{{\em Nucl. Phys. B}
  {\bfseries 518} (1998) 689--713},
  \href{http://arxiv.org/abs/hep-th/9710239}{{\ttfamily arXiv:hep-th/9710239}}.

\bibitem{Aharony:1997bh}
O.~Aharony, A.~Hanany, and B.~Kol, ``{Webs of (p,q) five-branes,
  five-dimensional field theories and grid diagrams},''
  \href{http://dx.doi.org/10.1088/1126-6708/1998/01/002}{{\em JHEP} {\bfseries
  01} (1998) 002}, \href{http://arxiv.org/abs/hep-th/9710116}{{\ttfamily
  arXiv:hep-th/9710116}}.

\bibitem{Kim:2014nqa}
S.-S. Kim and F.~Yagi, ``{5d E$_{n}$ Seiberg-Witten curve via toric-like
  diagram},'' \href{http://dx.doi.org/10.1007/JHEP06(2015)082}{{\em JHEP}
  {\bfseries 06} (2015) 082}, \href{http://arxiv.org/abs/1411.7903}{{\ttfamily
  arXiv:1411.7903 [hep-th]}}.

\bibitem{ggd2011}
E.~Girondo and G.~González-Diez,
  \href{http://dx.doi.org/10.1017/CBO9781139048910}{{\em Introduction to
  Compact Riemann Surfaces and Dessins d’Enfants}}.
\newblock London Mathematical Society Student Texts. Cambridge University
  Press, 2011.

\bibitem{10.1112/blms/28.6.561}
G.~Jones and D.~Singerman, ``{Belyi Functions, Hypermaps and Galois Groups},''
  \href{http://dx.doi.org/10.1112/blms/28.6.561}{{\em Bulletin of the London
  Mathematical Society} {\bfseries 28} no.~6, (11, 1996) 561--590}.

\bibitem{sketch}
A.~Grothendieck, {\em Esquisse d'un Programme}.
\newblock 1984.

\bibitem{mp}
M.~Mulase and M.~Penkava, ``Ribbon graphs, quadratic differentials on riemann
  surfaces, and algebraic curves defined over $\bar q$,''
  \href{http://dx.doi.org/10.4310/AJM.1998.v2.n4.a11}{{\em Asian J. Math.}
  {\bfseries 2} (11, 1998) }.

\bibitem{MS}
J.~McKay and A.~Sebbar, ``{J-invariants of arithmetic semistable elliptic
  surfaces and graphs},'' {\em CRM Proceedings and Lecture Notes} (2001)
  119--130.

\bibitem{francesco2012conformal}
P.~Francesco, P.~Mathieu, and D.~S{\'e}n{\'e}chal, {\em Conformal field
  theory}.
\newblock Springer Science \& Business Media, 2012.

\bibitem{Huang:2013yta}
M.-X. Huang, A.~Klemm, and M.~Poretschkin, ``{Refined stable pair invariants
  for E-, M- and $[p, q]$-strings},''
  \href{http://dx.doi.org/10.1007/JHEP11(2013)112}{{\em JHEP} {\bfseries 11}
  (2013) 112}, \href{http://arxiv.org/abs/1308.0619}{{\ttfamily arXiv:1308.0619
  [hep-th]}}.

\bibitem{Huang:2010kf}
M.-X. Huang and A.~Klemm, ``{Direct integration for general $\Omega$
  backgrounds},'' \href{http://dx.doi.org/10.4310/ATMP.2012.v16.n3.a2}{{\em
  Adv. Theor. Math. Phys.} {\bfseries 16} no.~3, (2012) 805--849},
  \href{http://arxiv.org/abs/1009.1126}{{\ttfamily arXiv:1009.1126 [hep-th]}}.

\bibitem{Huang:2011qx}
M.-X. Huang, A.-K. Kashani-Poor, and A.~Klemm, ``{The $\Omega$ deformed B-model
  for rigid $\mathcal{N}=2$ theories},''
  \href{http://dx.doi.org/10.1007/s00023-012-0192-x}{{\em Annales Henri
  Poincare} {\bfseries 14} (2013) 425--497},
  \href{http://arxiv.org/abs/1109.5728}{{\ttfamily arXiv:1109.5728 [hep-th]}}.

\bibitem{Witten:1997sc}
E.~Witten, ``{Solutions of four-dimensional field theories via M theory},''
  \href{http://dx.doi.org/10.1016/S0550-3213(97)00416-1}{{\em Nucl. Phys. B}
  {\bfseries 500} (1997) 3--42},
  \href{http://arxiv.org/abs/hep-th/9703166}{{\ttfamily arXiv:hep-th/9703166}}.

\bibitem{Becker:1995kb}
K.~Becker, M.~Becker, and A.~Strominger, ``{Five-branes, membranes and
  nonperturbative string theory},''
  \href{http://dx.doi.org/10.1016/0550-3213(95)00487-1}{{\em Nucl. Phys. B}
  {\bfseries 456} (1995) 130--152},
  \href{http://arxiv.org/abs/hep-th/9507158}{{\ttfamily arXiv:hep-th/9507158}}.

\bibitem{sagemath}
{The Sage Developers}, {\em {S}ageMath, the {S}age {M}athematics {S}oftware
  {S}ystem ({V}ersion 9.1)}, 2020.
\newblock {\tt https://www.sagemath.org}.

\bibitem{Connell1999}
I.~Connell, ``{Elliptic Curve Handbook},''.
  \url{https://webs.ucm.es/BUCM/mat/doc8354.pdf}.

\end{thebibliography}\endgroup

\end{document}